\documentclass[pra,showpacs]{revtex4}
\usepackage{amsfonts}
\usepackage{amssymb}
\usepackage{amsmath}
\usepackage{bm}
\usepackage[dvips]{graphicx}
\usepackage{graphicx}
\usepackage[ansinew]{inputenc}
\usepackage{array}
\usepackage{color}
\usepackage{amsmath}
\usepackage{amsxtra}
\usepackage{amstext}
\usepackage{amssymb}
\usepackage{latexsym}
\usepackage{dsfont}

\begin{document}

\title{Lie transformation method on quantum state evolution
of a general time-dependent driven and damped parametric oscillator}
\author{Lin Zhang}
\affiliation{School of physics and information technology, Shaanxi Normal University,
Xi'an 710119, China}
\author{Weiping Zhang}
\affiliation{Department of Physics, East China Normal University, Shanghai 200062, China}
%\date{}
%\today

\begin{abstract}
A variety of dynamics in nature and society can be approximately treated as a
driven and damped parametric oscillator. An intensive investigation of this
time-dependent model from an algebraic point of view provides a consistent
method to resolve the classical dynamics and the quantum evolution in order
to understand the time-dependent phenomena that occur not only in the macroscopic
classical scale for the synchronized behaviors but also in the microscopic quantum
scale for a coherent state evolution. By using a Floquet U-transformation on a
general time-dependent quadratic Hamiltonian, we exactly solve the dynamic
behaviors of a driven and damped parametric oscillator to obtain the optimal
solutions by means of invariant parameters of $K$s to combine with Lewis-Riesenfeld
invariant method. This approach can discriminate the external dynamics from the
internal evolution of a wave packet by producing independent parametric equations that dramatically
facilitate the parametric control on the quantum state evolution in a dissipative
system. In order to show the advantages of this method, several time-dependent
models proposed in the quantum control field are analyzed in details.
\end{abstract}

\maketitle

\section{Introduction}

The classical dynamics of a driven and damped harmonic oscillator (DDHO) is a
fundamental problem discussed in many textbooks and all its behaviors are well
known \cite{Fitzpatrick,Thomsen}. However, this simple model can be used to
understand the dynamics of more general systems embedded in different dissipative
environments and driven by arbitrary external forces. A more complicated system
possesses a richer frequency structure of internal dynamics but only one frequency
window plays the dominant role in a certain parametric region. The external driving
force often maintains an energy input to stimulate or to control the dynamics of the system.
A more complex driving force often induces similar dynamics as that does by a simple
periodic driving due to the limited bandwidth of a responsive window. Up to now,
some behaviors of the damped harmonic oscillator driven by a simple periodic force
still exhibit exciting dynamics, especially due to the rapidly developing fields
of the optomechanical systems \cite{Marqu} and the quantum shortcut control problem \cite{Chen}.
In this paper, we want to reconsider this model in a unified framework from both
classical and quantum mechanical points of view to exactly solve the dynamical equation
for the sake of dynamic controls based on a general time-dependent model. In many
control problems, DDHO is the simplest but fundamental model to reveal the main properties
of a controlled system. Although the dynamics of a classically forced and damped harmonic
oscillator will finally follow the external driving force and its behavior is completely
deterministic, its transient behavior to a final state or its dynamical response to
an external force is still an interesting topic to be explored in the quantum regime
from a control perspective. For any classical or quantum control problems, the controlled
systems are definitely nonconservative and the corresponding theories for an arbitrary time-dependent
Hamiltonian, which should exhibit rich and novel behaviors beyond the perturbation theory
and the adiabatic theory, are still under development \cite{Tannor}. In this paper,
we will extensively investigate a general time-dependent model, a driven and damped
parametric oscillator (DDPO), in a unified view of Lie transformation method to obtain
not only the classical and quantum dynamics, but also the parametric connections to the
Lewis-Riesenfeld invariant method. In order to provide a complete description of this model,
we first give a brief review on the classical dynamics of DDHO to lay down a basic knowledge
for the classical motion of DDPO, and then completely solve the problem in the quantum regime
by using a time-dependent transformation method based on Lie algebra.

\section{Classical dynamics of DDHO}

In order to identify dynamic differences between classical and quantum behaviors, we
first briefly review the main properties of the classical dynamics of a DDHO.
The equation of motion to govern a classical DDHO with constant parameters is
($m=1$)
\begin{equation}
\ddot{x}(t)+2\gamma \dot{x}(t)+\omega _{0}^{2}x(t)=F\left( t\right) ,
\label{Geq}
\end{equation}%
where $\omega _{0}$ is the internal frequency, $\gamma$ is the damping rate related
to the $Q$ factor by $Q=\omega _{0}/2\gamma $, and $F(t)$ is the external driving force.
Eq.(\ref{Geq}) is a fundamental equation to understand a general damped and driven
oscillators emerging in many physical systems. Besides the traditional weak vibration
in a mechanical system, one important model in the plasma, called Lorentz oscillator,
is used to describe the motion of a charged particle (trapped ion) driven by the
electromagnetic fields. The other typical model is about the electronic current in
the RLC circuits or the electromagnetic field in a finesse cavity pumping by an input
field. Surely, many other phenomena in chemical, biological or economic fields can also
be explained or understood by DDHO model \cite{Fitzpatrick,Thomsen}.

\begin{figure}[t]
\begin{center}
\includegraphics[width=0.32 \textwidth]{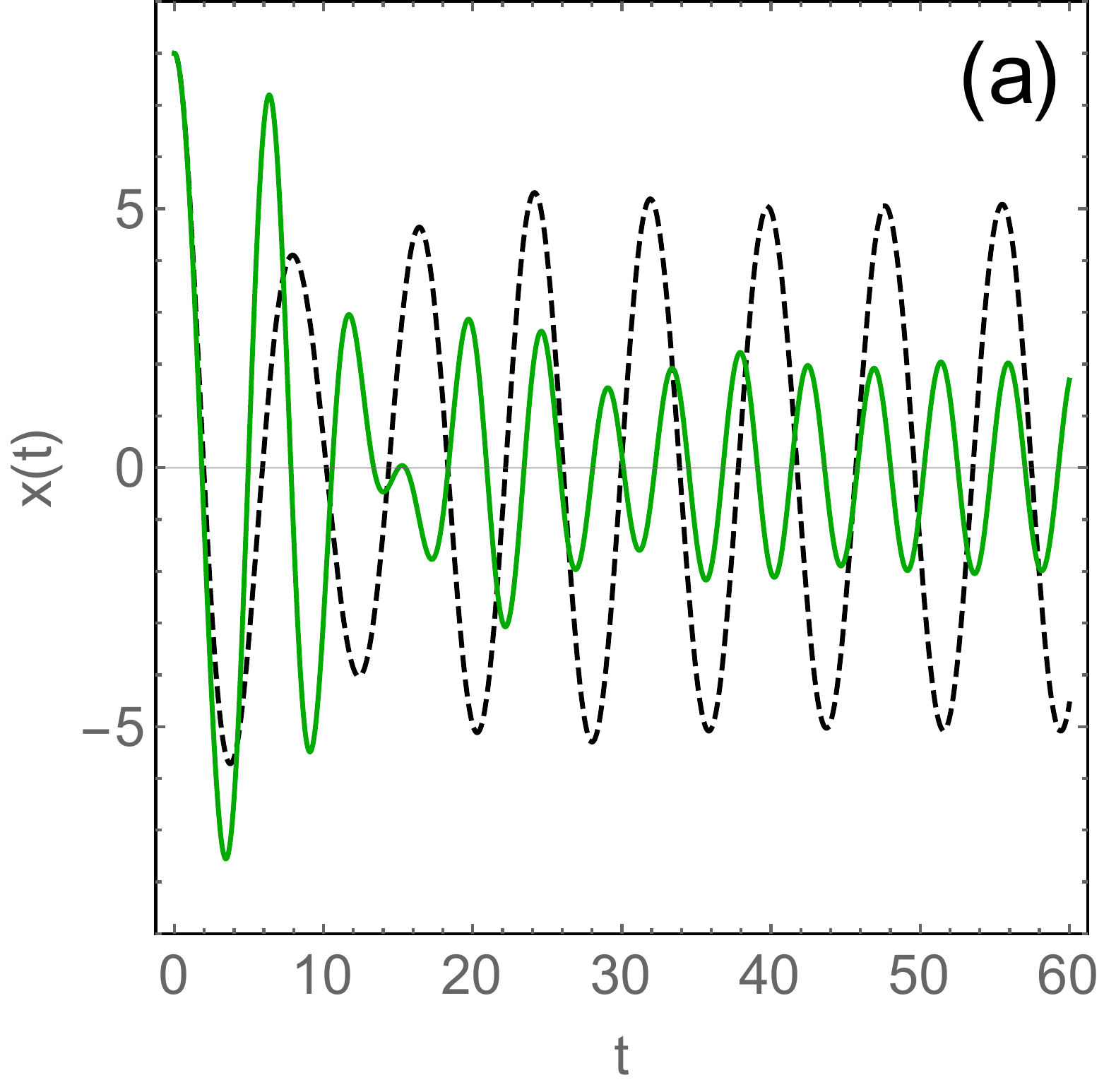}%
\includegraphics[width=0.32 \textwidth]{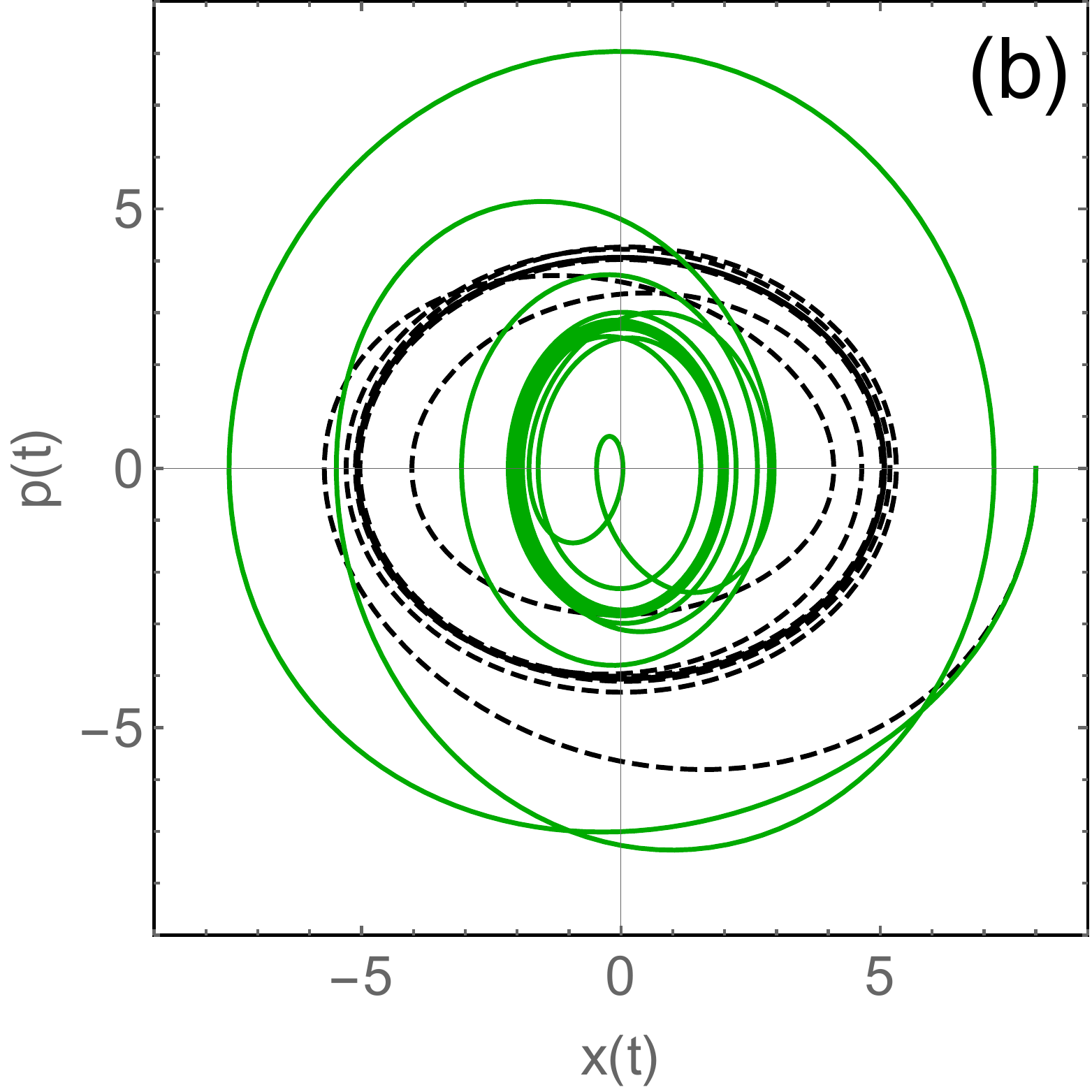}%
\includegraphics[width=0.33 \textwidth]{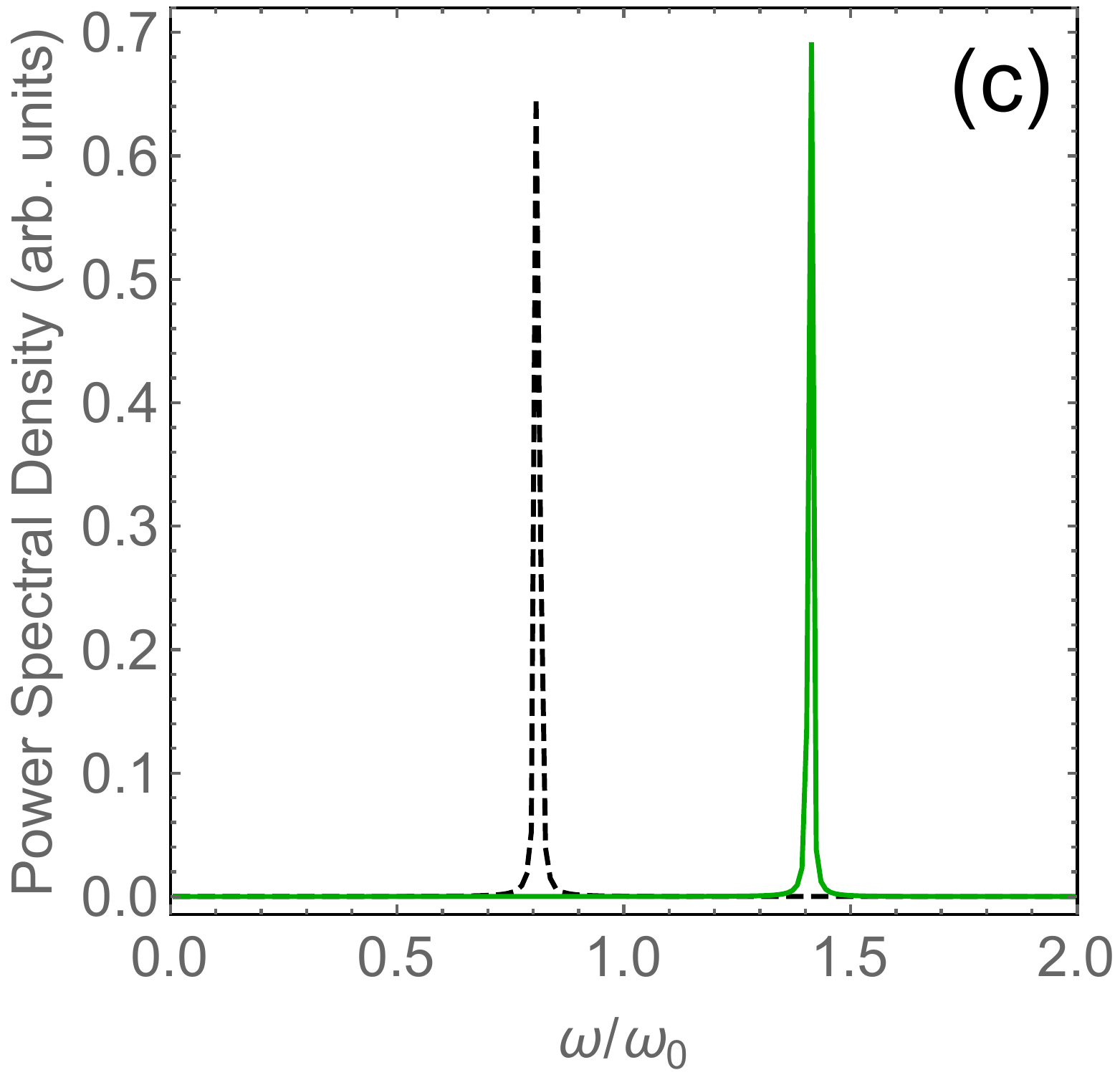}%
\end{center}
\caption{(a) Two typical motions, (b) the corresponding phase-space trajectories and (c) power
spectra of the damped harmonic oscillator driven by a red-detuned force ($\Omega=0.8\omega_0$,
the dashed line) and a blue-detuned force ($\Omega=1.4\omega_0$, the solid line).
The other parameters are $\gamma=0.1, F_{0}=2$ scaled by $\omega _{0}$ and the initial conditions are $x(0)=8, p(0)=0$.}
\label{figure1}
\end{figure}

The dynamics of a damped oscillator under a driving force exhibits a very important
phenomenon: resonance, and many behaviors in this world can be explained or controlled
by the resonant effect. As most of the driving or control signals can be decomposed into
harmonic components, most studies of DDHO focus on a harmonic driving force and
Eq.(\ref{Geq}) simplifies to%
\begin{equation}
\ddot{x}+2\gamma \dot{x}+\omega _{0}^{2}x=F_{0}\cos \left( \Omega t+\phi
\right) ,  \label{FDh}
\end{equation}%
where $F_{0}$, $\Omega $ and $\phi $ are the driving strength, frequency and
initial phase, respectively. Eq.(\ref{FDh}) can be analytically solved and
Fig.\ref{figure1}(a) demonstrates two typical motions of the oscillator driven by
two different forces with the same initial conditions. After a transient motion,
the oscillator finally locks itself to the driving force as shown by two closed orbits
in the phase space (Fig.\ref{figure1}(b)) and the corresponding frequency spectra are
given in Fig.\ref{figure1}(c). That the final motion of a driven oscillator
follows the pace of driving force is actually a universal ``synchronized" dynamics for a damped
system induced by a robust external driving. Although the force-induced oscillation will
disappear and the oscillator will return back immediately to a damping oscillation when
the driving force is removed, this kind of synchronized motion is still a universal
phenomenon induced by an active driving source. Strictly, the force-induced oscillation
is not a self-sustained oscillation and thus the final closed trajectory in the phase space
is not a limit circle but a passive orbit of driving force. However this
synchronized behavior induced by a robust external force occurs in many low-energy dynamic
processes, especially in a classically driven nanomechianical system.

The analytical solution of Eq.(\ref{FDh}) is%
\begin{equation}
x\left( t\right) =x_{H}\left( t\right) +x_{I}\left( t\right) ,  \label{Sfdh}
\end{equation}%
where $x_{H}(t)$, the general solution to the homogeneous equation of Eq.(\ref{FDh}), is%
\begin{equation*}
x_{H}\left( t\right) =A_{1}e^{-\gamma t}\cos \left( \omega t+\phi
_{1}\right)
\end{equation*}%
and $x_I(t)$ is a particular solution. The damping shifted frequency $\omega$ is defined
by $\omega =\sqrt{\omega _{0}^{2}-\gamma ^{2}}$. Because $x_{H}(t)$ will damp out,
the final dynamics is due to $x_{I}(t)$.

A normal form of $x_{I}\left( t\right) $ is
\begin{equation}
x_{I}\left( t\right) =A_{2}\cos \left( \Omega t+\phi _{2}\right) ,
\label{x1}
\end{equation}%
where the amplitude and the relative phase are
\begin{equation}
A_{2}=\frac{F_{0}}{\sqrt{\left( 2\gamma \Omega \right) ^{2}+\left( \omega
_{0}^{2}-\Omega ^{2}\right) ^{2}}},\quad \tan \left( \phi _{2}-\phi \right) =%
-\frac{2\gamma \Omega }{\omega _{0}^{2}-\Omega ^{2}}.
\label{Ap}
\end{equation}%
Eq.(\ref{Sfdh}) reveals that the final frequency of DDHO is only
determined by the external driving force because its internal oscillation
will disappear due to the damping. While the final amplitude (energy) depends
on all the parameters of the system, such as the intrinsic frequency, the damping
rate, the driving strength and frequency. Surely, the final oscillation
doesn't depend on the initial conditions.
\begin{figure}[t]
\begin{center}
\includegraphics[width=0.32 \textwidth]{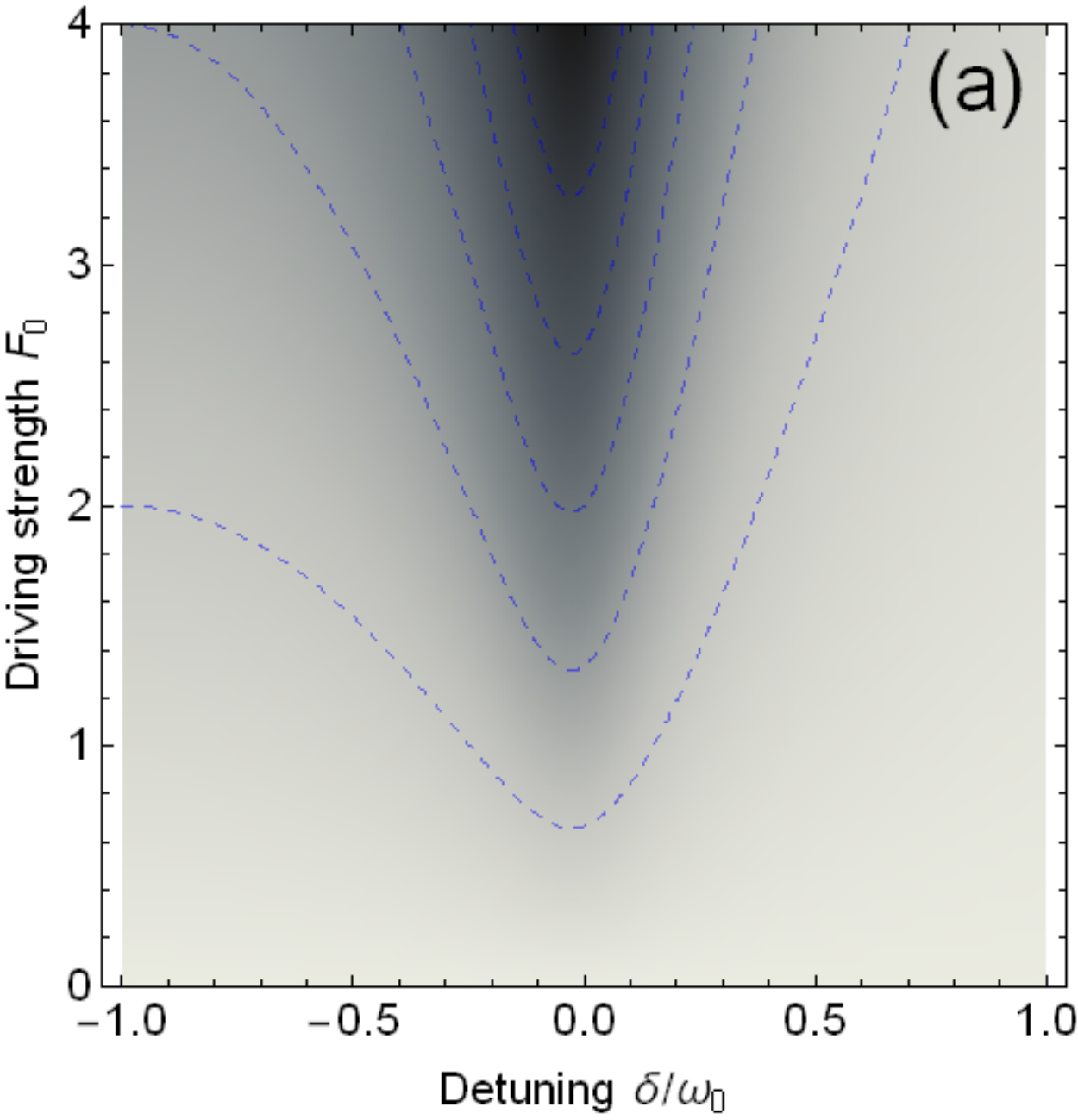}%
\includegraphics[width=0.32 \textwidth]{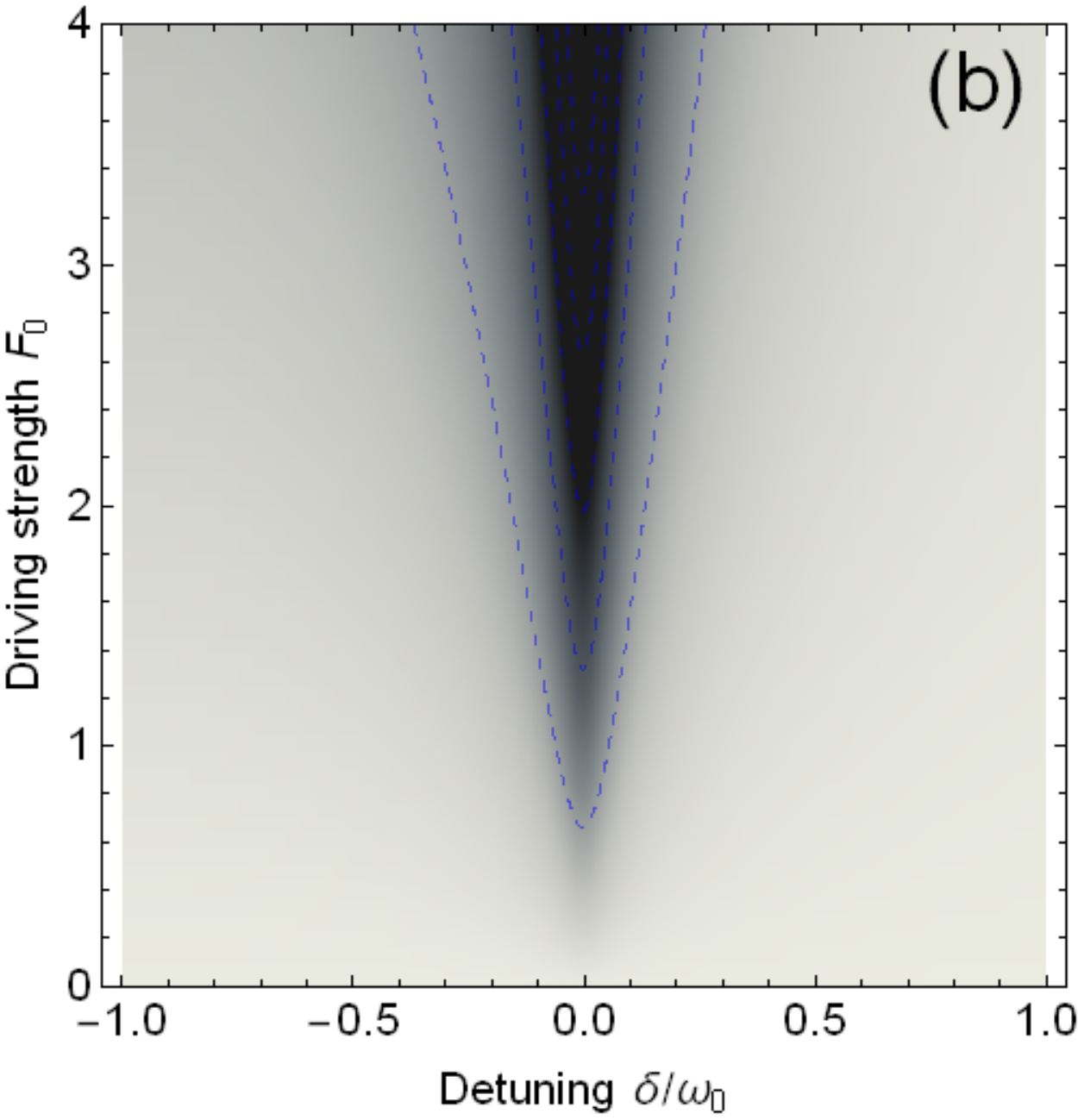}%
\includegraphics[width=0.32 \textwidth]{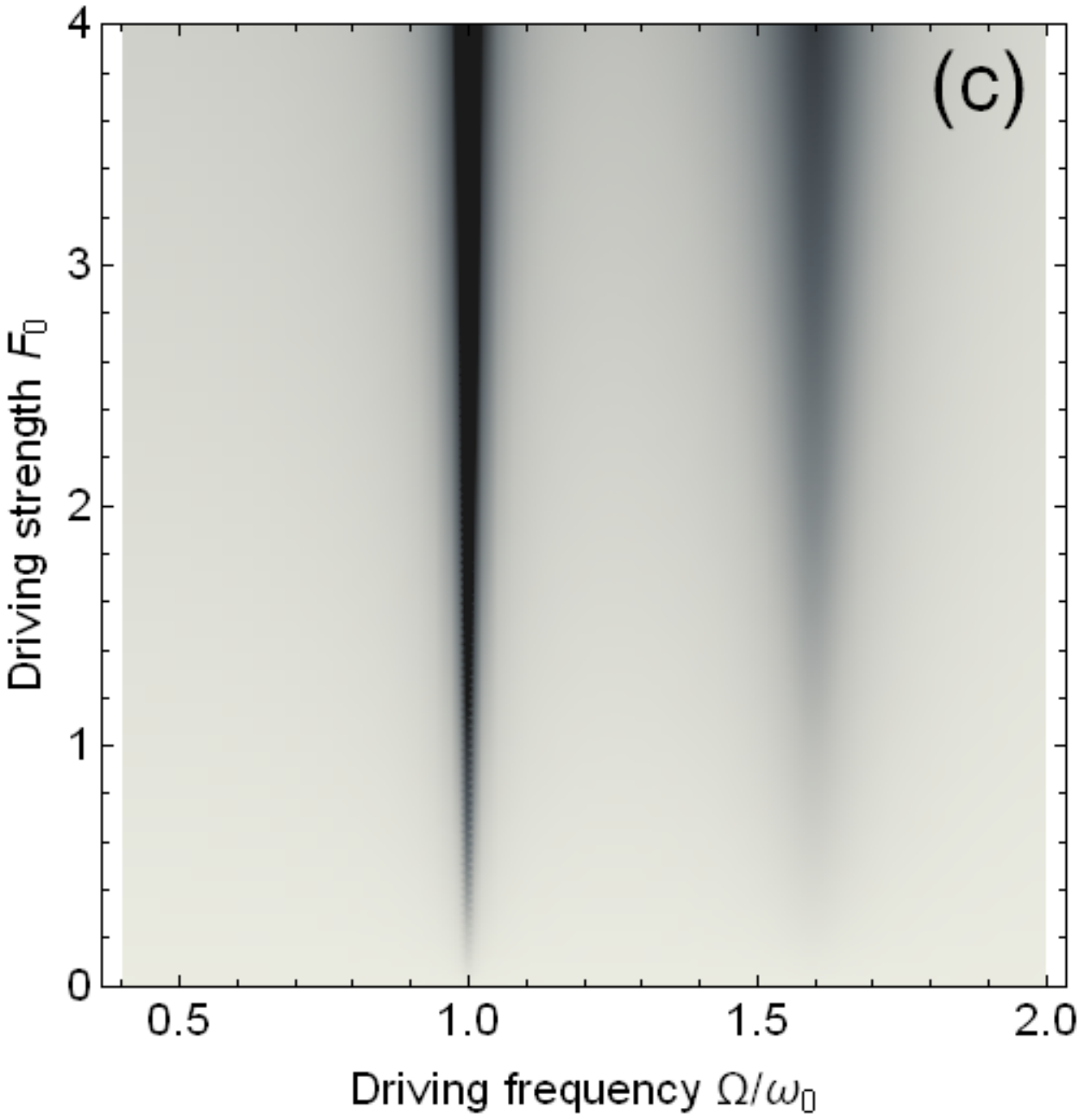}%
\end{center}
\caption{The amplitude responses of DDHO driven by
the external forces with (a) $Q=3$ and (b) $Q=10$; (c) The response of an anharmonic
oscillator with two resonant windows at $\Omega=\omega_0$ ($Q=100$) and
$1.6\omega_0$ (Q=10). The dashed contour lines are samples for reference.}
\label{figure2}
\end{figure}

Now let us take a closer look at the most important phenomenon in a forced system:
resonance. Simply, the response of an oscillator to a driving force can be measured
by its final amplitude. For $\Omega >0$, the maximum amplitude occurs at a resonant
driving frequency of $\Omega _{R}=\sqrt{\omega _{0}^{2}-2\gamma ^{2}}$. Fig.\ref{figure2}
shows the amplitude responses with respect to the driving strength $F_0$
(power) and the frequency detuning $\delta=(\Omega-\omega_0)/\gamma$. The V-shaped profiles
displayed in Fig.\ref{figure2}(a) and Fig.\ref{figure2}(b) reveal a major characteristic
of resonance which can also be detected in a more complicated system. Fig.\ref{figure2}(c)
displays a responsive graph of an anharmonic oscillator with two independent resonant
windows of different widths, indicating two well-separated (without coupling) internal
modes of the system.

\begin{figure}[t]
\begin{center}
\includegraphics[width=0.32 \textwidth]{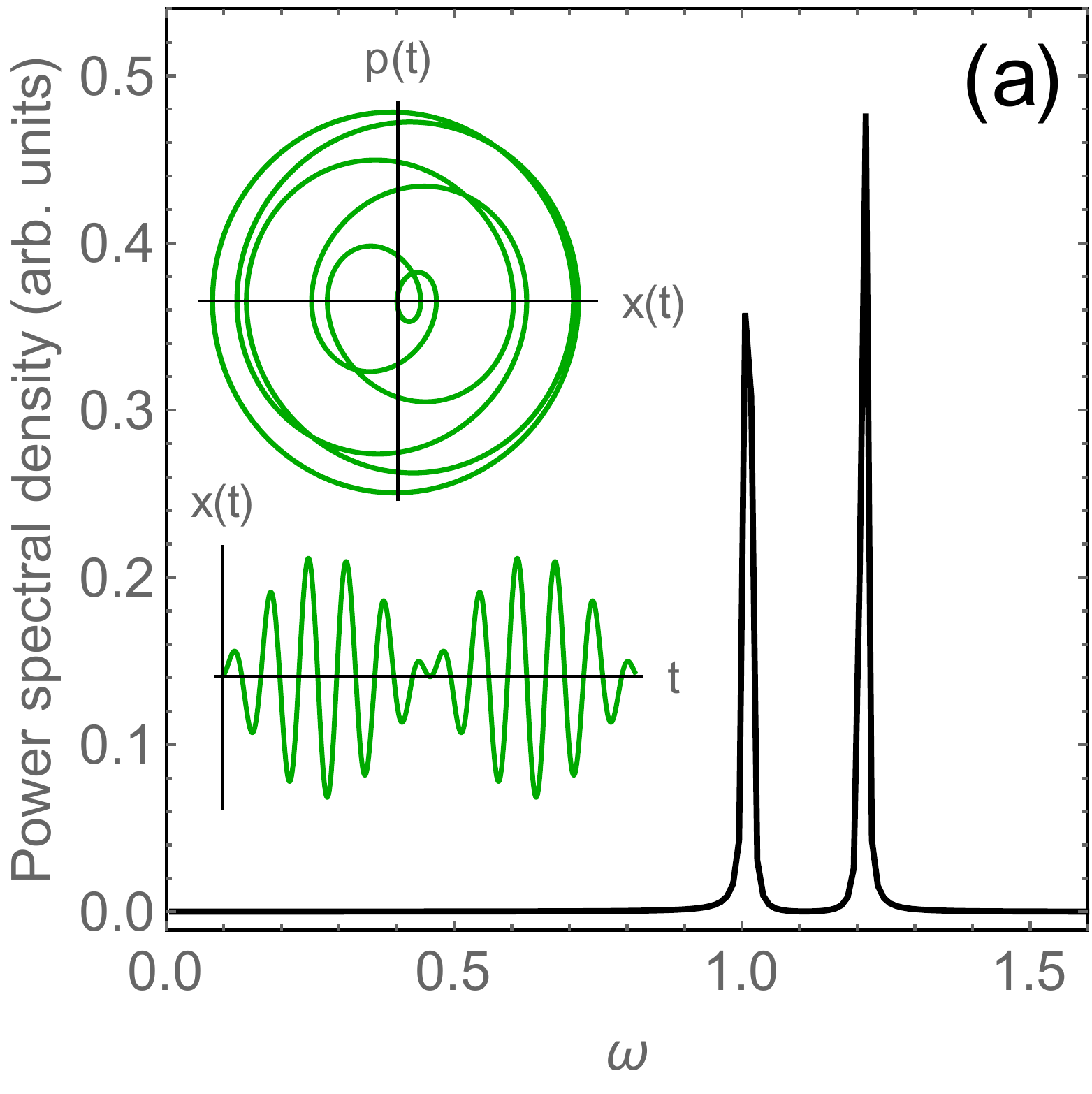}%
\includegraphics[width=0.32 \textwidth]{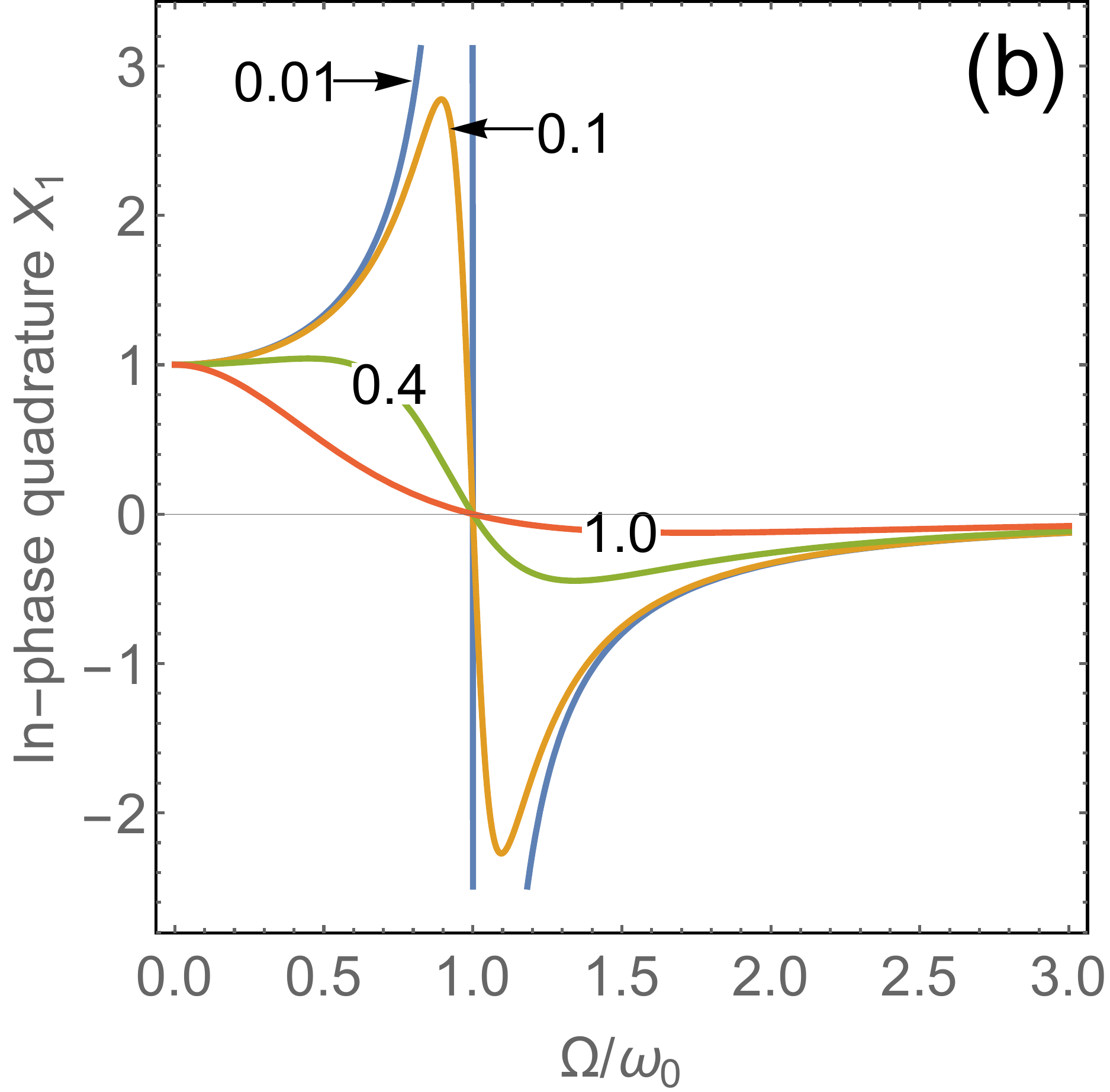}%
\includegraphics[width=0.32 \textwidth]{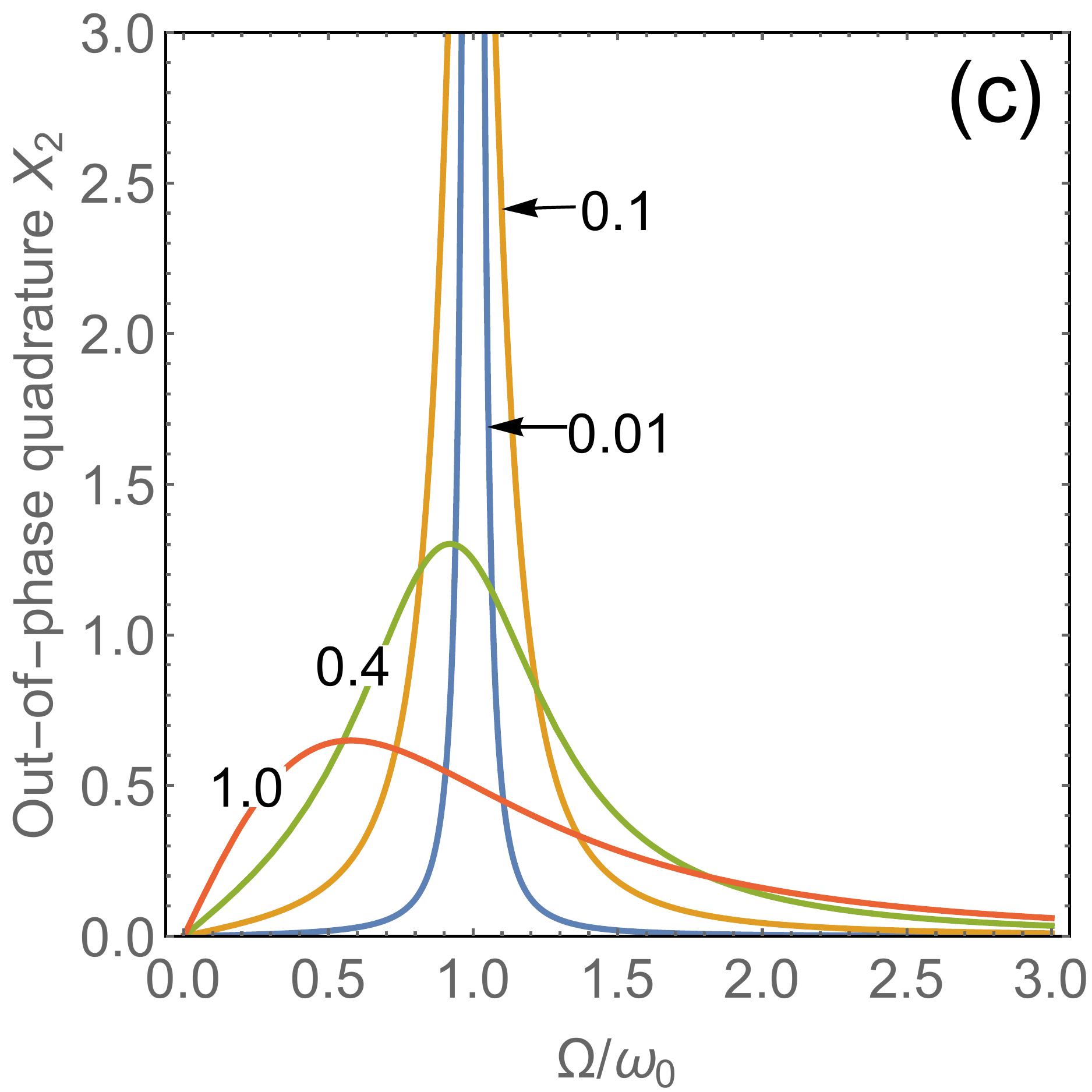}%
\end{center}
\caption{(a) The power spectrum of the driven harmonic oscillator
with an extremely small damping rate $\gamma=10^{-5}$. Inset: the
phase trajectory (upper) and the beating behavior (lower) of
$x(t)$. The parameters are $F_0=1, \Omega=1.2$ and $x(0)=0, p(0)=0.2$.
The quadratures of (b) $X_1$ and (c) $X_2$ versus the driving frequency
with different damping rates labeled by the numbers in the figures. The
force strength is $F_0=1$. }
\label{figure3}
\end{figure}

In order to clarify the important role of damping in the classical dynamics, we calculate
a long-time behavior of DDHO with a very small damping rate ($\gamma=10^{-5}$) in
Fig.\ref{figure3}(a). The upper inset of Fig.\ref{figure3}(a) demonstrates a nearly closed
phase trajectory which never seems to converge towards a closed orbit as shown in
Fig.\ref{figure1}(b). The double-peak spectrum of $x(t)$ (compared with Fig.\ref{figure1}(c))
indicates a long-time survival of the internal oscillation with a shifted frequency
of $\omega\approx \omega_0$.
%--Figure--
If the damping is weak, the oscillator can not quickly stabilize its motion to
the driving trajectory due to its intrinsic oscillation. In this case, the intrinsic
oscillation and the force-induced oscillation will mix together by
\begin{equation*}
x\left( t\right) =A_{1}\cos \left( \omega_0 t+\phi _{1}\right) +A_{2}\cos
\left( \Omega t+\phi _{2}\right) ,
\end{equation*}%
which leads to a beating behavior of $x(t)$ within the resonant window of
$\Omega \sim \omega_0 $ (see the lower inset of Fig.\ref{figure3}(a)). Therefore,
the classical damping is a very important reason for a driven harmonic oscillator
to synchronize with the driving force.

An equivalent form of $x_{I}(t)$ is adopted in some occasions as%
\begin{equation}
x_{I}\left( t\right) =X_{1}\cos \left( \Omega t\right) +X_{2}\sin \left(
\Omega t\right) ,  \label{x2}
\end{equation}%
where
\begin{equation*}
\left(
\begin{array}{c}
X_{1} \\
X_{2}%
\end{array}%
\right) =\left[
\begin{array}{cc}
\frac{\omega _{0}^{2}-\Omega ^{2}}{\left( \omega _{0}^{2}-\Omega ^{2}\right)
^{2}+\left( 2\gamma \Omega \right) ^{2}} & \frac{2\gamma \Omega }{\left(
\omega _{0}^{2}-\Omega ^{2}\right) ^{2}+\left( 2\gamma \Omega \right) ^{2}}
\\
\frac{2\gamma \Omega }{\left( \omega _{0}^{2}-\Omega ^{2}\right) ^{2}+\left(
2\gamma \Omega \right) ^{2}} & -\frac{\omega _{0}^{2}-\Omega ^{2}}{\left(
\omega _{0}^{2}-\Omega ^{2}\right) ^{2}+\left( 2\gamma \Omega \right) ^{2}}%
\end{array}%
\right] \left[
\begin{array}{c}
F_{0}\cos\phi \\
F_{0}\sin\phi%
\end{array}%
\right].
\end{equation*}%
$X_{1}$ and $X_{2}$ are the amplitudes of in-phase and out-of-phase quadratures
with respect to force components of $F_{0}\cos \phi $ and $F_{0}\sin \phi $, respectively.
The responsive curves for $X_1$ and $X_2$ as a function of driving frequency with different
damping rates are shown in Fig.\ref{figure3}(b) and (c), respectively.
When the driving frequency passes through the intrinsic frequency
of an oscillator, the in-phase quadrature $X_1$ undergoes a dynamic
transition from positive to negative oscillation, and the out-of-phase quadrature $X_2$ exhibits
a positive resonant nature. This indicates that, within a resonant window, the final phase of the forced oscillation,
$\phi_2 =\arctan (X_{2}/X_{1})$, for a high-$Q$ oscillator is very sensitive to the frequency of
the driving force. If the driving force is blue detuned ($\Omega >\omega _{0}$), the motion of the
oscillator is lagged and tends to keep up with the driving force, and if the force is red detuned
($\Omega <\omega _{0}$), the motion of the harmonic oscillator will be pulled back by the driving
force and tends to be out of the driving phase.

From the above analysis, we know that the classical behavior of DDHO exhibits a resonant
dynamics following Eq.(\ref{FDh}). We should notice that all the classical behaviors of DDHO
are deterministic at a macroscopic scale and include no internal fluctuations
by only introducing an effective damping rate. Including external noises into the system will lead no
new features to the classical dynamics because any external noises can be safely treated as external
driving forces. However, for a quantum system at the microscopic scale, the internal fluctuations
will definitely bring new dynamic features to a system beyond its classical dynamics.

\section{Quantum dynamics of DDPO}

\subsection{A brief note on dissipative quantum system}

Similar to classical systems, the quantum systems are inevitably coupled to their
surrounding environments which definitely lead to energy dissipations. On a classical
time scale, the dissipation can safely be described by phenomenologically introducing
an overall damping rate. However, on the quantum scale, the details of dissipation process are
important and the dynamics of dissipation becomes difficult to deal with (see the review paper \cite{PR,PR2}).
One reason for the trouble of studying a dissipative system in quantum regime is that the
system doesn't admit a standard Hamiltonian for quantization \cite{Musielak,Kanai}.
The traditional theory on this problem is somehow to construct a non-standard time-dependent
Hamiltonian and the quantization of the damped oscillator leads to a dual Bateman Hamiltonian
by including an auxiliary oscillator (time-reversed counterpart) to valid a total standard
Hamiltonian \cite{PR2,Kanai}. As a brief example, we can easily check that the following simple
equation of motion,%
\begin{equation*}
\ddot{x}+2\gamma \dot{x}=0,
\end{equation*}%
does not admit a standard Hamiltonian because no invariant energy (Hamiltonian) of a dissipative
system exists due to energy losses. But a non-standard Hamiltonian can still be developed in the
framework of the Euler-Lagrange equation \cite{Musielak}, such as by using a \emph{time-dependent}
Lagrange function of \cite{Bateman}
\begin{equation*}
\mathcal{L}=e^{2\gamma t}\frac{1}{2}m \dot{x}^2.
\end{equation*}%
Therefore, for a DDHO
described by Eq.(\ref{Geq}), no standard Hamiltonian is available. However, we
can formally construct the time-dependent Hamiltonians from which Eq.(\ref{Geq})
can be derived \cite{Torres}. A well-known Caldirola-Kanai Hamiltonian to describe
the DDHO is \cite{Kanai}
\begin{equation}
H(t)=e^{-2\gamma t}\frac{p^{2}}{2m}+e^{2\gamma t}\left[ \frac{1}{2}m\omega
_{0}^{2}x^{2}-xF\left( t\right) \right] ,  \label{Hdd}
\end{equation}%
which safely recovers the classical Eq.(\ref{Geq}) of%
\begin{equation*}
\ddot{x}+2\gamma \dot{x}+\omega _{0}^{2}x=\frac{F(t)}{m}.
\end{equation*}

Therefore, this treatment is formally correct from a classical point of review
but a complete investigation on a quantum dissipative system should resort to the
open quantum theory by including a continuum reservoir which can be modeled
by an ensemble of harmonic oscillators with infinite degree of freedom \cite%
{Philbin1}. However, the open quantum theory is somehow cumbersome and heavily depends
on the statistical properties of the reservoirs that we should know in advance.
Alternatively, a simple but equivalent method to deal with a damped quantum
system is to use a non-Hermitian effective Hamiltonian as \cite{Eva}%
\begin{equation}
\hat{H}=\hat{H}_{0}-i\hat{\Gamma},  \label{nonH}
\end{equation}%
where $\hat{H}$ is decomposed into a Hermitian part and an anti-Hermitian part
with $\hat{H}_{0}^{\dag }=\hat{H}_{0}$ and $\hat{\Gamma}^{\dag }=\hat{\Gamma}$.
Then the dynamics of a pure state $\left\vert \Psi \right\rangle $ is still
governed by the conventional Schr\"{o}dinger equation of%
\begin{equation*}
i\hbar \frac{\partial }{\partial t}\left\vert \Psi \right\rangle =\hat{H}%
\left\vert \Psi \right\rangle .
\end{equation*}%
However, the Hermitian Hamiltonian leads to a non-unitary state evolution and
the norm of the state is not conserved. One well-known non-Hermitian
Hamiltonian for a damped oscillator is%
\begin{equation}
\hat{H}=\frac{1}{2}\left( \hat{P}^{2}+\hat{X}^{2}\right) -i\frac{\Gamma }{2}%
\left( \hat{X}\hat{P}+\hat{P}\hat{X}\right) =\left( \hat{a}^{\dag }\hat{a}+%
\frac{1}{2}\right) -\frac{\Gamma }{2}\left( \hat{a}^{\dag 2}-\hat{a}%
^{2}\right) ,  \label{nonH1}
\end{equation}%
and which can recover the classical equation of motion for a damped harmonic
oscillator as%
\begin{eqnarray*}
\dot{x} =p-\Gamma x, \quad
\dot{p} =-x-\Gamma p,
\end{eqnarray*}%
where $x$ and $p$ are defined by $\alpha =\left( x+ip\right) /\sqrt{2}$ for
a coherent state $\left\vert \alpha \right\rangle $. Here, the classical
Hamiltonian reads%
\begin{equation*}
H_{c}=\left\langle \alpha \right\vert \hat{H}\left\vert \alpha \right\rangle
=\left( \alpha ^{\ast }\alpha +\frac{1}{2}\right) -\frac{\Gamma }{2}\left(
\alpha ^{\ast 2}-\alpha ^{2}\right) =\frac{1}{2}\left( x^{2}+p^{2}\right)
-i\Gamma xp\equiv H_{c0}-i\Gamma _{c},
\end{equation*}%
and the classical equation of motion in the phase-space follows \cite{Eva}%
\begin{equation*}
\left(
\begin{array}{c}
\dot{x} \\
\dot{p}%
\end{array}%
\right) =\Omega ^{-1}\nabla H_{c0}-G^{-1}\nabla \Gamma _{c},
\end{equation*}%
where $\nabla $\ is the phase space gradient operator, $\Omega $ is the symplectic
unit matrix
\begin{equation*}
\Omega =\left(
\begin{array}{cc}
0 & -1 \\
1 & 0%
\end{array}%
\right) ,\quad \Omega ^{-1}=\left(
\begin{array}{cc}
1 & 0 \\
0 & -1%
\end{array}%
\right) ,
\end{equation*}%
and the phase space metric $G$ is%
\begin{equation*}
G=\left(
\begin{array}{cc}
1 & 0 \\
0 & 1%
\end{array}%
\right) .
\end{equation*}%
The spectrum of Hamiltonian Eq.(\ref{nonH1}) can be obtained by a similarity
transformation of%
\begin{equation*}
\hat{H}^{\prime }=\hat{R}\hat{H}\hat{R}^{-1}=\frac{\omega }{2}\left( \hat{P}%
^{2}+\hat{X}^{2}\right) ,\quad \hat{R}=e^{-\frac{\theta }{2}\left( \hat{P}%
^{2}-\hat{X}^{2}\right) },
\end{equation*}%
where $\theta =-\Gamma$ and $\omega =\sqrt{1+\Gamma ^{2}}$.

However, there are many non-Hermitian Hamiltonians which can be used to describe the damped
oscillators under different damping mechanisms. For example, if $\hat{\Gamma}=\gamma \hat{X}%
^{2}/2$, another Hamiltonian for a damped oscillator is
\begin{equation}
\hat{H}=\frac{1}{2}\left( \hat{P}^{2}+\hat{X}^{2}\right) -i\frac{\gamma }{2}%
\hat{X}^{2}=\frac{1}{2}\hat{P}^{2}+\frac{1}{2}g\hat{X}^{2},  \label{qdH}
\end{equation}%
where $g=1-i\gamma $ is a complex number. Generally, the non-Hermitian Hamiltonian Eq.(\ref{qdH})
is related to a class of $\mathcal{PT}$-symmetric Hamiltonian of \cite{Carl}%
\begin{equation*}
\hat{H}=\frac{1}{2}\hat{P}^{2}-\frac{1}{2}( i\hat{X}) ^{\epsilon }\hat{X}^{2},
\end{equation*}%
where the parameter $\epsilon $ is real. The non-Hermitian Hamiltonian is
$\mathcal{PT}$-symmetric if $\mathcal{\hat{P}\hat{T}}\hat{H}%
\mathcal{\hat{P}\hat{T}}=\hat{H}$, which means
$[ \hat{H},\mathcal{\hat{P}\hat{T}}] =0$,
where $\hat{\mathcal{P}}\hat{X}\hat{\mathcal{P}}=-\hat{X}, \hat{\mathcal{P}}%
\hat{P}\hat{\mathcal{P}}= -\hat{P}$, $\hat{\mathcal{T}}\hat{X}\hat{\mathcal{T%
}}= \hat{X},\hat{\mathcal{T}}\hat{P}\hat{\mathcal{T}}= -\hat{P}, \mathcal{%
\hat{T}}i\mathcal{\hat{T}}= -i$, and $\mathcal{\hat{P}}^{2}=1, \mathcal{\hat{%
T}}^{2}=1, \mathcal{\hat{P}\hat{T}=\hat{T}\hat{P}}$. The study on the above non-Hermitian
Hamiltonian shows that, when $\epsilon \geq 0$, all the eigenvalues of
the Hamiltonian are real and positive (unbroken $\mathcal{PT}$%
-symmetric parametric region), but when $\epsilon <0$, there are complex
eigenvalues (broken $\mathcal{PT}$-symmetric parametric region) leading to
the damping processes \cite{Bender}.

\subsection{Lie transformation method on DDPO}

\subsubsection{The general quadratic Hamiltonian and its Lie algebra}

Now we consider the complete dynamics of a damped and driven oscillator
in a unified algebraic framework based on the time-dependent algebraic method.
A general DDPO can be described by the following time-dependent quadratic
Hamiltonian \cite{Choi,Choi2,Yeon2}%
\begin{equation}
\hat{H}=\frac{1}{2}\left[ A(t)\hat{p}^{2}+B(t)\hat{x}^{2}+C(t)\left( \hat{x}%
\hat{p}+\hat{p}\hat{x}\right) \right] +D(t)\hat{p}+E(t)\hat{x}+F(t),
\label{H0}
\end{equation}%
where all the time-dependent functions $A(t)\cdots F(t)$ are continuous functions
of time which can be used to describe the classical driving signals or the parametric
controlling on a specific quantum system. We consider for simplicity the one-dimensional
model and a $d$-dimensional generalization is theoretically straightforward \cite{Lohe}.
Totally, this Hamiltonian owns a well known real symplectic group of $Sp(2d, \mathbb{R})$ \cite{sp}
and we will consider it in a decomposed space with a Lie algebraic structure of $su(2)\bigoplus
h(4)$ \cite{Cheng}. Therefore six independent generators can be separately defined by \cite{zeng}
\begin{equation*}
\hat{J}_{+}=\frac{1}{2\hbar }\hat{x}^{2},\quad \hat{J}_{-}=\frac{1}{2\hbar }%
\hat{p}^{2},\quad \hat{J}_{0}=\frac{i}{4\hbar }\left( \hat{x}\hat{p}+\hat{p}%
\hat{x}\right) ,
\end{equation*}%
which satisfy the commutation relations of $su(2)$ algebra%
\begin{equation*}
\left[ \hat{J}_{+},\hat{J}_{-}\right] =2\hat{J}_{0},\quad \left[ \hat{J}_{0},%
\hat{J}_{\pm }\right] =\pm \hat{J}_{\pm },
\end{equation*}%
and
\begin{equation*}
\hat{T}_{1}=\frac{\hat{x}}{\hbar },\quad \hat{T}_{2}=\frac{\hat{p}}{\hbar }, \quad \hat{T}%
_{0}=\frac{i}{\hbar },
\end{equation*}%
which meet the commutation relations of Heisenberg-Weyl algebra%
\begin{equation*}
\left[ \hat{T}_{1}, \hat{T}_{2}\right] =\hat{T}_{0}, \left[ \hat{T}_{1},\hat{T}%
_{0}\right] =0, \left[ \hat{T}_{2},\hat{T}_{0}\right] =0.
\end{equation*}%
Then the Hamiltonian can be written as%
\begin{equation*}
\hat{H}=\hbar \left( A\hat{J}_{-}+B\hat{J}_{+}-i2C\hat{J}_{0}\right) +\hbar
\left( D\hat{T}_{2}+E\hat{T}_{1}-iF\hat{T}_{0}\right) .
\end{equation*}%
The above time-dependent Hamiltonian has been extensively studied \cite%
{Choi,Yeon} and we will follow the algebraic method firstly proposed by Wei
and Norman \cite{Wei,Cheng} and, recently, summarized in Ref.\cite{Sierra}.
Actually, the method we use here is a reverse procedure proposed by Lohe in Ref.%
\cite{Lohe}. For simplify, we adopt the units of zero-point
fluctuations of
\begin{equation}
\quad x_{\mathrm{zpf}}=\sqrt{\frac{\hbar }{m_0\omega _{0}}},\quad p_{\mathrm{%
zpf}}=\sqrt{\hbar m_0\omega _{0}},  \label{zpf0}
\end{equation}%
to scale the Hamiltonian as
\begin{equation}
\hat{\mathcal{H}}(t) =\frac{1}{2}a\left( t\right) \hat{P}^{2}+\frac{1}{2}%
b\left( t\right) \hat{X}^{2}+\frac{1}{2}c\left( t\right) \left( \hat{X}\hat{P%
}+\hat{P}\hat{X}\right) +d\left( t\right) \hat{P}+e\left( t\right) \hat{X}%
+f\left( t\right) ,  \label{Hs}
\end{equation}%
where $a\left( t\right) $, $b\left( t\right) $, $c\left( t\right) $, $%
d\left( t\right) $, $e\left( t\right) $ and $f\left( t\right) $ are scaled
time-dependent functions and the basic commutation relation between the scaled
position and momentum becomes $[\hat{X},\hat{P}] =i$. The mass $m_0$ and the
frequency $\omega _{0}$ used in Eq.(\ref{zpf0}) are the intrinsic or characteristic
parameters which can be properly chosen according to the initial conditions for a
controlled system. Then the six generators of $su(2)$ and $h(4)$ algebras are%
\begin{equation*}
\hat{J}_{+}=\frac{1}{2}\hat{X}^{2},\quad \hat{J}_{-}=\frac{1}{2}\hat{P}%
^{2},\quad \hat{J}_{0}=\frac{i}{4}\left( \hat{X}\hat{P}+\hat{P}\hat{X}%
\right) ,\quad \hat{T}_{1}=\hat{X},\quad \hat{T}_{2}=\hat{P},\quad \hat{T}%
_{0}=i.
\end{equation*}%
The merit of an algebraic method is that the generators defined above can be
easily generalized to other form of realization which enables the study of the
time-dependent Hamiltonian beyond a quadratic style \cite{Dong}.

If all the operators in Eq.(\ref{Hs}) are treated with their corresponding mean-value $c$
numbers such as $\hat{X}\rightarrow X$, the classical equation of motion
to describe a DDPO reads
\begin{equation}
\ddot{X}-\frac{\dot{a}}{a}\dot{X}+\left( ab+\frac{\dot{a}}{a}c-c^{2}-\dot{c}%
\right) X=cd-ae+\dot{d}-\frac{\dot{a}}{a}d.  \label{d1}
\end{equation}
Based on the above closed Lie algebra, the time-dependent quantum system can be parameterized by
six time-dependent parametric functions of $a(t)\cdots f(t)$ and the state evolution can be decomposed into
two groups of dynamical equations in 6-dimensional parametric space.

\subsubsection{Time-dependent transformations on Lie algebra}

Now we give the main idea of the Lie transformation method \cite{Cheng} to
solve or control the general quadratic Hamiltonian of Eq.(\ref{Hs})
by adopting proper transformation parameters. If an original Schr\"{o}dinger equation is%
\begin{equation}
i\hbar \frac{\partial }{\partial t}\Psi \left( t\right) =\hat{\mathcal{H}}\left(
t\right) \Psi \left( t\right) ,
\label{sch0}
\end{equation}%
and a time-dependent transformation on $\Psi \left( t\right) $ is introduced by%
\begin{equation}
\Psi \left( t\right) =\hat{U}\left( t\right) \psi \left( t\right),
\label{sol}
\end{equation}%
then the Schr\"{o}dinger equation for the new state $\psi \left( t\right) $
will be
\begin{equation}
i\hbar \frac{\partial \psi \left( t\right) }{\partial t}
=\mathcal{\hat{H}}_{U}(t)\psi \left(
t\right) ,
\label{sch1}
\end{equation}%
where%
\begin{equation}
\mathcal{\hat{H}}_{U}(t)=\hat{U}^{-1}\left( t\right) \hat{\mathcal{H}}(t)
\hat{U}\left( t\right) -i\hbar \hat{U}^{-1}\left( t\right) \frac{\partial
\hat{U}\left( t\right) }{\partial t}.  \label{Ut}
\end{equation}%
The transformation of Eq.(\ref{Ut}) can be called Floquet $U$-transformation (FUT)
following Ref.\cite{Sierra}, and the transformed Hamiltonian $\mathcal{\hat{H}}_{U}$,
generally, becomes non-Hermitian for that $\mathcal{\hat{H}}_{U}\neq \mathcal{\hat{H}}_{U}^{\dag}$.
Properly, if we select a successive transformation of $\hat{U}\left( t\right)$
based on a closed Lie algebra, we can finally simplify the original Hamiltonian
$\mathcal{\hat{H}}(t)$\ to a solvable $\mathcal{\hat{H}}_{U}(t)$ whose eigenstates $\psi(t)$
can be easily obtained, then the original wave function for system $\mathcal{\hat{H}}(t)$ will
be determined by Eq.(\ref{sol}) if the initial state is given.%

If a control is applied at $t=0$ to a quantum system, we can set all the
transformation parameters of $\hat{U}\left( t\right) $ to be zero at  $t=0$ by
giving $\hat{U}\left( 0\right) =1$. Therefore the initial state of a time-dependent
controlled system can be determined by its initial Hamiltonian of $\hat{\mathcal{H}}(0)$
with the time-independent eigenequation of%
\begin{equation*}
\hat{\mathcal{H}}(0)\varphi _{n}=E_{n}\varphi _{n}.
\end{equation*}%
Therefore any initial state of the system can then be expanded by the above basis as%
\begin{equation*}
\psi_0(t) =\sum_{n}C_{n}\varphi _{n}e^{-iE_{n}t/\hbar},
\end{equation*}%
where $C_{n}$ are complex constants, and%
\begin{equation}
\mathcal{\hat{H}}_{U}(0)=\hat{\mathcal{H}}(0),\quad \psi_0 \left( 0\right)
=\sum_{n}C_{n}\varphi _{n}.
\label{ini}
\end{equation}

Therefore, for quadratic Hamiltonian of Eq.(\ref{Hs}), we
can define six independent operators on $su(2)\bigoplus h(4)$
algebra by
\begin{equation*}
e^{-is}, e^{-i\beta \hat{X}},e^{-i\alpha \hat{P}},e^{-i\theta _{-}\hat{J}%
_{-}},e^{-2\theta _{0}\hat{J}_{0}},e^{-i\theta _{+}\hat{J}_{+}},
\end{equation*}
where $s, \alpha, \beta, \theta_{+}, \theta_0, \theta_{-}$ are six transformation
parameters. Then the transformation of $\hat{U}(t)$ can be a combination of
the above six operators in a specific order. For simplicity, we choose
the following successive transformations separated by $h(4)$ algebra and $su(2)$
algebra as \cite{Cheng}%
\begin{equation}
\hat{U}\left( t\right) =\hat{U}_{1}\left( t\right) \hat{U}_{2}\left(
t\right) ,
\label{U}
\end{equation}%
where%
\begin{eqnarray*}
\hat{U}_{1}(t) &=&e^{-is\left( t\right) }e^{-i\alpha \left( t\right) \hat{P}%
}e^{-i\beta \left( t\right) \hat{X}}
\end{eqnarray*}%
is defined on $h(4)$ algebra, and
\begin{eqnarray*}
\hat{U}_{2}(t) &=&e^{-i\theta
_{+}\hat{J}_{+}}e^{-2\theta _{0}\hat{J}_{0}}e^{-i\theta _{-}\hat{J}_{-}}
=e^{-i\theta _{+}\hat{X}^{2}/2}e^{-i\theta _{0}\left( \hat{X}%
\hat{P}+\hat{P}\hat{X}\right) /2}e^{-i\theta _{-}\hat{P}^{2}/2}
\end{eqnarray*}%
is defined on $su(2)$ algebra. Surely, Eq.(\ref{U}) is a specific
combination of six operators and the relations to other combinations arranged in
different orders can be found by using the Baker-Campbell-Hausdorff formula for
the relevant algebras \cite{Truax}. Generally, the six parameters $\alpha \left( t\right)$,
$\beta \left( t\right) $, $s\left( t\right)$ and $\theta _{+}\left( t\right)$,
$\theta_{0}\left( t\right) $, $\theta _{-}\left( t\right)$ ($\theta$s) are piecewise continuous
functions of time defined within a control time interval. The real parametric functions
will keep the system under a unitary evolution while the complex functions will break
it in order to effectively describe a damping process. Anyway, according to Eq.(\ref{sol}),
the Lie transformations defined above not only induce transitionless evolution of the wave
function (adiabatic process), but also produce non-adiabatic transitions controlled by the
parametric functions. Substituting the successive transformations of Eq.(\ref{U}) into Eq.(\ref{Ut}),
we have%
\begin{equation*}
\hat{\mathcal{H}}_{U}=\hat{U}_{2}^{-1}\left[ \hat{%
U}_{1}^{-1}\hat{\mathcal{H}}\hat{U}_{1}-i\hat{U}_{1}^{-1}\frac{\partial \hat{U}_{1}}{%
\partial t}\right] \hat{U}_{2}-i\hat{U}_{2}^{-1}\frac{\partial \hat{U}_{2}}{%
\partial t},
\end{equation*}%
where%
\begin{eqnarray*}
\hat{\mathcal{H}}_{1} &\equiv &\hat{U}_{1}^{-1}\hat{\mathcal{H}}\hat{U}_{1}-i\hat{U}%
_{1}^{-1}\frac{\partial \hat{U}_{1}}{\partial t}
=\frac{1}{2}a\hat{P}^{2}+%
\frac{1}{2}b\hat{X}^{2}+\frac{1}{2}c\left( \hat{X}\hat{P}+\hat{P}\hat{X}%
\right) \\
&&+\left( c\alpha -a\beta +d-\dot{\alpha}\right) \hat{P}+\left( b\alpha
-c\beta +e-\dot{\beta}\right) \hat{X} \\
&&+\frac{1}{2}a\beta ^{2}+\frac{1}{2}b\alpha ^{2}-c\alpha \beta
-d\beta +e\alpha +f +\dot{\alpha}\beta -\dot{s},
\end{eqnarray*}%
and $\hat{\mathcal{H}}$ is in the unit of $\hbar \omega _{0}$. In order to simplify
the Hamiltonian, we set the coefficients before the generators of $h(4)$ to be
zero. Then the dynamical equation of motion for the transformation parameters are%
\begin{eqnarray}
\dot{\alpha} &=&c\alpha -a\beta +d, \notag \\
\dot{\beta} &=&b\alpha -c\beta +e, \label{h4}\\
\dot{s} &=&\frac{1}{2}a\beta ^{2}+\frac{1}{2}b\alpha ^{2}-c\alpha
\beta -d\beta +e\alpha +f +\dot{\alpha}\beta , \notag
\end{eqnarray}%
which can reduce the original Hamiltonian Eq.(\ref{Hs}) to a standard quadratic form of
$su(2)$ algebra as%
\begin{equation}
\hat{\mathcal{H}}_{1}=\frac{1}{2}a\hat{P}^{2}+\frac{1}{2}b\hat{X}^{2}+\frac{1%
}{2}c\left( \hat{X}\hat{P}+\hat{P}\hat{X}\right) =a\hat{J}_{-}+b\hat{J}%
_{+}-2ic\hat{J}_{0}.  \label{H1}
\end{equation}%
The first two parametric equations of Eq.(\ref{h4}) lead to%
\begin{equation}
\ddot{\alpha}-\frac{\dot{a}}{a}\dot{\alpha}+\left( ab+\frac{\dot{a}c}{a}%
-c^{2}-\dot{c}\right) \alpha =cd-ae+\dot{d}-\frac{\dot{a}}{a}d,  \label{d2}
\end{equation}%
which exactly recovers the classical mean-value dynamics of Eq.(\ref{d1}). The last
equation of Eq.(\ref{h4}) leads to a classical dynamical action of
\begin{equation*}
s\left( t\right) =\int_{0}^{t}\mathcal{L}\left( \alpha ,\beta ,t^{\prime}\right) dt^{\prime},
\end{equation*}%
where the classical Lagrangian is defined by
\begin{equation}
\mathcal{L}\left( \alpha ,\beta ,t\right)=
\frac{1}{2}b\alpha ^{2}-\frac{1}{2}a\beta ^{2}+e\alpha +f.
\label{Lag0}
\end{equation}%
It should be noted that the above Lagrangian is not a unique one
corresponding to the dynamical equation of Eq.(\ref{d2}) due to the
existence of many linear canonical transformations on this system \cite{PR,sp}. As the
parameter $\alpha $ is a solution of the classical dynamical equation of Eq.(%
\ref{d1}), we can denote it as a classical
position by $\alpha \rightarrow X_{c}=\langle \hat{X}\rangle $ and Eq.(\ref%
{d2}) can be written in a standard form of DDPO as%
\begin{equation}
\ddot{X}_{c}+\chi \left( t\right) \dot{X}_{c}+\xi \left( t\right) X_{c}=\eta
\left( t\right),  \label{fm}
\end{equation}%
where%
\begin{eqnarray*}
\chi \left( t\right) &=&-\frac{\dot{a}}{a}, \\
\xi \left( t\right) &=&ab+\frac{\dot{a}c}{a}-c^{2}-\dot{c}, \\
\eta \left( t\right) &=&cd-ae+\dot{d}-\frac{\dot{a}}{a}d.
\end{eqnarray*}

After the transformations of $\hat{U}_1$ on Heisenberg-Weyl
algebra $h(4)$, the original Hamiltonian Eq.(\ref{Hs}) becomes a standard quadratic
time-dependent Hamiltonian which has been exactly solved by the Lewis-Riesenfeld
invariant method \cite{Yeon,Lima}. Although the dynamical invariant method
is powerful, there is no simple procedure to find an explicit
invariant for a certain time-dependent Hamiltonian (the invariant for Eq.(\ref{H1})
is not unique). Subsequently, in our
present method, a second group of independent FUTs on $su(2)$ algebra gives%
\begin{equation}
\hat{\mathcal{H}}_{U}=\hat{U}_{2}^{-1}\hat{\mathcal{H}}_{1}\hat{U}_{2}-i\hat{%
U}_{2}^{-1}\frac{\partial \hat{U}_{2}}{\partial t},  \label{U2}
\end{equation}%
where (see Appendix \ref{appA})%
\begin{eqnarray*}
\hat{\mathcal{H}}_{U} &=&\frac{1}{2}e^{2\theta _{0}}\left( a\theta
_{+}^{2}-2c\theta _{+}-\dot{\theta}_{+}+b\right) \hat{X}^{2} \\
&&+\frac{1}{2}\left[ ae^{-2\theta _{0}}+e^{2\theta _{0}}\left( a\theta
_{+}^{2}-2c\theta _{+}-\dot{\theta}_{+}+b\right) \theta _{-}^{2}+2\theta
_{-}\left( c-a\theta _{+}-\dot{\theta}_{0}\right) -\dot{\theta}_{-}\right]
\hat{P}^{2} \\
&&+\frac{1}{2}\left[ e^{2\theta _{0}}\left( a\theta _{+}^{2}-2c\theta _{+}-%
\dot{\theta}_{+}+b\right) \theta _{-}+\left( c-a\theta _{+}-\dot{\theta}%
_{0}\right) \right] \left( \hat{X}\hat{P}+\hat{P}\hat{X}\right) .
\end{eqnarray*}

Similarly, if we set all the coefficients before $su(2)$ generators ($\hat{X}^2$, $\hat{P}^2$ and $%
\hat{X}\hat{P}+\hat{P}\hat{X}$) in the transformed Hamiltonian $\hat{\mathcal{H}}_{U}$ to be zero,
we get the dynamical equations for $\theta$s as
\begin{eqnarray}
\dot{\theta}_{+} &=&a\theta _{+}^{2}-2c\theta _{+}+b,  \notag \\
\dot{\theta}_{-} &=&ae^{-2\theta _{0}},  \label{dyth} \\
\dot{\theta}_{0} &=&c-a\theta _{+}.  \notag
\end{eqnarray}%
That means if all the parameters $\theta $s are controlled by Eq.(\ref{dyth}),
the transformed Hamiltonian $\hat{\mathcal{H}}_{U}$ will shift to a zero point of
energy and the system will stay on its initial state $%
|\psi_0\rangle$ beyond any quantum evolution. According to Eq.(\ref%
{U2}), we can see, in this case,
\begin{equation}
i\frac{\partial \hat{U}_{2}}{\partial t} =\hat{\mathcal{H}}_{1}\hat{U}_{2},
\end{equation}%
and the solution of $\hat{\mathcal{H}}_{1}$ can be exactly solved by the
transformation of $\hat{U}_2$. Combined with transformation $\hat{U}_1$,
a complete solution of the initial Hamiltonian of Eq.(\ref{Hs}) can be easily
obtained
\begin{equation}
\Psi \left(X, t\right) =e^{-is\left( t\right) }e^{-i\alpha \left( t\right) \hat{P}%
}e^{-i\beta \left( t\right) \hat{X}}e^{-i\theta _{+}\hat{X}%
^{2}/2}e^{-i\theta _{0}\left( \hat{X}\hat{P}+\hat{P}\hat{X}\right)
/2}e^{-i\theta _{-}\hat{P}^{2}/2}\psi_0 \left( X,t\right),
\label{solution}
\end{equation}%
where $\psi_0\left( X,t\right)$ is the initial state of the system.
Therefore, if all the six transformation parameters $\alpha, \beta, s$ and $%
\theta_{+},\theta_{0},\theta_{-}$ are solved or controlled by the
corresponding parametric equations of Eq.(\ref{fm}) and Eq.(\ref%
{dyth}), the wave function of the system described by Eq.(\ref{Hs}) can be
exactly solved by this method. The solution of Eq.(\ref{solution}) accompanied
with two systems of nonlinear dynamical equations can perfectly resolve the complete evolution
of an initial state into a classical and a quantum part, and can naturally
discriminate the dynamic phase from the geometric phase in the parametric space
during a quantum evolution. However, by considering the physical meaning
of the six parameters in Eq.(\ref{solution}), we can see some problems of
solving the parametric equation of Eq.(\ref{dyth}) for a dissipative system.

According to the mathematical properties of $\hat{X}$ and $\hat{P}$, we notice
that $e^{-i\beta \left( t\right) \hat{X}}$ and $e^{-i\alpha \left( t\right) \hat{P}}$
are displacement operators which lead to translations of a wave packet
in momentum and real space, respectively. The operators $e^{-i\theta _{+}\hat{X}^{2}/2}$
and $e^{i\theta _{-}\hat{P}^{2}/2}$ are dispersive ones which induce
width modifications of a wave packet in momentum and real space, respectively,
and the transformation $e^{-i\theta _{0}\left( \hat{X}\hat{P}+\hat{P}\hat{X}\right) /2}$
will result in squeezing (or dilation) of a wave packet. The geometric effects of the
transformation parameters can be seen clearly in a Heisenberg picture. The position and
momentum operators in the Heisenberg picture evolve as \cite{Sierra}%
\begin{eqnarray*}
\hat{X}_{h}(t) &=&\hat{U}^{-1}\hat{X}\hat{U}=e^{\theta _{0}}\hat{X}+\theta
_{-}e^{\theta _{0}}\hat{P}+\alpha , \\
\hat{P}_{h}(t) &=&\hat{U}^{-1}\hat{P}\hat{U}=\left( e^{-\theta
_{0}}-e^{\theta _{0}}\theta _{+}\theta _{-}\right) \hat{P}-e^{\theta
_{0}}\theta _{+}\hat{X}-\beta,
\end{eqnarray*}
and, in a matrix form, the transformation $\hat{U}(t)$ gives
\begin{equation*}
\left(
\begin{array}{c}
\hat{X}_{h} \\
\hat{P}_{h}%
\end{array}%
\right) =\left(
\begin{array}{cc}
1 & 0 \\
-\theta _{+} & 1%
\end{array}%
\right) \left(
\begin{array}{cc}
e^{\theta _{0}} & 0 \\
0 & e^{-\theta _{0}}%
\end{array}%
\right) \left(
\begin{array}{cc}
1 & \theta _{-} \\
0 & 1%
\end{array}%
\right) \left(
\begin{array}{c}
\hat{X} \\
\hat{P}%
\end{array}%
\right) +\left(
\begin{array}{c}
\alpha \\
-\beta%
\end{array}%
\right).
\end{equation*}%
The above equation indicates a successive operation of translation, rotation and dilation on the
operators $\hat{X}$ and $\hat{P}$ and it is actually equivalent to a canonical
transformation of $\hat{X}$ and $\hat{P}$. Specifically, if the initial wave
function is in a number state of $\vert n\rangle$, we can see the average values of
\begin{eqnarray*}
X_{h}(t) =\left\langle n\right\vert \hat{X}_{h}\left\vert n\right\rangle
=\alpha(t), \quad
P_{h}(t) =\left\langle n\right\vert \hat{P}_{h}\left\vert n\right\rangle
=-\beta(t),
\end{eqnarray*}%
which directly show the total displacements of a wave packet in real
and momentum space, respectively. The standard deviations of $\hat{X}$
and $\hat{P}$ will be
\begin{eqnarray}
\Delta X_{h} &=&e^{\theta _{0}}\sqrt{1+\theta _{-}^{2}}\sqrt{n+\frac{1}{2}},
\label{dX} \\
\Delta P_{h} &=&e^{\theta _{0}}\sqrt{\theta _{+}^{2}+\left( e^{-2\theta
_{0}}-\theta _{+}\theta _{-}\right) ^{2}}\sqrt{n+\frac{1}{2}},  \label{dP}
\end{eqnarray}%
which indicate the profile modifications of a wave packet in real and
momentum space, respectively. Therefore, back to the Schr\"{o}dinger
picture, the transformation $\hat{U}(t)$ shows that the average position and the
width of the wave packet will be controlled by the parametric equations of Eq.(\ref{fm})
and Eq.(\ref{dyth}) during a quantum evolution. Furthermore, based on the transformation
of $\hat{U}(t)$, the propagator of the wave function will be easily obtained as shown in
Ref.\cite{Sierra}.

As the classical dynamics of Eq.(\ref{fm}) has been discussed in section II, now,
we try to find solutions of Eq.(\ref{dyth}) in order to finally
determine the wave function of Eq.(\ref{solution}). We can easily find
that the first equation of Eq.(\ref{dyth}) is decoupled from others by
\begin{equation}
\dot{\theta}_{+}=a\theta _{+}^{2}-2c\theta _{+}+b.
\label{Riccati}%
\end{equation}%
Eq.(\ref{Riccati}) is the Riccati differential equation \cite{Hans} and
its real solution can be solved by a Riccati transformation of
\begin{equation}
\theta _{+}=-\frac{\dot{u}}{a u},  \label{theta}
\end{equation}%
where $u$ should satisfy the following linear second-order ordinary differential
equation%
\begin{equation}
\ddot{u}+\left( 2c-\frac{\dot{a}}{a}\right) \dot{u}+abu=0.
\label{Rati}
\end{equation}%
Therefore, if $\theta _{+}$ is determined, the other two parameters are%
\begin{eqnarray*}
\theta _{0}\left( t\right) &=& \int_{0}^{t}c\left( \tau \right) d\tau +\ln
\frac{u\left( t\right) }{u\left( 0\right) }, \\
\theta _{-}\left( t\right) &=&a\left( t\right) \left[ \frac{u\left( 0\right)
}{u\left( t\right) }\right] ^{2}e^{-2\int_{0}^{t}c\left( \tau \right) d\tau
},
\end{eqnarray*}%
where we can set the initial condition of $\theta _{+}\left( 0\right)=\theta _{-}\left( 0\right)
=\theta _{0}\left( 0\right) =0$ for a control problem. Finally, as all the six parameters
are solved according to Eq.(\ref{fm}) and Eq.(\ref{dyth}), the quantum wave function of
Eq.(\ref{solution}) will be easily obtained through this method.

\subsubsection{Quantum driven harmonic oscillator without damping}

In order to verify the efficiency of the time-dependent FUT method, we first consider a solved problem: a driven
harmonic oscillator without damping, which is described by a Hermite Hamiltonian
\cite{Breuer}%
\begin{equation}
\hat{H}\left( t\right)=\frac{\hat{p}^{2}}{2m_0}+\frac{1}{2}m_0\omega _{0}^{2}%
\hat{x}^{2}-\hat{x}F\left( t\right),  \label{Hforce}
\end{equation}%
where $F\left( t\right) =m_0\omega _{0}^{2}f\left( t\right) $ is an external linear force
with $f(t)$ having a dimension of length \cite{Griffth}.
Usually, the parameters of the harmonic oscillator are time-independent,
but many works have been done on a
general time-dependent harmonic oscillator with the time varying mass and
frequency \cite{Lima}.

Now we use the algebraic transformation method to solve
the schrödinger equation
\begin{equation}
i\hbar \frac{\partial }{\partial t}\Psi \left( x,t\right) =\hat{H}(t)\Psi \left(
x,t\right).  \label{Hfq}
\end{equation}%
A solution based on Lie
transformation of Heisenberg algebra $h(4)$
\cite{Heis} is
\begin{equation}
\Psi \left( x,t\right)=\hat{U}(t)\psi \left( x,t\right)\equiv e^{-is\left( t\right) /\hbar } e^{-i\alpha \left(
t\right) \hat{p}/\hbar }e^{-i\beta \left( t\right) \hat{x}/\hbar }\psi \left( x,t\right) ,
\label{ut0}
\end{equation}%
which can transform Eq.(\ref{Hfq}) into%
\begin{equation*}
i\hbar \frac{\partial }{\partial t}\psi \left( x,t\right) =\hat{H}_{U}\psi \left( x,t\right),
\end{equation*}
where
\begin{equation}
\hat{H}_{U}\left( t\right) =\frac{\hat{p}^{2}}{2m_0}+\frac{1}{2}m_0\omega
_{0}^{2}\hat{x}^{2}+\left( m_0\omega _{0}^{2}\alpha-\dot{\beta}-F\right) \hat{%
x}-\left( \frac{\beta }{m_0}+\dot{\alpha}\right) \hat{p}+\left( \frac{\beta
^{2}}{2m_0}+\frac{1}{2}m_0\omega _{0}^{2}\alpha ^{2}-\alpha F+\dot{\alpha}\beta -%
\dot{s}\right) .  \label{Hu}
\end{equation}%

Usually, the transformation parameters $s(t), \alpha(t)$ and $\beta(t)$ are real
in order to keep the transformation unitary. However, in some cases, especially in
a dissipative system described by non-Hermitian Hamiltonian, the parameters can be
generalized to complex functions. Removing all the linear terms in the Hamiltonian
of Eq.(\ref{Hu}), we have a specific case of Eq.(\ref{h4}) as
\begin{eqnarray}
\dot{\beta} &=&m_{0}\omega _{0}^{2}\alpha -F,  \label{peq1} \\
\dot{\alpha} &=&-\frac{\beta }{m_{0}},  \label{peq2} \\
\dot{s} &=&\frac{\beta ^{2}}{2m_{0}}+\frac{1}{2}m_{0}\omega _{0}^{2}\alpha
^{2}-\alpha F+\dot{\alpha}\beta ,  \label{peq3}
\end{eqnarray}%
and $\hat{H}_U$ becomes a time-independent Hamiltonian of harmonic oscillator
\begin{equation*}
\hat{H}_{U}\left( t\right) =\frac{1}{2m_0}\hat{p}^{2}+\frac{1}{2}m_0\omega
_{0}^{2}\hat{x}^{2}.
\end{equation*}%
The solutions of the harmonic oscillator are well-known as
\begin{equation}
\varphi _{n}\left( x,t\right) =\varphi
_{n}\left( x\right)e^{-i E_n t/\hbar},
\label{hwave}
\end{equation}%
where $\varphi _{n}\left( x\right) $ is the stationary function of the harmonic
oscillator with the quantum energy of $E_{n}=(n+1/2)\hbar \omega _{0}$.

Now let's find out why the transformation parameters of $\hat{U}(t)$
can convert a driven harmonic oscillator into a free one. From Eq.(\ref{peq1})
and Eq.(\ref{peq2}), we find that the parameter $\alpha(t)$ satisfies
\begin{equation}
\ddot{\alpha}+\omega _{0}^{2}\alpha =\frac{F}{m_0},  \label{xc0}
\end{equation}%
which clearly obeys the classical dynamical equation of Eq.(\ref{Geq}) without damping.
In this case, we set $\alpha (t)=\langle \hat{x}\rangle=x_{c}(t)$
and
\begin{equation*}
\beta (t)=-m_0\dot{\alpha}=-m_0\dot{x}_{c}=-p_c,
\end{equation*}%
which corresponds a classical momentum but in a reverse direction. The
parameter $s(t)$ determined by Eq.(\ref{peq3}) is
\begin{equation*}
s\left( t\right) = \int_{0}^{t}%
\mathcal{L}\left( x_{c},\dot{x}_{c},t^{\prime }\right) dt^{\prime },
\end{equation*}%
where
\begin{equation}
\mathcal{L}\left( x_{c},\dot{x}_{c},t\right) =-\left( \frac{1}{2}m_0\dot{x}%
_{c}^{2}-\frac{1}{2}m\omega _{0}^{2}x_{c}^{2}+x_{c}F\right)  \label{Lag}.
\end{equation}%
We can see that $\mathcal{L}$ is the classical Lagrangian for a driven harmonic
oscillator with a negative sign and thus the parameter $s(t)$ is a negative classical
action. For this reason, we can see that the parameters $\alpha, \beta$ and $s$
actually enable a time reversal transformation on the system to convert a time-dependent
problem into a stationary one. Here we should notice that different forms of $\hat{U}(t)$
will definitely lead to different classical Lagrangian functions of
$\mathcal{L}\left( x_{c},\dot{x}_{c},t\right)$. As Lagrangian functions producing the same
dynamical equation are not unique \cite{Gold}, different Lagrangian functions actually correspond
to different types of linear canonical transformations on Eq.(\ref{Lag}) \cite{PR}.

Therefore, if the initial state (before the force is applied) is
$\varphi _{n}(x,t)$, the solution of Eq.(\ref{ut0}) at time $t$ will be
\begin{equation}
\Psi _{n}\left( x,t\right)=e^{-\frac{i}{\hbar }\int_{0}^{t}\mathcal{L}\left(
x_{c},\dot{x}_{c},t^{\prime }\right) dt^{\prime }}e^{i m_0\dot{x}_{c}\left(
x-x_{c}\right) /\hbar }\varphi _{n}\left( x-x_{c},t\right) ,  \label{psi1}
\end{equation}%
where the displacement operator $e^{-ix_{c}\hat{p}/\hbar }$ gives
$e^{-ix_{c}\hat{p}/\hbar }\varphi\left( x,t\right) =\varphi\left(
x-x_{c},t\right)$.
Clearly, we can easily check that the solution of Eq.(\ref{psi1}) is
dynamically equivalent to that given in Ref.\cite{Nogami,Husimi}.
\begin{figure}[t]
\begin{center}
\includegraphics[width=0.25 \textwidth]{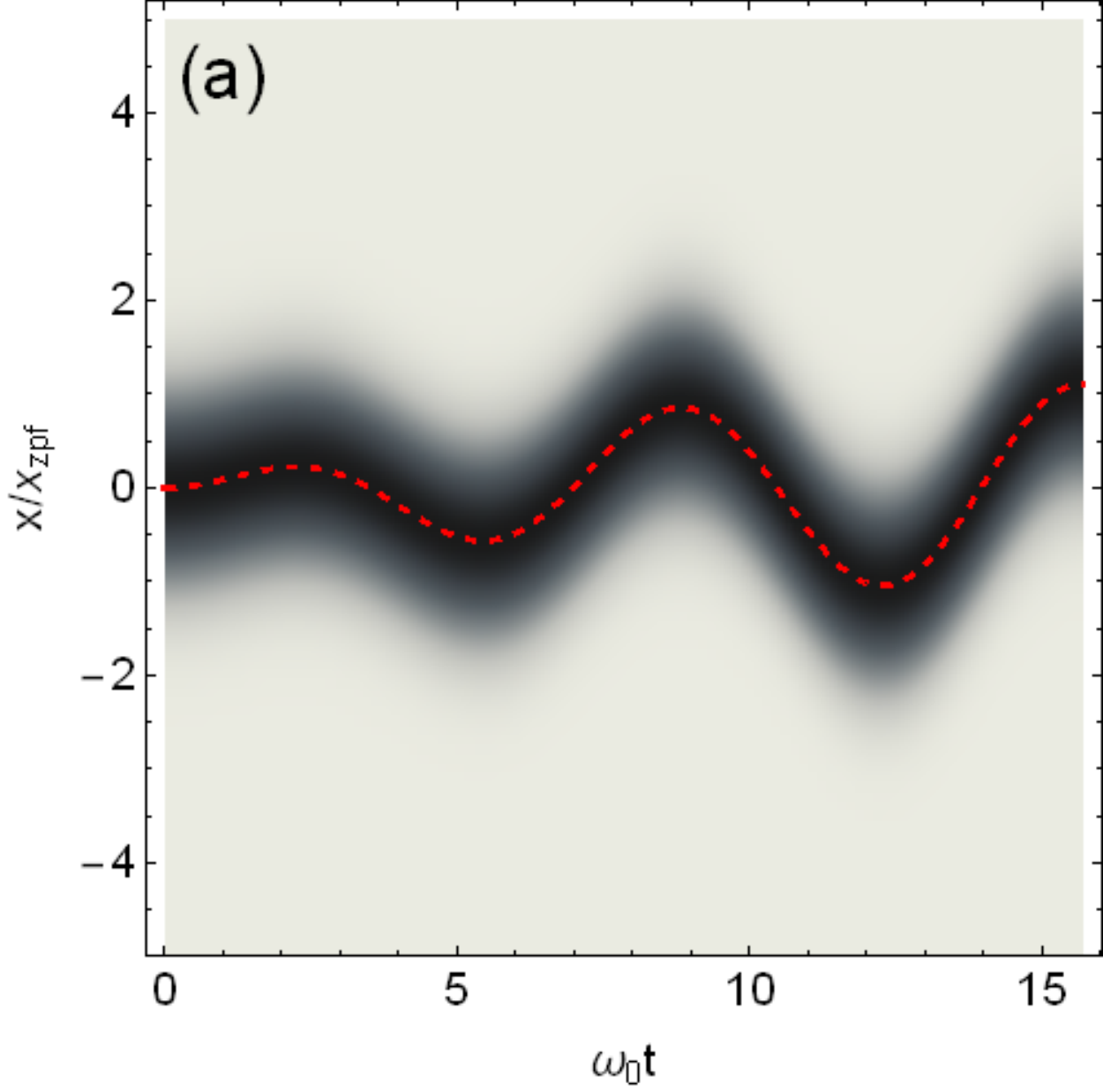}%
\includegraphics[width=0.25 \textwidth]{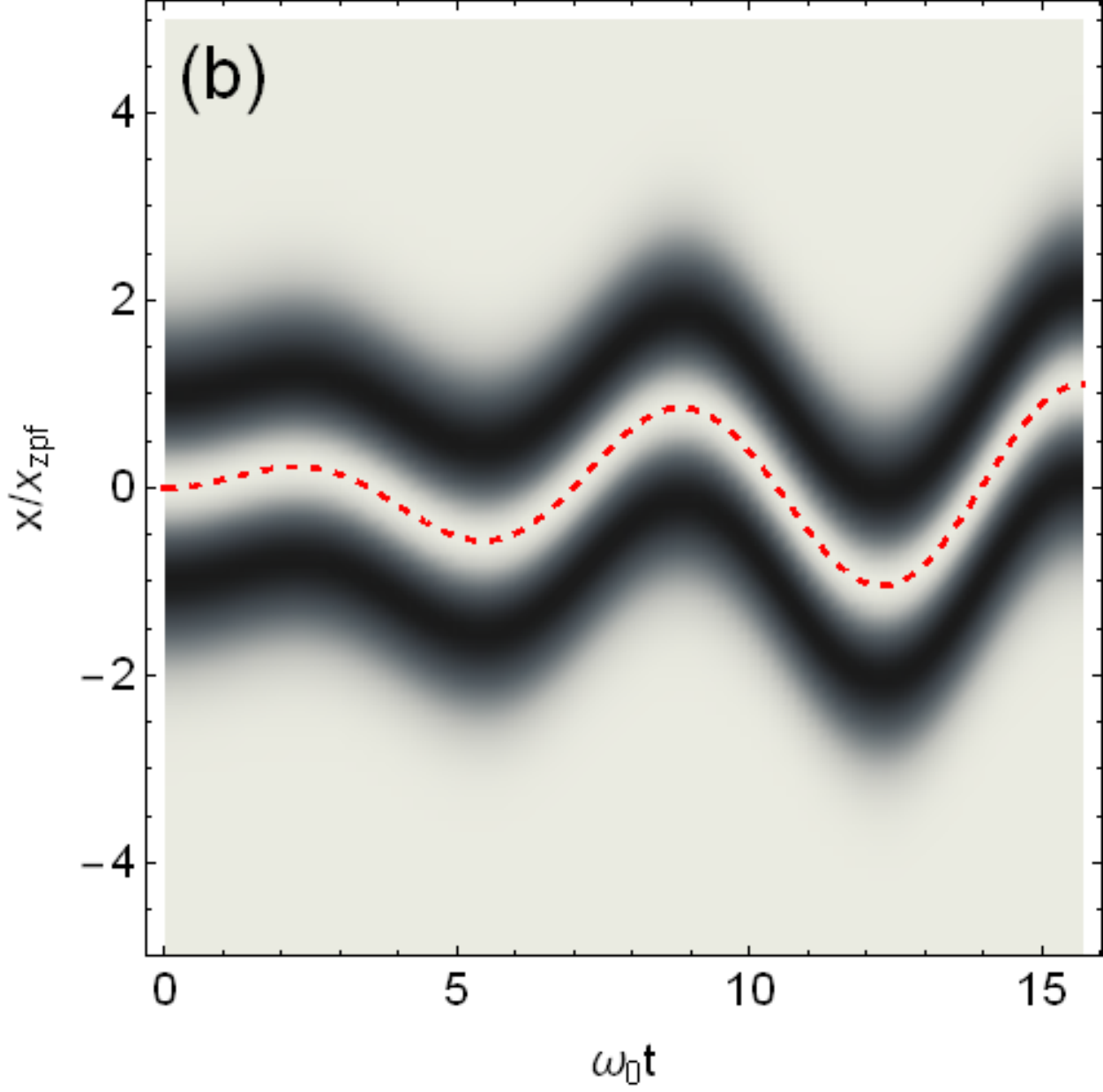}%
\includegraphics[width=0.25 \textwidth]{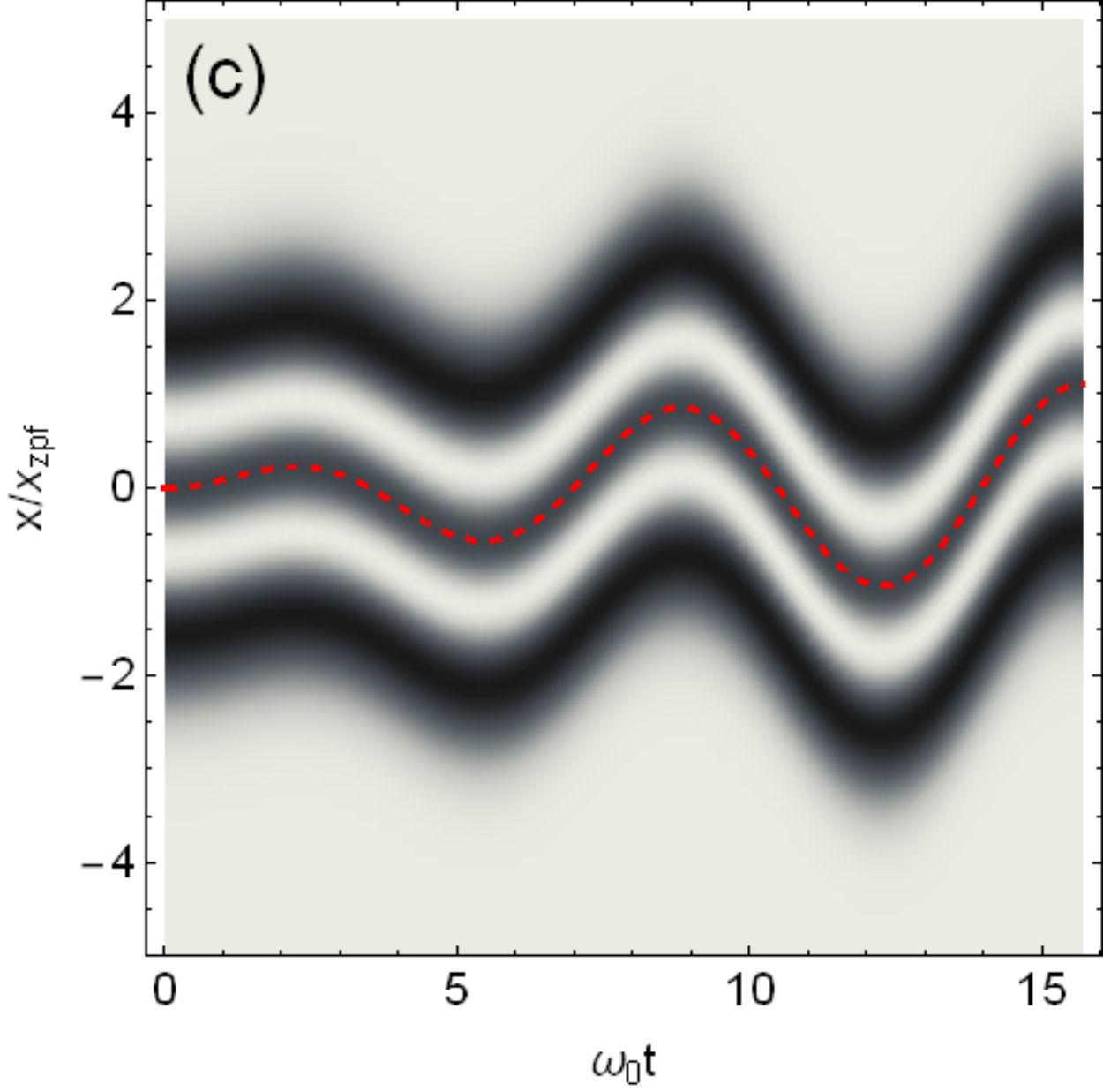}%
\includegraphics[width=0.25 \textwidth]{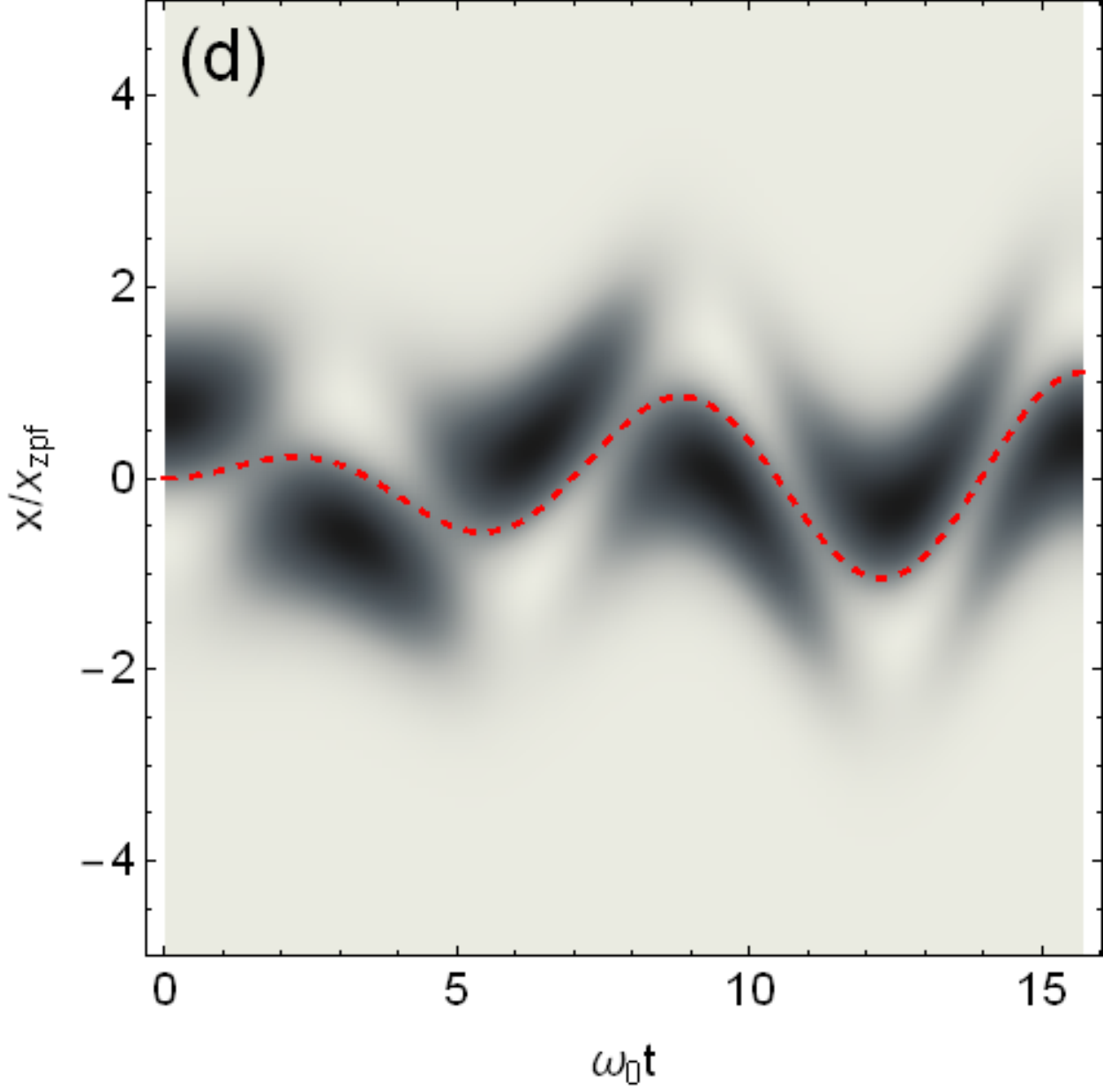}%
\end{center}
\caption{The quantum evolution of the probability distributions and their classical
orbits (red dashed lines) of the driven harmonic oscillator starting from initial
states of (a) $|0\rangle$, (b) $|1\rangle$, (c) $|2\rangle$ and (d) $(|0\rangle+|1\rangle)/\sqrt{2}$.
The driving parameters are $f=0.2, \Omega=0.8$ and the classical parameters are set
to be $x_c(0)=p_c(0)=0$ in order to highlight the quantum distribution.}
\label{figure4}
\end{figure}

Based on Eq.(\ref{psi1}), we calculate the quantum evolutions of the driven harmonic oscillator
starting from three different initial states in Fig.\ref{figure4}. We can see a clear connection
between the classical and quantum dynamics by Lie transformation method. The solution reveals that
the classical and quantum dynamics are dynamically separated (for the different dynamic scales) and the
parameters related to the classical motion become the external parameters of the quantum wave function.
Fig.\ref{figure4} demonstrates that the probability distribution of a driven harmonic oscillator
maintains its profile shape during a pure state evolution and the driving force only
adjusts the mean position of the probability distribution to follow the classical orbit (dashed line).
The translation operation on a quantum state controlled by an external force can be revealed by the
properties of a displaced number state \cite{Oliveira}. With Dirac notation, Eq.(\ref{psi1}) becomes%
\begin{equation}
\left\vert n,\alpha _{c},t\right\rangle =e^{-iE_n  t /\hbar }e^{-i s\left( t\right) /\hbar }e^{-i%
\frac{x_c p_c}{2\hbar }}\hat{D}\left( \alpha _{c}\right) \left\vert
n\right\rangle  \label{psi2}
\end{equation}%
where $\hat{D}(\alpha_c)$ is the displacement operator with the parameter%
\begin{equation*}
\alpha _{c}=\frac{1}{\sqrt{2}}\left( \frac{x_{c}}{x_{\text{zpf}}}+i\frac{%
p_{c}}{p_{\text{zpf}}}\right),
\end{equation*}%
and $|n\rangle$ is the number state of the harmonic oscillator.
Usually, the displaced number state $\left\vert n,\alpha\right\rangle $ is
defined by%
\begin{equation*}
|n,\alpha \rangle =\hat{D}(\alpha )|n\rangle=e^{\left( \alpha\hat{a}^{\dag
}-\alpha^{\ast }\hat{a}\right)}|n\rangle,
\end{equation*}
and it is also called generalized coherent
state for a Heisenberg algebra \cite{Oliveira}. The displaced number state
can be expanded by \cite{Oliveira,Philbin}%
\begin{equation*}
\left\vert n,\alpha \right\rangle =\sum_{m}\left\vert m\right\rangle
\left\langle m\right\vert n,\alpha \rangle =\sum_{m}c_{m,\alpha }e^{i\varphi
_{m}}\left\vert m\right\rangle ,
\end{equation*}%
where $\alpha =\left\vert \alpha \right\vert e^{i\varphi_0 }$ and
\begin{equation*}
c_{m,\alpha }=\sqrt{\frac{n!}{m!}}\left( -1\right) ^{n-m}\left\vert \alpha
\right\vert ^{n-m}e^{-\left\vert \alpha \right\vert ^{2}/2}L_{m}^{n-m}\left(
\left\vert \alpha \right\vert ^{2}\right) ,\quad \varphi _{m}=\left(
n-m\right) \varphi_0 .
\end{equation*}
Then the survival probability of a driven harmonic oscillator in the initial
state $\left\vert n\right\rangle$ will be
\begin{equation}
P_{n}=\left\vert \left\langle n\right\vert \hat{U}(t)\left\vert
n\right\rangle \right\vert ^{2}=\left\vert c_{n,\alpha _{c}}\right\vert
^{2}=e^{-E_{c}\left( t\right) }\left[ L_{n}\left(
E_{c}\left( t\right) \right) \right] ^{2},
\label{pn}
\end{equation}%
where $L_{n}\left( x\right) $ is Laguerre polynomial function and
$E_c=|\alpha_c|^2=(x_{c}^{2}+p_{c}^{2})/2$ is the classical energy.
Eq.(\ref{pn}) means that the probability of the system remaining in
its initial state depends on the classical energy of $E_{c}\left( t\right) $,
which is only related to the classical motion induced by the external driving
force. If a simple harmonic driving force is exerted, the final classical
motion is $x_{c}\left( t\right) =A_{2}\cos \left( \Omega t+\phi _{2}\right)$
(see Eq.(\ref{Sfdh})) and the probability in the initial state will be%
\begin{equation*}
P_{n}(t)=e^{-E_{f}^{c}(t)}\left[ L_{n}\left( E_{f}^{c}\right) \right] ^{2},
\end{equation*}%
where $E_{f}^{c}(t)$ is the final classical energy stored in the harmonic oscillator.
The above result indicates that the survival probability of a wave function in state
$|n\rangle$ is determined only by the classical dynamics due to an overall shift of
the wave function relative to its initial position. As the probability distribution
of the quantum state remains unchanged (see Fig.\ref{figure4}) except for an overall
translation, the quantum resonant transition between different quantum state of $|n\rangle$
can not be excited by a resonant linear force of $F(t)$. However, as there is no damping
in this case, the overall shift of the wave packet will continuously increase as a result
of the classical resonant dynamics and the quantum distribution will be totally masked by an
increasing displacement of the wave packet.

\subsubsection{Problems for Lie transformation method}

Although the Lie transformation seems simple and perfect, the solutions given by this method in some
cases exhibit bad behaviors, especially for an open quantum system with dissipations,
because the real parameters $\theta$s often become divergent and no physical solution can survive
then (this will be demonstrated in \ref{ks0}). Therefore, more control parameters should be
introduced by performing further transformations in order to avoid the singular dynamics of
Eq.(\ref{dyth}). Another way to smooth the divergent behavior for a damping case without
increasing the parametric dimension is to use a non-Hermitian effective Hamiltonian (the
parameters $a(t), b(t)$ or $c(t)$ may be complex functions) to find a complex solution for
the Riccati Eq.(\ref{Riccati}) by taking a form of \cite{Hans}
\begin{equation}
\theta _{+}=-\frac{\dot{\rho}}{a\rho}-i\frac{1}{\rho^{2}},  \label{cu}
\end{equation}%
and $\rho$ will satisfy the following
differential equation instead%
\begin{equation}
\ddot{\rho}+\left( 2c-\frac{\dot{a}}{a}\right) \dot{\rho}+ab\rho=\frac{a^{2}%
\label{Ev}
}{\rho^{3}},
\end{equation}%
where we suppose $a\left( t\right) $ is still a real function in this case.
The complex solution of Eq.(\ref{cu}) has a better dynamical behavior and Eq.(\ref{Ev})
is the famous Ermakov equation often used in the problems of time-dependent
parametric harmonic oscillator \cite{Hans}. In this case, the other two parameters become%
\begin{eqnarray*}
\theta _{0}\left( t\right) &=&\int_{0}^{t}c\left( \tau \right) d\tau +\ln
\frac{\rho\left( t\right) }{\rho\left( 0\right) }+i\int_{0}^{t}\frac{a}{%
\rho^{2}}d\tau , \\
\theta _{-}\left( t\right) &=&a\left[ \frac{\rho\left( 0\right) }{\rho\left(
t\right) }\right] ^{2}e^{-2\int_{0}^{t}c\left( \tau \right) d\tau
}e^{-2i\int_{0}^{t}\frac{a}{\rho^{2}}d\tau }.
\end{eqnarray*}%
However, we should notice that the parameters $\theta$s are complex numbers, and
the unitary evolution of the wave function will be broken as expected
for a dissipative system.

Because of the pathological behavior of the dynamical equation of Eq.(\ref{dyth}) for
a dissipative system, we can not always obtain optimal solutions for \emph{real} $\theta$s
through Eq.(\ref{dyth}) to find a physical solution of Eq.$(\ref{solution})$. Anyway,
we can always choose a transformation of $\hat{U}(t)$ to convert the original Hamiltonian
into a solvable one by generally setting the corresponding
coefficients before $su(2)$ generators in Eq.(\ref{U2}) to be arbitrary functions, say
$K_1(t), K_2(t), K_3(t)$ ($K$s). Then, for the Lie transformation of Eq.(\ref{U}), a
general parametric equation will be got
\begin{eqnarray}
\dot{\theta}_{+} &=&a\theta _{+}^{2}-2c\theta _{+}-K_{2}e^{-2\theta _{0}}+b,
\notag \\
\dot{\theta}_{-} &=&ae^{-2\theta _{0}}-K_{2}\theta _{-}^{2}+2K_{3}\theta
_{-}-K_{1},  \label{gtheta} \\
\dot{\theta}_{0} &=&K_{2}\theta _{-}-a\theta _{+}+c-K_{3},  \notag
\end{eqnarray}%
and the final transformed Hamiltonian becomes
\begin{equation}
\hat{\mathcal{H}}_{U}(t)=\frac{1}{2}K_{1}(t)\hat{P}^{2}+\frac{1}{2}K_{2}(t)\hat{X}%
^{2}+\frac{1}{2}K_{3}(t)\left( \hat{X}\hat{P}+\hat{P}\hat{X}\right),  \label{Hf}
\end{equation}
which again takes the same form as Eq.(\ref{H1}). In principle, the repeating
FUT transformations of $\hat{U}_2$ will connect all the relevant quadratic
Hamiltonians in the form of Eq.($\ref{H1}$) and finally give a closed group \cite{Leach}.
As different functions assigned to $K$s will lead to different parametric
equations and thus yield different wave functions, now, the key problem is how
to determine or design the time-dependent functions of $K$s. The freedom
to choice $K$s indicates a trick that the FUTs should transform or simplify the
original Hamiltonian to a solvable form according to specific problems with special
initial conditions, which we will show in the next section by solving some typical
models of DDPO.

As the Lewis-Riesenfeld invariant method is a powerful tool to solve Hamiltonian
of $\hat{\mathcal{H}}_1$ and has been extensively discussed in many works
\cite{Yeon,Lewis,Pedrosa}, we can easily connect Lie transformation with this
method to determine the proper functions of $K$s. As, in principle,
there exist many dynamical invariants of $\hat{I}$ for a given Hamiltonian \cite{Lewis,Lohe},
so, generally, we suppose that the Hamiltonian $\hat{\mathcal{H}}_{1}$ can be transformed
into a real function of the dynamical invariant of $\hat{\mathcal{H}}_{1}$ \cite{Xu}, i.e.,
\begin{equation}
\hat{\mathcal{H}}_{U}(t)=F(\hat{I}_1)
\end{equation}
where the invariant $\hat{I}_1(t)$ of $\hat{\mathcal{H}}_{1}$ satisfies
\begin{equation}
\frac{d}{dt}\hat{I}_1(t)=\frac{\partial \hat{I}_1}{\partial t}+\frac{1}{i}\left[
\hat{I}_1,\hat{\mathcal{H}}_{1}\right] =0.  \label{LS}
\end{equation}%
Surely, if the real function $F(x)$ is properly designed for a specific control problem,
all the time-dependent quadratic Hamiltonians connected by FUT on $su(2)$ algebra can
be exactly solved by considering only the invariants of the system. Among different invariants
of $\hat{I}_1(t)$, we only focus on the simplest one, such as the basic invariant \cite{Gao},
or on a suitable one chosen for a specifically controlled system. Incidently, if
$K_1(t)=a(t), K_2(t)=b(t), K_3(t)=c(t)$, the Hamiltonian is invariant under the FUT and
the corresponding transformation operator $\hat{U}_2$ itself will be an invariant operator
following Eq.(\ref{LS}). This indicates a convergent case for the repeated FUT transformations
on $\hat{\mathcal{H}}_1$.

Simply, if Eq.(\ref{Hf}) can be reduced to \emph{one} of the invariants of $\hat{\mathcal{H}}_{1}$,
the invariant requirement of Eq.(\ref{LS}) will lead to a dynamical equation for the parameter
$K$s as
\begin{eqnarray}
\dot{K}_{1} &=&2cK_{1}-2aK_{3},  \notag \\
\dot{K}_{2} &=&2bK_{3}-2cK_{2},  \label{Kt} \\
\dot{K}_{3} &=&bK_{1}-aK_{2}.  \notag
\end{eqnarray}%
As shown in Ref.\cite{Yeon}, an exact solution of Eq.(\ref{Kt}) can be constructed by setting
\begin{eqnarray}
K_{1}(t) &=&\rho ^{2},  \notag \\
K_{2}(t) &=&\frac{\Omega ^{2}}{\rho ^{2}}+\frac{1}{a^{2}}\left( c\rho -\dot{%
\rho}\right) ^{2}, \label{Ks} \\
K_{3}(t) &=&\frac{\rho }{a}\left( c\rho -\dot{\rho}\right), \notag
\end{eqnarray}%
where an auxiliary function $\rho(t)$ is introduced to solve Eq.(\ref{Kt}) and
it should obey a dynamical equation of
\begin{equation}
\ddot{\rho}+\chi \left( t\right) \dot{\rho}+\xi \left( t\right)%
\rho =\Omega ^{2}\frac{a^{2}}{\rho ^{3}}, \label{rho}
\end{equation}
where $\Omega$ is an arbitrary real constant, $\chi(t)$ and $\xi(t)$ are defined in
Eq.(\ref{fm}). We can see that Eq.(\ref{rho}) is a different equation from the classical
dynamic equation of Eq.(\ref{fm}) but is similar to the Ermakov equation of Eq.(\ref{Ev})
obtained by a complex solution for Riccati equation. The new dynamical equation of
Eq.(\ref{rho}) coming up here is due to the requirement of keeping the transformed
$\hat{\mathcal{H}}_{U}$ to be a dynamical invariant of Hamiltonian (\ref{H1}).
For the solution of Eq.(\ref{Ks}), Eq.(\ref{Hf}) gives an invariant of
\begin{equation}
\hat{\mathcal{H}}_U=\frac{1}{2}\left( \frac{\Omega }{\rho }\right) ^{2}%
\hat{X}^{2}+\frac{1}{2}\left[ \rho \hat{P}%
+\frac{1}{a}\left( c\rho -\dot{\rho}\right) \hat{X}\right] ^{2}. \label{It}
\end{equation}%

As the functions of $K$s are determined by Eq.(\ref{Ks}), the transformation parameters
of $\theta$s in Eq.(\ref{gtheta}) can provide a general Lie transformation to exactly
solve the wave function of Eq.(\ref{solution}). Surely, for a control problem to obtain
a target state, we can ``design" the functions of $K$s for a controlled system from its
initial Hamiltonian $\hat{\mathcal{H}}(0)$ to construct a final $\hat{\mathcal{H}}(T)$
by an inverse Lie transformation during a finite controlling time interval of $T$ \cite{Lohe,Lejarreta}.

Certainly, the simplest case to determine Lie transformations is to set all the three
parameters of $K$s to be constant (the constant $K$s are the special solutions of Eq.(\ref{Kt})).
In this case, $\hat{\mathcal{H}}_U$ will naturally become an invariant and the initial
time-dependent system $\hat{\mathcal{H}}(t)$ can be turned into a time-independent system
(conservative system) by the time-dependent FUT transformation of $\hat{U}(t)$. Theoretically,
the constant $K$s can always be permitted by this method but, unfortunately, sometimes the
constant $K$s will lead to pathological solutions of $\theta$s which fail to determine the
wave function. In this case, a constructed invariant Hamiltonian from $\hat{\mathcal{H}}_U(0)$
by an inverse Lie transformation of $\hat{\mathcal{H}}(t)=\hat{U}(t)\hat{\mathcal{H}}_U\hat{U}^{-1}(t)$
is needed. In the following part, we will apply this Lie transformation method to different
models of DDPO and show how to solve or construct an invariant Hamiltonian with different
choice of $K$s. A general case of $K$s is considered in the end.

\subsection{Some typical models of DDPO}

In this section we will apply the above method to some
specific problems of DDPO by adopting different parametric
functions of $a(t), \cdots, f(t)$ in Hamiltonian of Eq.(\ref{Hs}).
A typical Hamiltonian widely used in the literature to described a DDPO
is \cite{Kanai}%
\begin{equation}
\hat{H}(t)=e^{-2\gamma t}\frac{\hat{p}^{2}}{2m}+e^{2\gamma t}\left[ \frac{1}{2}m\omega
^{2}\hat{x}^{2}-\hat{x}F\left( t\right) \right] .  \label{Hddd}
\end{equation}%
If the mass $m$ and the frequency $\omega$ of the parametric oscillator are
time-independent ($m\rightarrow m_0, \omega\rightarrow \omega_0$), this
Hermitian Hamiltonian with real parameters has a scaled form of%
\begin{equation}
\hat{\mathcal{H}}(t)=\frac{1}{2}e^{-2\gamma t}\hat{P}^{2}+\frac{1}{2}e^{2\gamma t}\hat{X}%
^{2}-e^{2\gamma t}f_c\left( t\right)\hat{X},
\label{Hd}%
\end{equation}%
where the system parameters are $a(t)=e^{-2\gamma t}$, $b(t)=e^{2\gamma t}$ and
$e(t)=-e^{2\gamma t}f_c(t)$.
The initial Hamiltonian of this model is
\begin{equation*}
\hat{\mathcal{H}}(0)=\frac{1}{2}\hat{P}^{2}+\frac{1}{2}\hat{X}%
^{2}-f_c\left( 0\right)\hat{X} ,
\end{equation*}%
which means initially the system is a harmonic oscillator under
a driving force of $f_c(t)$ and we can set $f_c(0)=0$
for simplicity.
By using the time-dependent Lie transformation method, we can
recover Eq.(\ref{Geq}) by Eq.(\ref{fm}) as
\begin{equation}
\ddot{X}_c+2\gamma \dot{X}_c+X_c =f_c\left( t\right) ,  \label{dy}
\end{equation}%
which is exactly the classical dynamical equation discussed in section II.
Here we consider again a simple driving case of
$f_c\left( t\right) =f_{0}\cos \left( \Omega t+\phi \right)$
and the classical Lagrangian given by Eq.(\ref{Lag0}) is
\begin{equation*}
\mathcal{L}\left( X_c ,\dot{X}_c ,t\right) =\frac{1}{2}e^{2\gamma
t}X_{c}^{2}-\frac{1}{2}e^{-2\gamma t}\dot{X}_{c}^{2}-f_{0}e^{2\gamma t}
\cos \left( \Omega t+\phi \right) X_c.
\end{equation*}
After the transformation of $\hat{U}_1$ on $h(4)$ algebra, the Hamiltonian
reduces to a pure quadratic form of
\begin{equation}
\hat{\mathcal{H}}_{1}=\frac{1}{2}e^{-2\gamma t}\hat{P}^{2}%
+\frac{1}{2}e^{2\gamma t}\hat{X}^{2},
\label{H11}%
\end{equation}%
and it becomes a well-known Hamiltonian for a free harmonic oscillator with
an exponentially increasing effective mass ($e^{2\gamma t}$). Then the followed transformation on
$su(2)$ algebra gives
\begin{equation}
\hat{\mathcal{H}}_{U}(t)=\frac{1}{2}K_{1}\hat{P}^{2}+\frac{1}{2}K_{2}\hat{X}%
^{2}+\frac{1}{2}K_{3}\left( \hat{X}\hat{P}+\hat{P}\hat{X}\right),
\label{Hf1}
\end{equation}
and the parametric equations are
\begin{eqnarray}
\dot{\theta}_{+} &=&e^{-2\gamma t}\theta _{+}^{2}-K_{2}e^{-2\theta _{0}}+e^{2\gamma t},
\notag \\
\dot{\theta}_{-} &=&e^{-2\gamma t}e^{-2\theta _{0}}-K_{2}\theta _{-}^{2}+2K_{3}\theta
_{-}-K_{1},  \label{theta1} \\
\dot{\theta}_{0} &=&K_{2}\theta _{-}-e^{-2\gamma t}\theta _{+}-K_{3}.  \notag
\end{eqnarray}%s
Surely, the solutions of $\theta$s depend on the choice of $K$s.
Now we combine different types of $K$s to determine transformation
parameters of $\theta $s by transforming the final Hamiltonian $\hat{\mathcal{H}}_{U}$
into differen solvable forms.

\subsubsection{Exact solvable case of $Ks=0$}
\label{ks0}
%--Figure---
\begin{figure}[htp]
\begin{center}
\includegraphics[width=0.43 \textwidth]{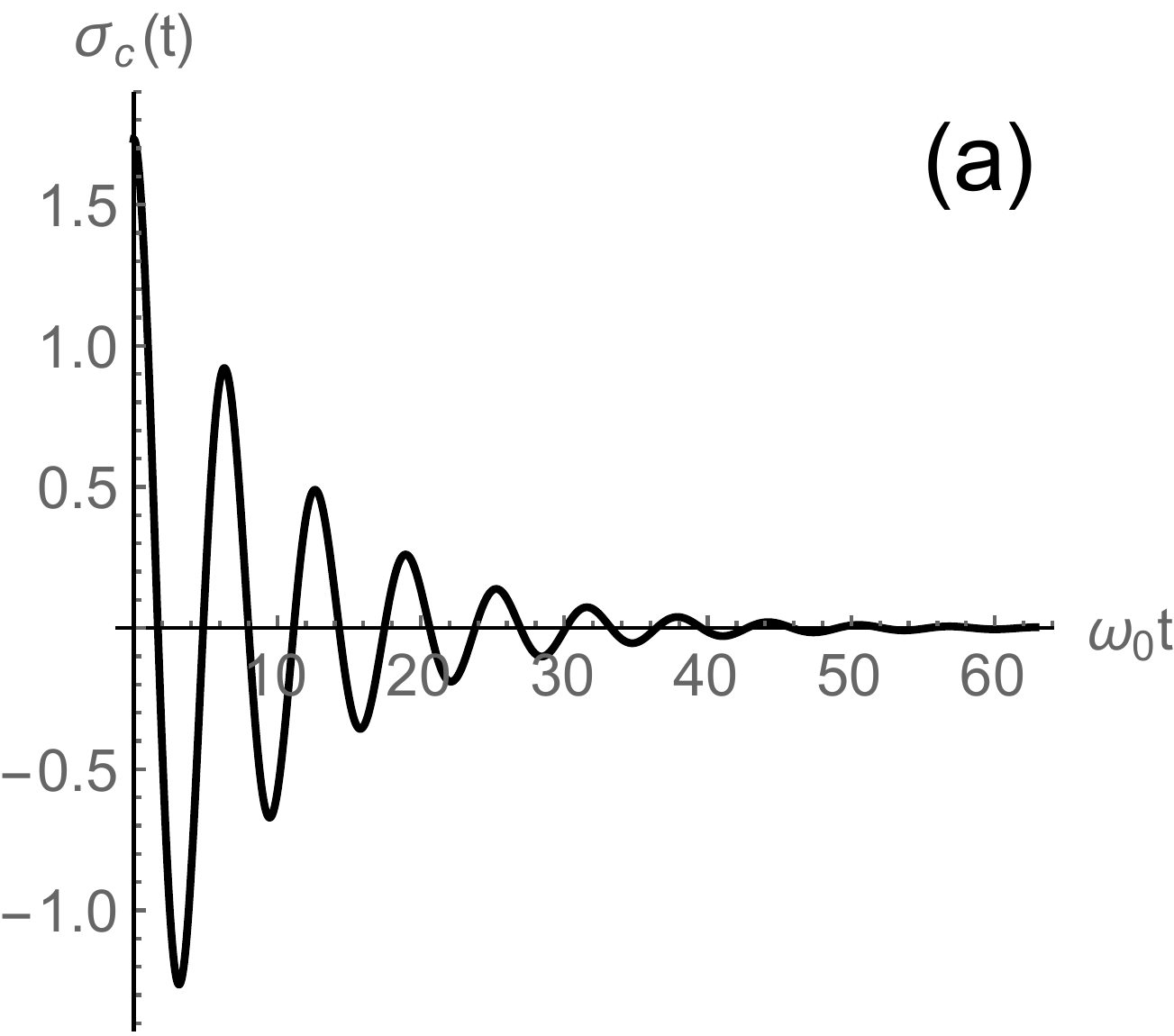}%
\includegraphics[width=0.43 \textwidth]{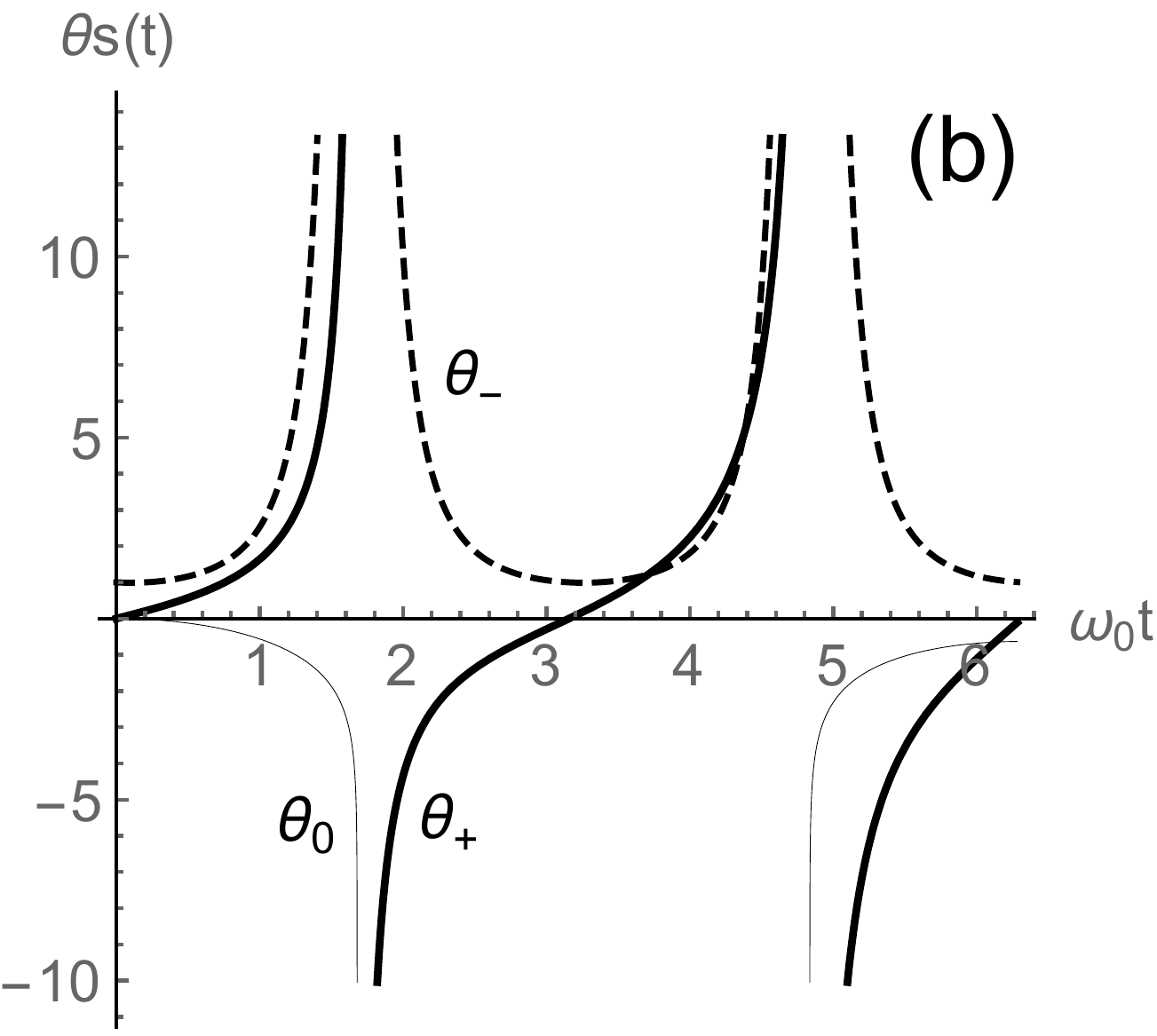}
\end{center}
\caption{The dynamics of parameters of (a) $\sigma(t)$ and (b)
$\theta_{+}(t), \theta_{0}(t)$, and $\theta_{-}(t)$. The other
parameters are $\gamma=0.1$ and $\sigma_0=\sqrt{3}$.}
\label{figure5}
\end{figure}
The simplest case is to set $K_1=K_2=K_3=0$ in Eq.(\ref{theta1}) and the
transformed Hamiltonian Eq.(\ref{Hf1}) will shift to a zero point of energy.
In principle, no constraints exclude this simplest case and the real solution
for the transformation parameter $\theta_{+}$ is%
\begin{equation}
\theta _{+}\left( t\right) =-e^{2\gamma t}\frac{\dot{\sigma}_{c}}{\sigma _{c}%
},\label{eq1}
\end{equation}%
where $\sigma_{c}$ satisfies the homogenous equation for a damped harmonic oscillator
as%
\begin{equation}
\ddot{\sigma}_{c}+2\gamma \dot{\sigma}_{c}+\sigma _{c}=0.  \label{dh}
\end{equation}%
The solution of Eq.(\ref{dh}) is well known
\begin{equation*}
\sigma _{c}\left( t\right) =\sigma _{0}e^{-\gamma t}\left[ \cos \left( \sqrt{%
1-\gamma ^{2}}t\right) +\frac{\gamma }{\sqrt{1-\gamma ^{2}}}\sin \left(
\sqrt{1-\gamma ^{2}}t\right) \right],
\end{equation*}%
where $\sigma_0$ is the initial value of $\sigma_c$. Then the other two transformation
parameters read%
\begin{eqnarray*}
\theta _{0}\left( t\right) =\ln \left( \frac{\sigma _{c}}{\sigma _{0}}%
\right) , \quad
\theta _{-}\left( t\right) =e^{-2\gamma t}\left( \frac{\sigma _{0}}{\sigma
_{c}}\right) ^{2}.  \notag
\end{eqnarray*}%
\begin{figure}[b]
\begin{center}
\includegraphics[width=0.33 \textwidth]{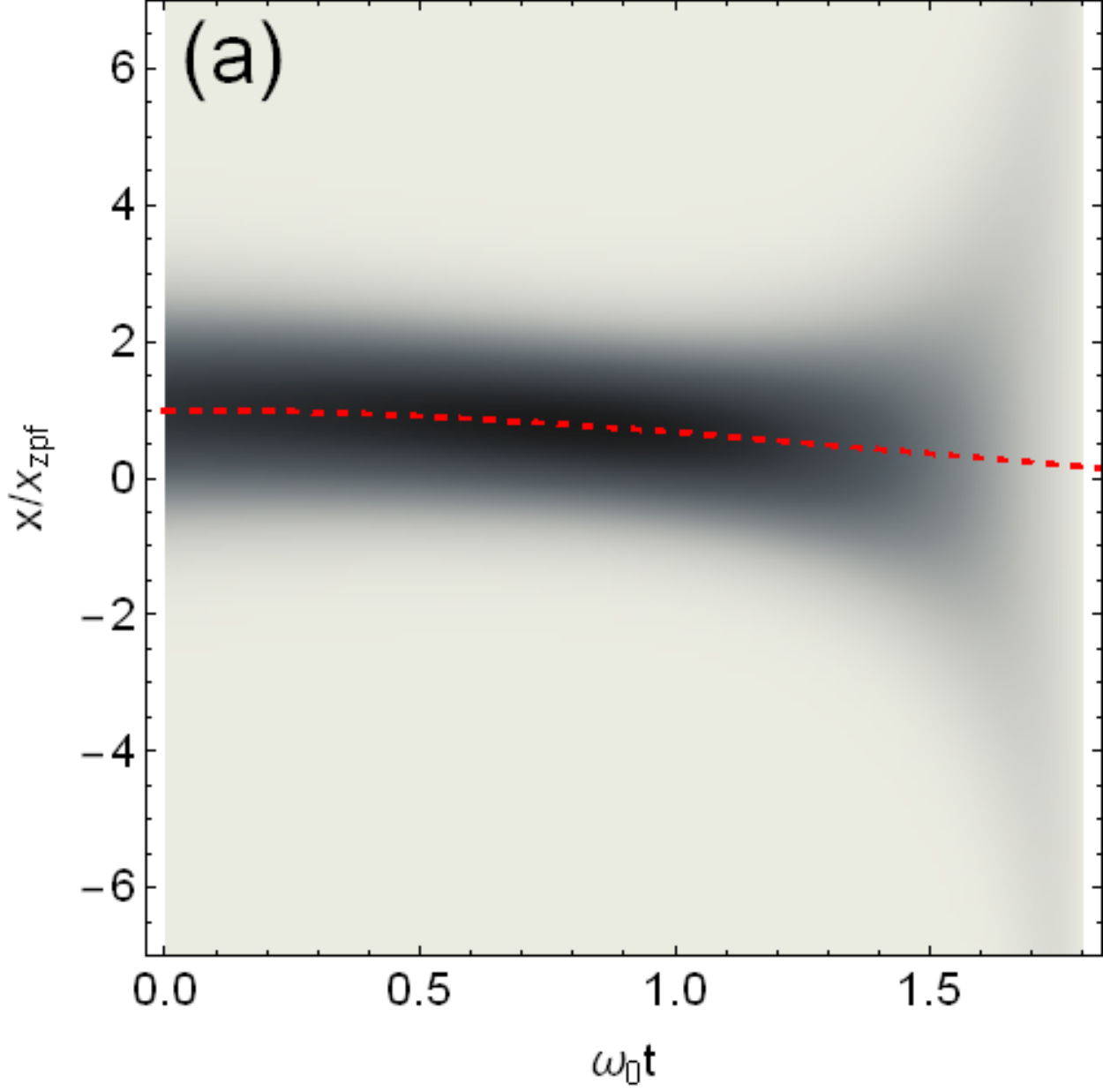}%
\includegraphics[width=0.33 \textwidth]{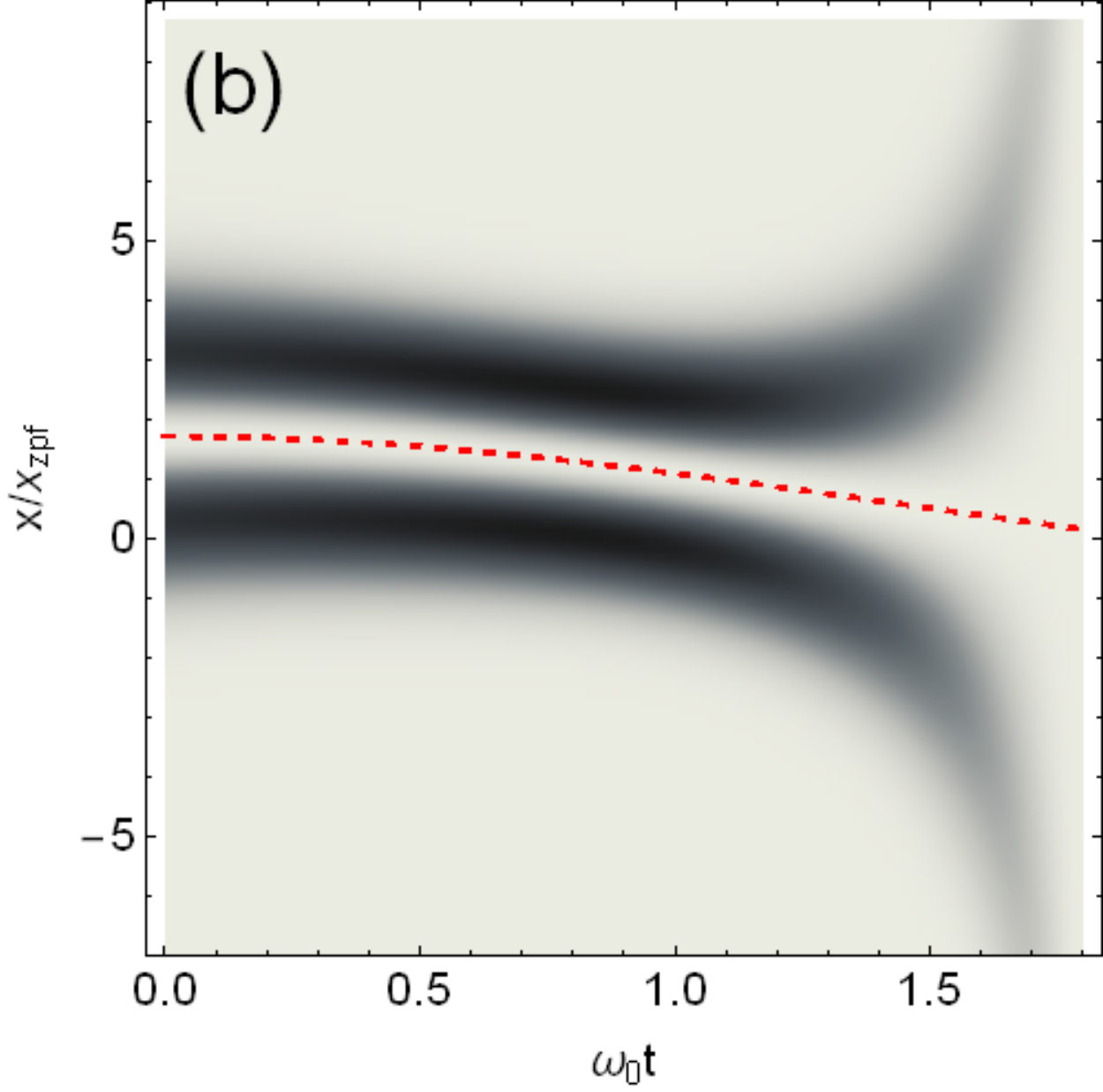}%
\includegraphics[width=0.33 \textwidth]{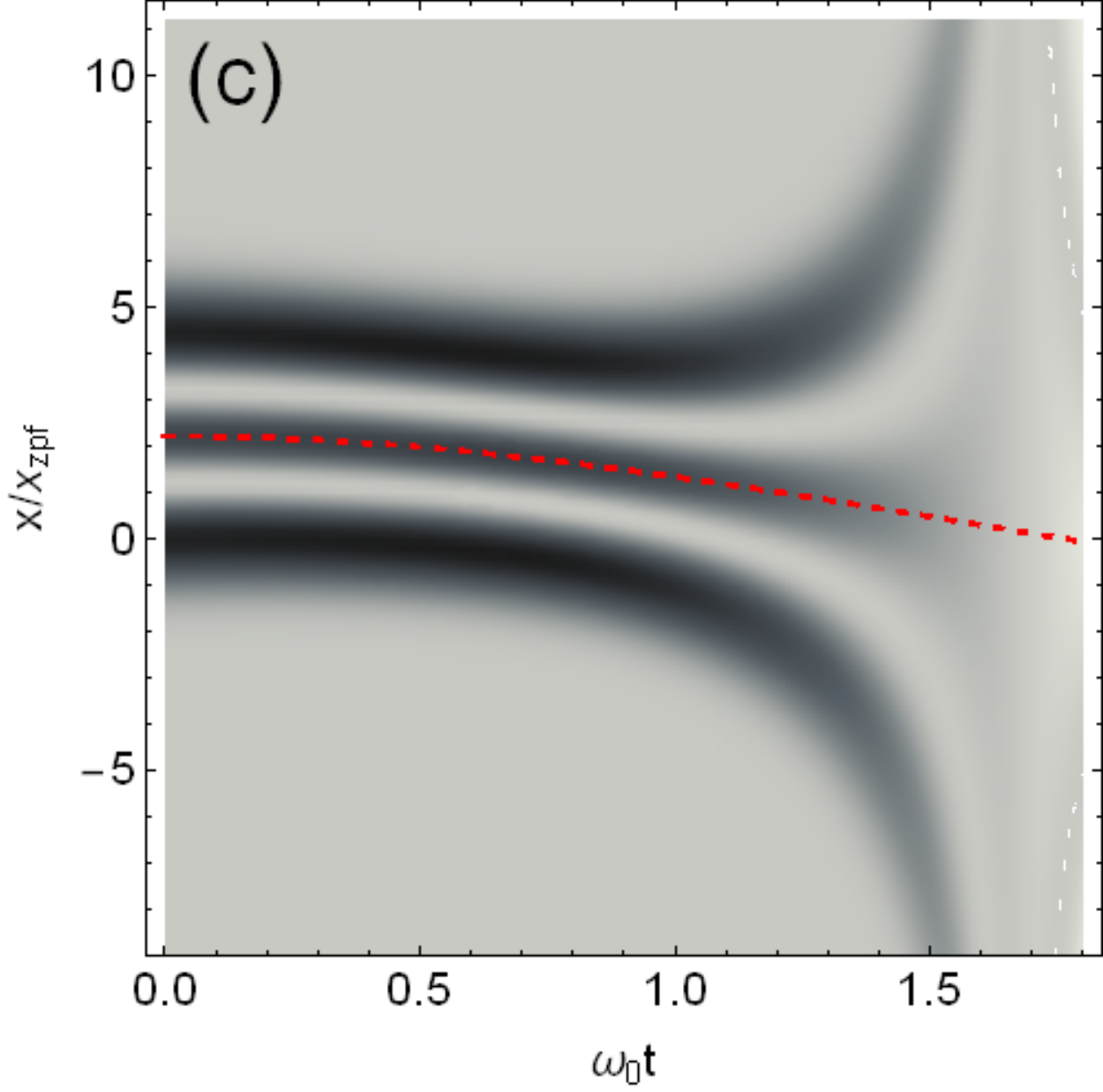}%
\end{center}
\caption{The evolution of probability distributions around the classical orbits
(red dashed lines) for the DDPO
starting from different initial states of (a) $|0\rangle$, (b) $|1\rangle$
and (c) $|2\rangle$ under a harmonic driving force of $f_c=0.2 \cos(0.8\omega_0 t)$.
The damping rate of the oscillator is $\gamma=0.2$ and the initial
conditions for (a) $X_c(0)=1, P_{c}(0)=0$, (b) $X_c(0)=\sqrt{3}, P_c(0)=0$ and
(c) $X_c(0)=\sqrt{5},P_c(0)=0$ according to the corresponding energies of the
number state.}
\label{figure6}
\end{figure}

Fig.\ref{figure5} shows the dynamics of the parameters of $\sigma_c(t)$ and $\theta $s
in this simplest case. As the parameter $\sigma _{c}\left( t\right) $ follows Eq.(\ref{dh}),
we can see clearly that $\theta$s become divergent when%
\begin{equation}
\cot \left( \sqrt{1-\gamma ^{2}}t\right) =-\frac{\gamma }{\sqrt{1-\gamma ^{2}%
}}.  \label{diver}
\end{equation}%
At the divergent time, $\theta$s become singular and the solution will loose its physical
meaning. If the initial wave function is in the eigenstate of the harmonic oscillator
\begin{equation*}
\varphi _{n}\left( X,0\right) =A_{n}e^{-\frac{1}{2}X^{2}}H_{n}\left(
X\right) ,\quad \left\vert A_{n}\right\vert ^{2}=\frac{1}{\sqrt{\pi }2^{n}n!},
\end{equation*}
then the wave function at time $t$ will be %%
\begin{equation}
\Psi \left( X,t\right) =e^{-iE_n t}e^{-i\int_{0}^{t}\mathcal{L}\left( X_{c},\dot{X}%
_{c},t\right) dt}e^{-\theta _{0}/2}e^{-iX_{c}\hat{P}}e^{i\left( e^{-2\gamma
t}\dot{X}_{c}\right) \hat{X}}e^{-i\theta _{+}\hat{X}^{2}/2}e^{-i\theta _{0}%
\hat{X}\hat{P}}e^{-i\theta _{-}\hat{P}^{2}/2}\varphi _{n}\left( X\right),
\label{psi3}
\end{equation}%
where the quantum energy $E_n$ is in the unit of $\hbar \omega_0$. In Fig.\ref{figure6},
we calculate the probability distributions of $\Psi \left(X,t\right)$ starting from
different initial states and clearly find the singular behaviors of Eq.(\ref{psi3}) compared
with the case without damping in Fig.\ref{figure4}. When the parameters $\theta $s become
divergent, the wave packet deviates from the classical orbit (dashed lines) and quickly
spreads out (see Eq.(\ref{dX}) or Eq.(\ref{dP})) at the time determined by Eq.(\ref{diver}).
We can see that the simplest case of $K_s=0$ can not give a convergent long-time solution
of the DDPO described by Caldirola-Kanai Hamiltonian of Eq.(\ref{Hddd}).
%--Figure---
\begin{figure}[htp]
\begin{center}
\includegraphics[width=0.32 \textwidth]{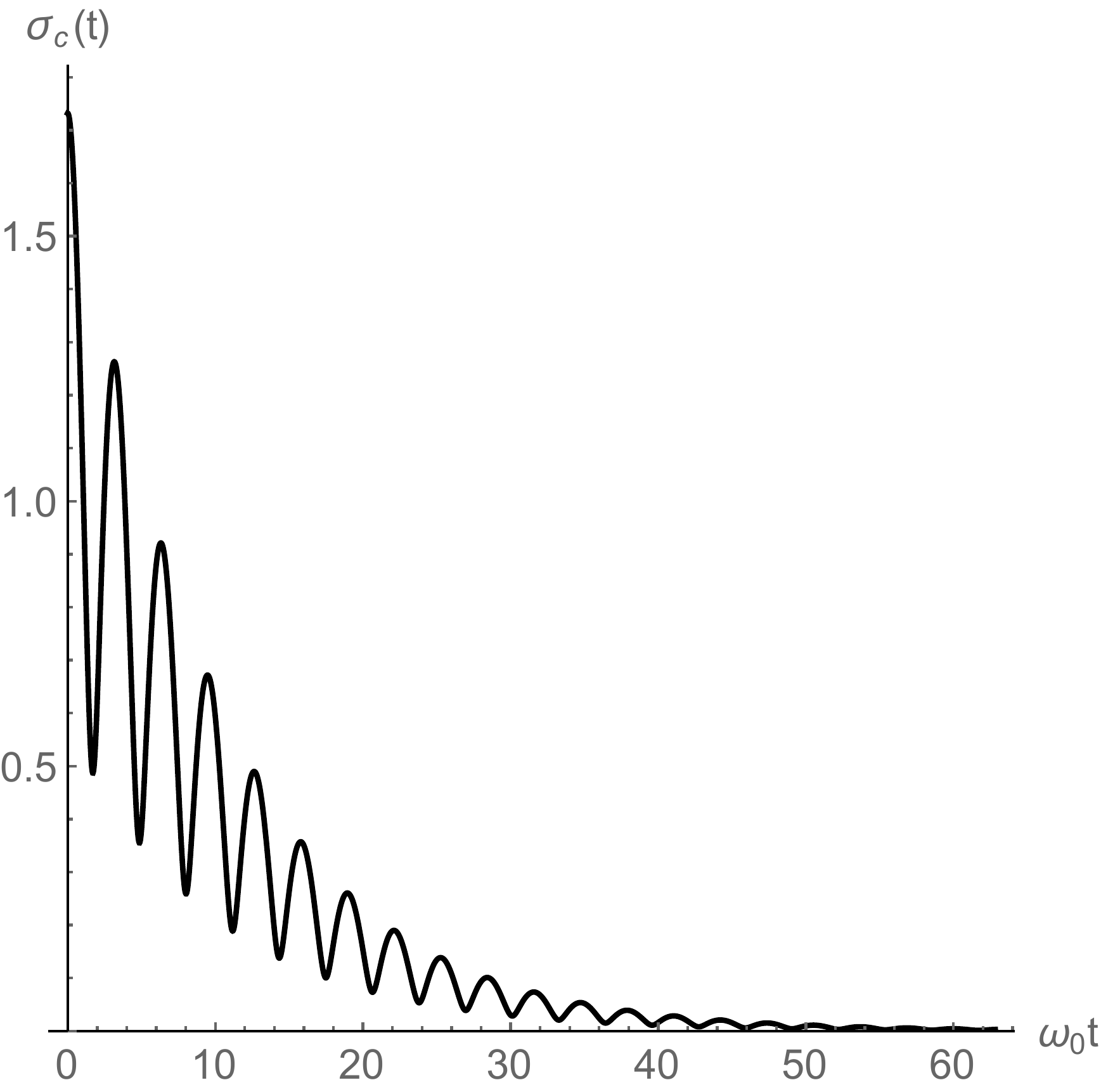}%
\includegraphics[width=0.32 \textwidth]{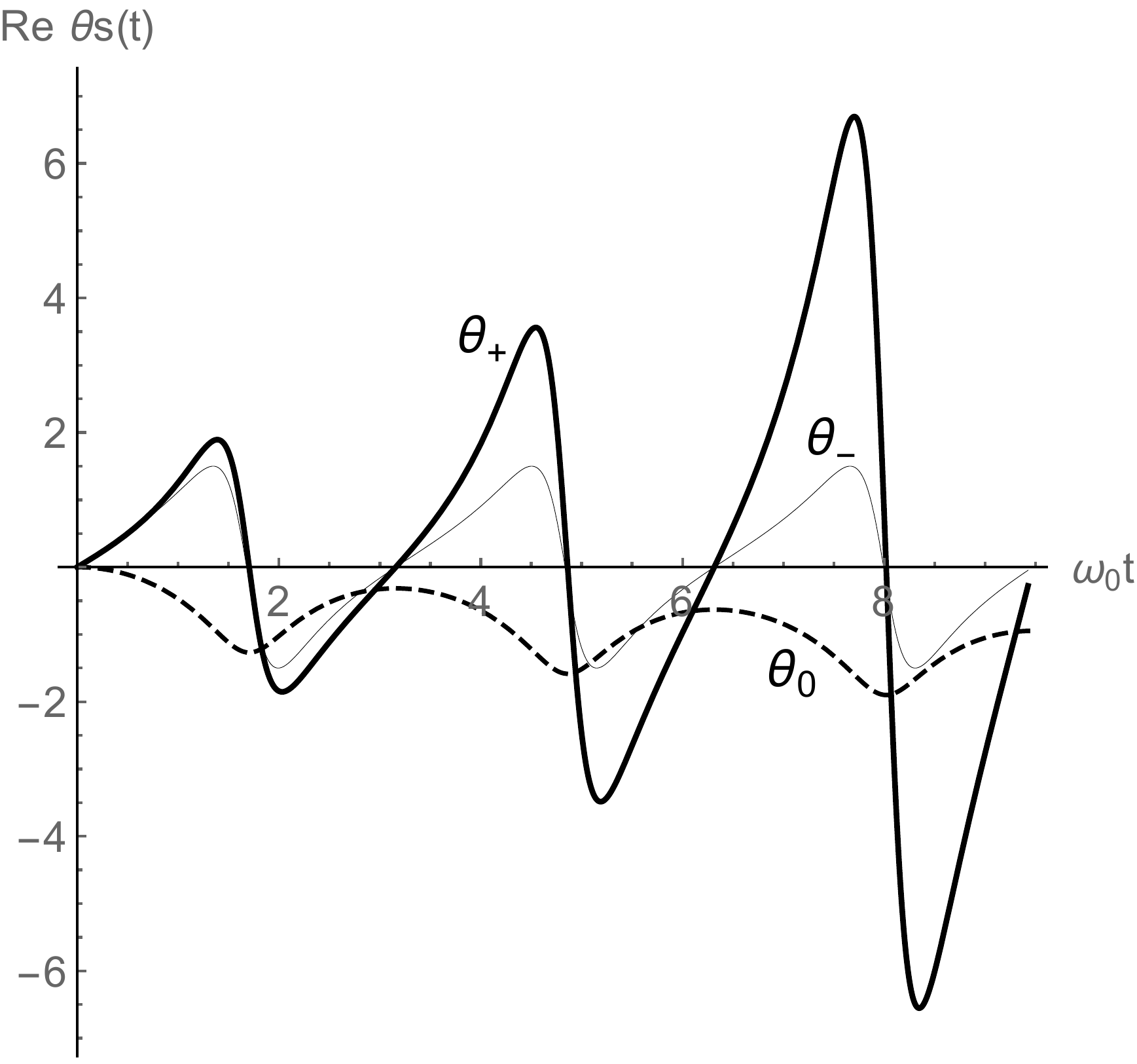}%
\includegraphics[width=0.32 \textwidth]{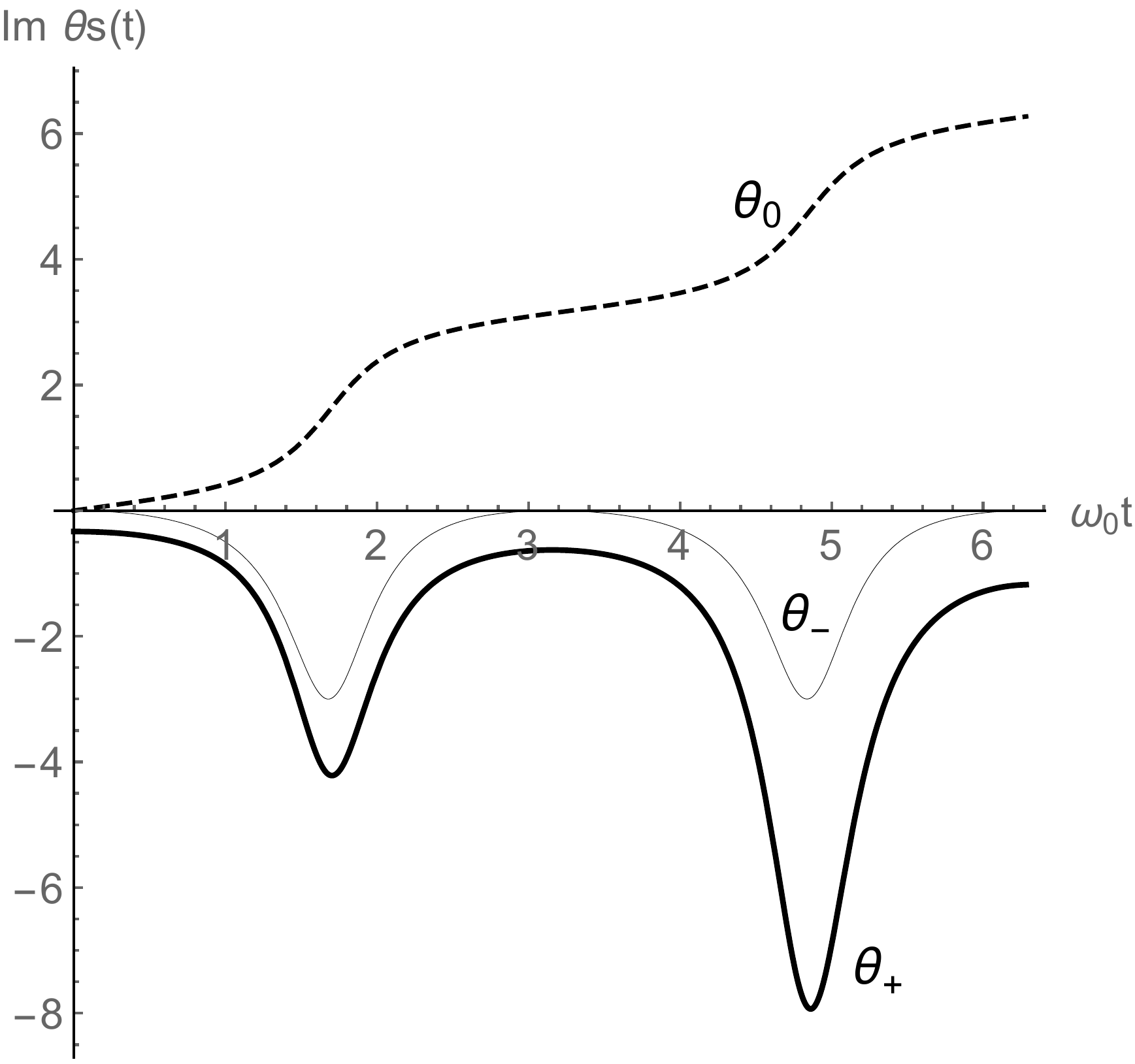}%
\end{center}
\caption{The dynamics of parameter $\protect\sigma _{c}(t)$ and the
dynamic behaviors of the real and imaginary parts of $\protect\theta %
_{+}, \protect\theta _{0}, \protect\theta _{-}$.
The other parameters are the same as that in Fig.\ref{figure5}.}
\label{figure7}%
\end{figure}%

Therefore, we can formally construct the complex $\theta$s of Eq.(\ref{cu}) for
a system described by a non-Hermitian Hamiltonian with complex coefficients \cite{Hans}.
Then the complex solution of $\theta_+$ is%
\begin{equation*}
\theta _{+}\left( t\right) =-e^{2\gamma t}\frac{\dot{\sigma}_{c}}{\sigma _{c}%
}-i\frac{1}{\sigma _{c}^{2}},
\end{equation*}%
and the classical dynamical equation changes to%
\begin{equation*}
\ddot{\sigma}_{c}+2\gamma \dot{\sigma}_{c}+\sigma _{c}=\frac{e^{-4\gamma t}}{%
\sigma _{c}^{3}}.
\end{equation*}%
The other two complex parameters are%
\begin{eqnarray*}
\theta _{0}\left( t\right) =\ln \left( \frac{\sigma _{c}}{\sigma _{0}}%
\right) +i\theta \left( t\right) , \quad
\theta _{-}\left( t\right) =e^{-2\gamma t}\left( \frac{\sigma _{0}}{\sigma
_{c}}\right) ^{2}e^{-2i\theta \left( t\right) },
\end{eqnarray*}%
where%
\begin{equation*}
\theta \left( t\right) =\int_{0}^{t}\frac{e^{-2\gamma \tau }}{\sigma _{c}^{2}%
}d\tau .
\end{equation*}%

In this case the dynamics of the parameters $\theta$s are improved at the divergent
times, as shown in Fig.\ref{figure7}, but, still, $\theta$s will go to infinity in
the long time limit. As the complex $\theta$s break the unitary evolution of the wave function,
the probability density of the wave function will be distorted by the imaginary parts
of $\theta $s compared with that of Eq.(\ref{psi3}) shown in Fig.\ref{figure6}.
%--Figure---
\begin{figure}[htp]
\begin{center}
\includegraphics[width=0.33 \textwidth]{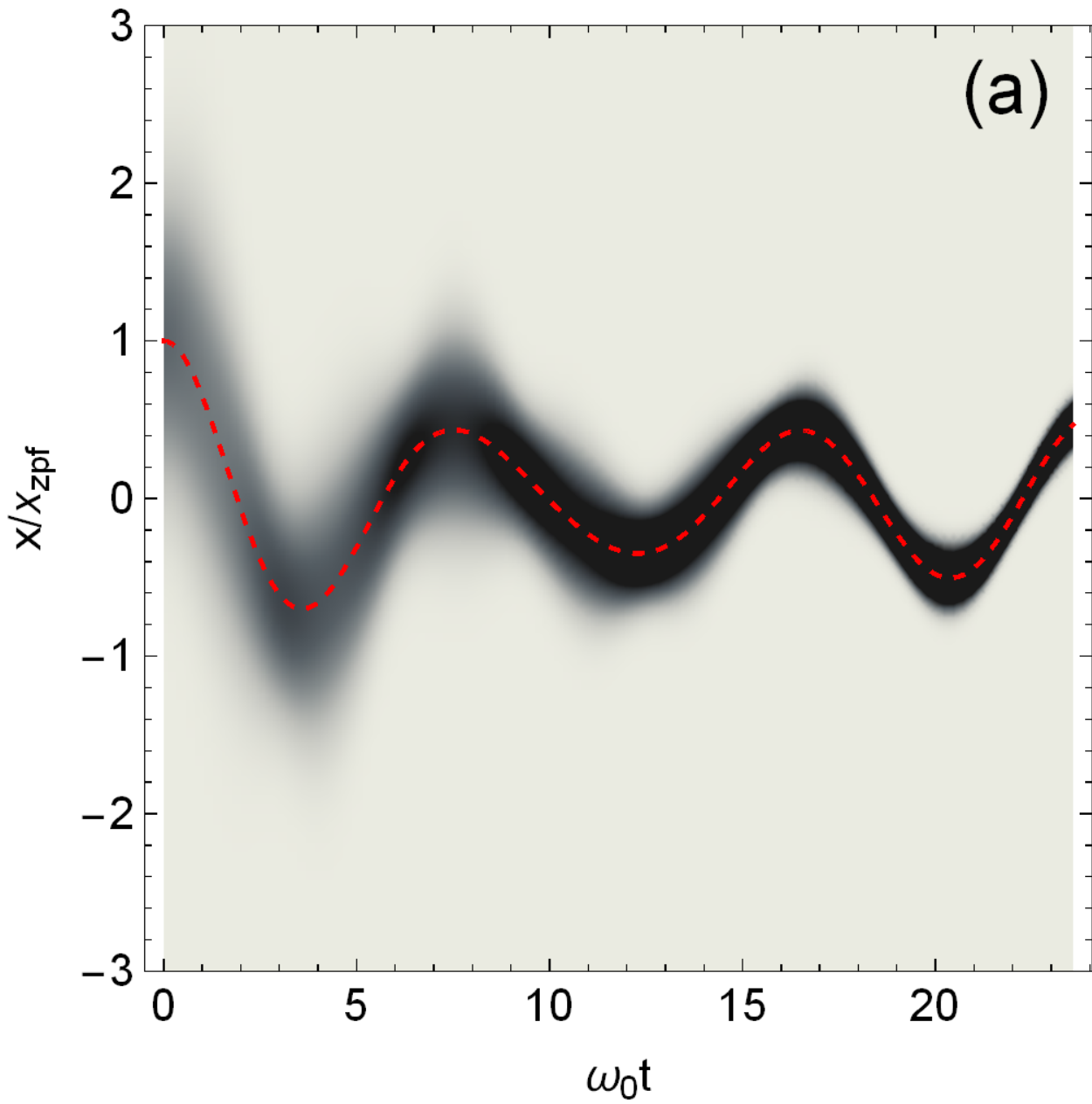}%
\includegraphics[width=0.33 \textwidth]{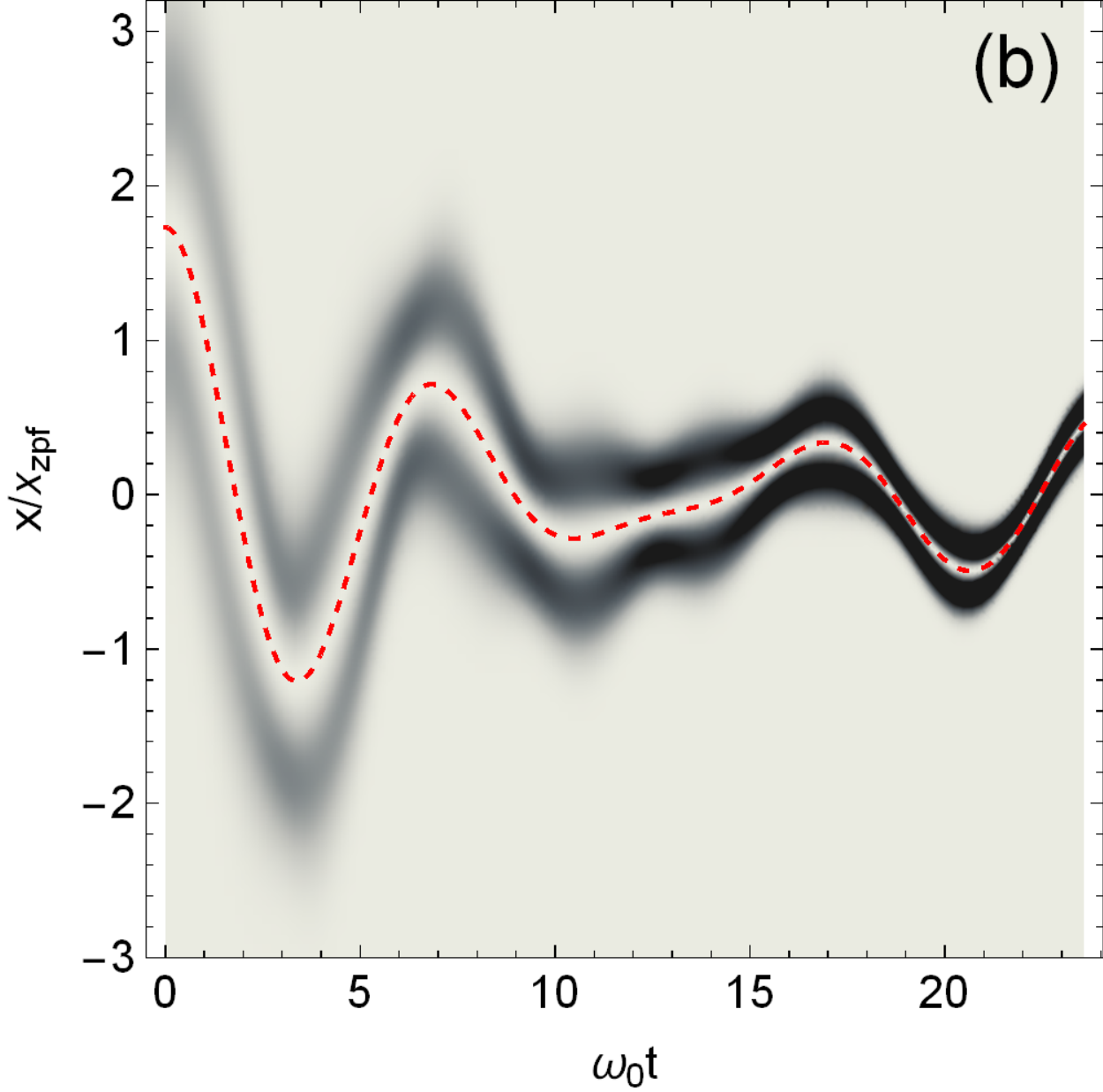}%
\includegraphics[width=0.33 \textwidth]{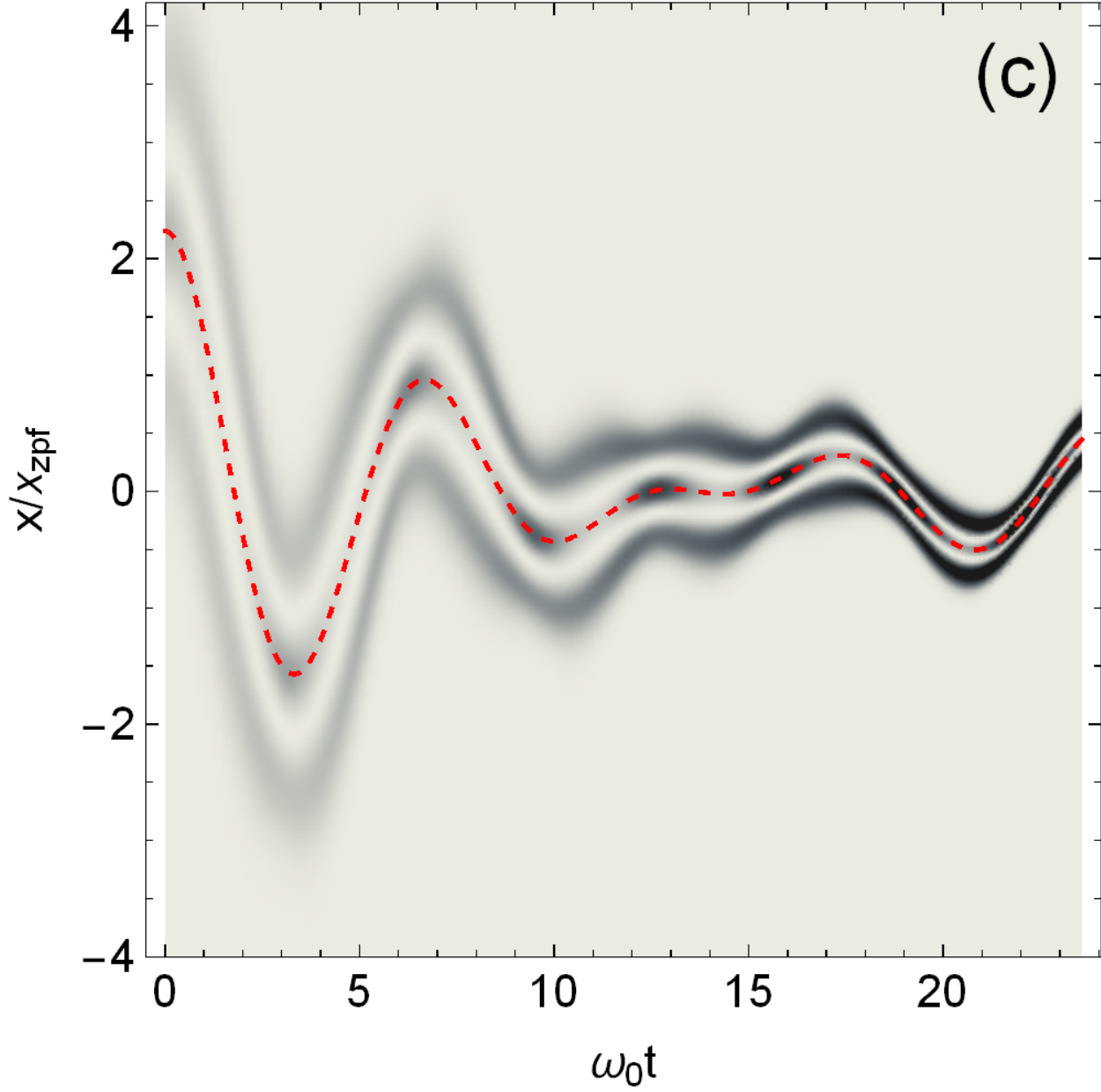}%
\end{center}
\caption{The evolution of the wave packets starting from number states of
(a) $|0\rangle$, (b) $|1\rangle$ and (c) $|2\rangle$
by using the complex parameter method.
The other parameters are the same as that in Fig.\ref{figure6}.}
\label{figure8}
\end{figure}
The evolutions of the probability distributions starting from different
number states are displayed in Fig.\ref{figure8}. We can see that the spread
of the wave packet is removed by a localization of the wave packet induced by
the imaginary parts of $\theta$s. As the amplitudes of the parameters $\theta$s
continuously increase with time, the local probability density will finally exceed
a value when the solution again becomes invalid. The above results demonstrate that,
although the Lie transformation method seems neat to get an exact quantum solution
if we set all the parameters of $K$s to be zero, we can't always find
physical wave functions for the time-dependent quantum system, especially, for a
dissipative one ($\sigma_c \rightarrow 0$). Certainly, a better solution can be
obtained if we choose other values of $K$s.

\subsubsection{The driven parametric harmonic oscillator}

If, initially, the system $\hat{\mathcal{H}}(0)$ is a free harmonic oscillator,
then we can naturally set $K_{1}=1$, $K_{2}=\Omega_{0}^2 $ and $K_{3}=0$ for a
better choice to improve the dynamical behavior of the transformation parameters.
In this case, the time-dependent Hamiltonian of Eq.(\ref{Hd}) can be
transformed into a time-independent harmonic oscillator as
\begin{equation}
\hat{\mathcal{H}}_{U}=\frac{1}{2}\hat{P}^{2}+\frac{1}{2}\Omega_{0}^{2} \hat{X}^{2},%
\label{ho}%
\end{equation}%
provided that the transformation parameters of $\theta$s
obey the following equations of motion
\begin{eqnarray}
\dot{\theta}_{+} &=&e^{-2\gamma t}\theta _{+}^{2}-\Omega_{0}^{2}e^{-2\theta
_{0}}+e^{2\gamma t},  \notag \\
\dot{\theta}_{-} &=&e^{-2\gamma t}e^{-2\theta _{0}}-\Omega_{0}^{2}\theta
_{-}^{2}-1,  \label{harmonic} \\
\dot{\theta}_{0} &=&\Omega_{0}^{2}\theta _{-}-e^{-2\gamma t}\theta _{+}.  \notag
\end{eqnarray}%
If the parameters $\theta$s are perfectly solved, the wave function will be
\begin{equation*}
\Psi \left( X,t\right) =\hat{U}\left( t\right) \psi_0 \left( X,t\right) =\hat{U%
}_{1}\left( t\right) \hat{U}_{2}\left( t\right) \psi_0 \left( X,t\right),
\end{equation*}%
and the initial wave function $\psi_0(X,t)$ can be expressed by
\begin{equation}
\psi_0 \left( X,t\right)=\sum_{n}C_n \varphi _{n}\left( X,t\right),
\label{psi0}
\end{equation}%
where $C_n$ are complex constants determined by the initial population distribution and
\begin{equation*}
\varphi _{n}\left( X,t\right)  =\frac{1}{\sqrt{\sqrt{\pi }2^{n}n!}}e^{-%
\frac{1}{2}X^{2}}H_{n}\left( X\right) e^{-iE_{n}t}, \quad
E_{n} =\Omega_0 \left( n+\frac{1}{2}\right) .
\end{equation*}%
%--Figure---
\begin{figure}[t]
\begin{center}
\includegraphics[width=0.40 \textwidth]{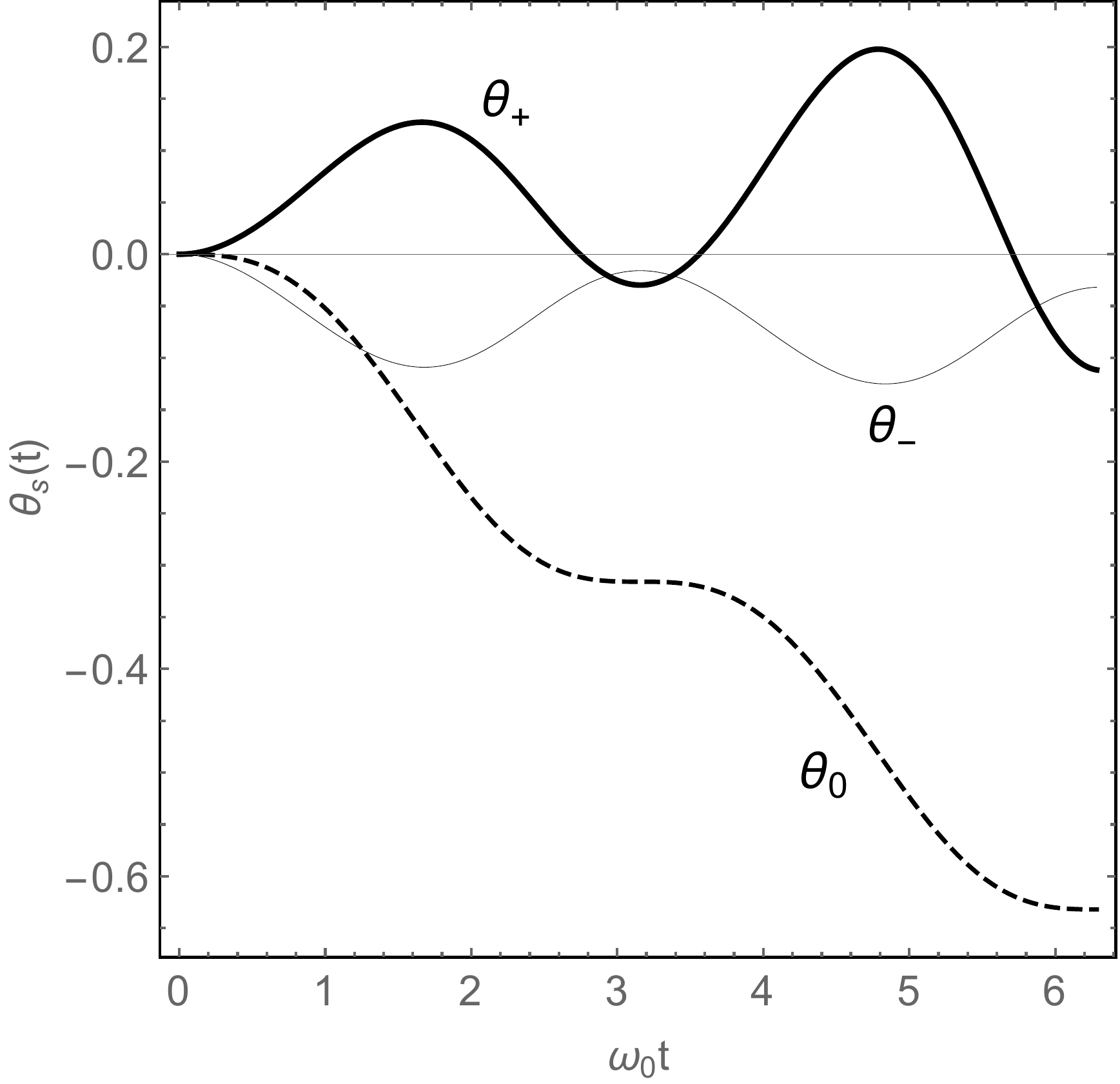}%
\includegraphics[width=0.40 \textwidth]{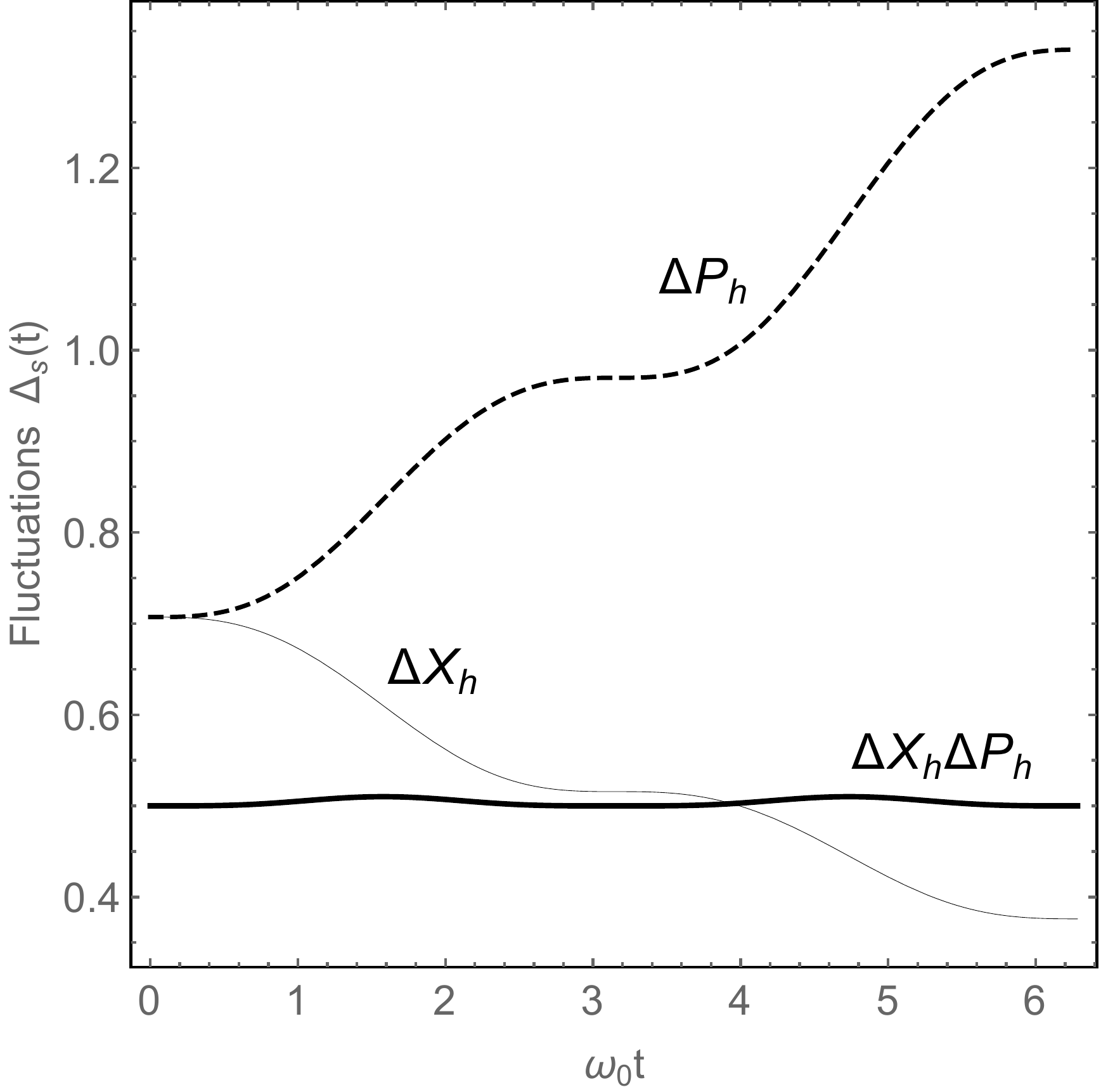}\\[0pt]
\includegraphics[width=0.33 \textwidth]{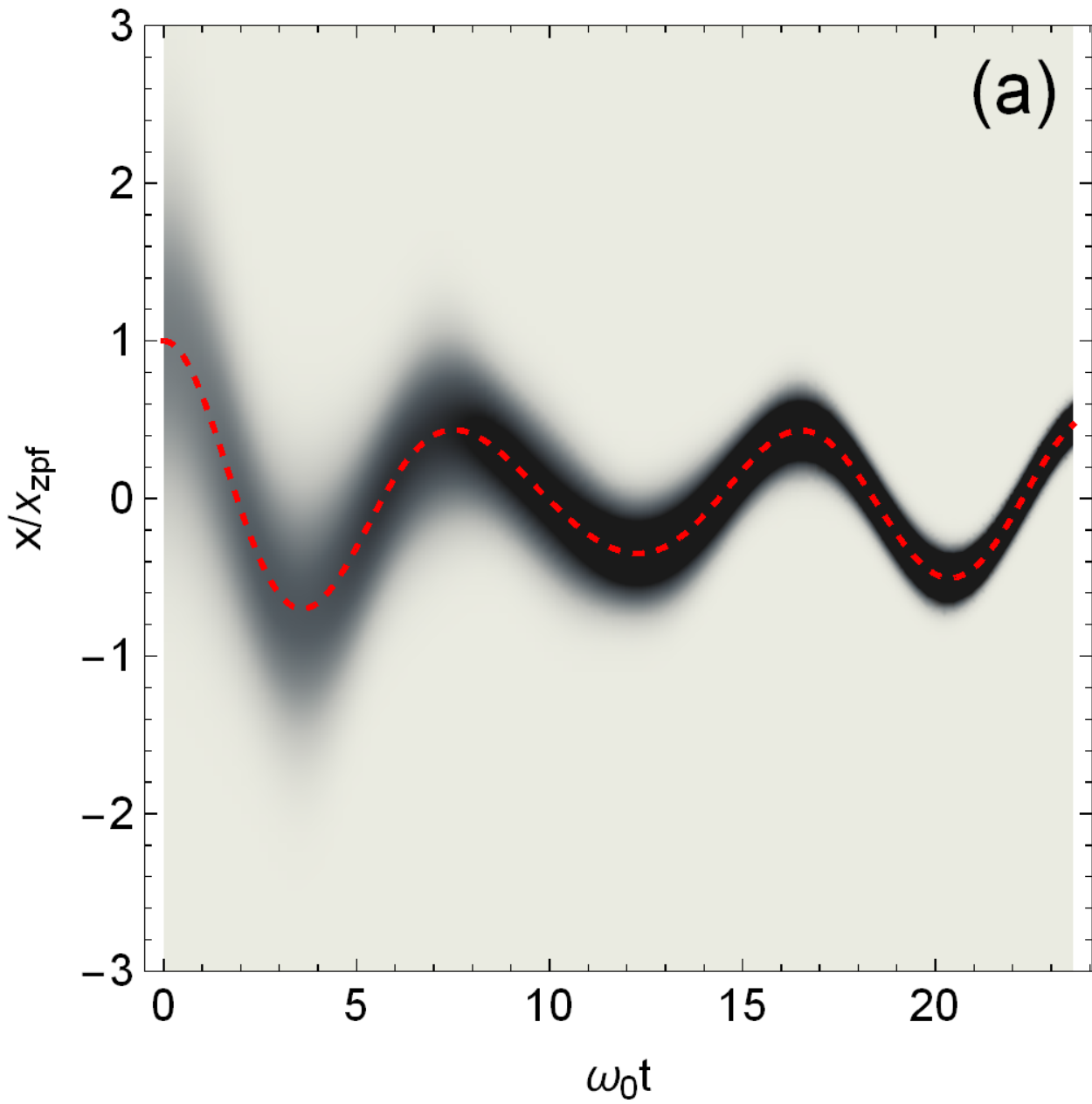}%
\includegraphics[width=0.33 \textwidth]{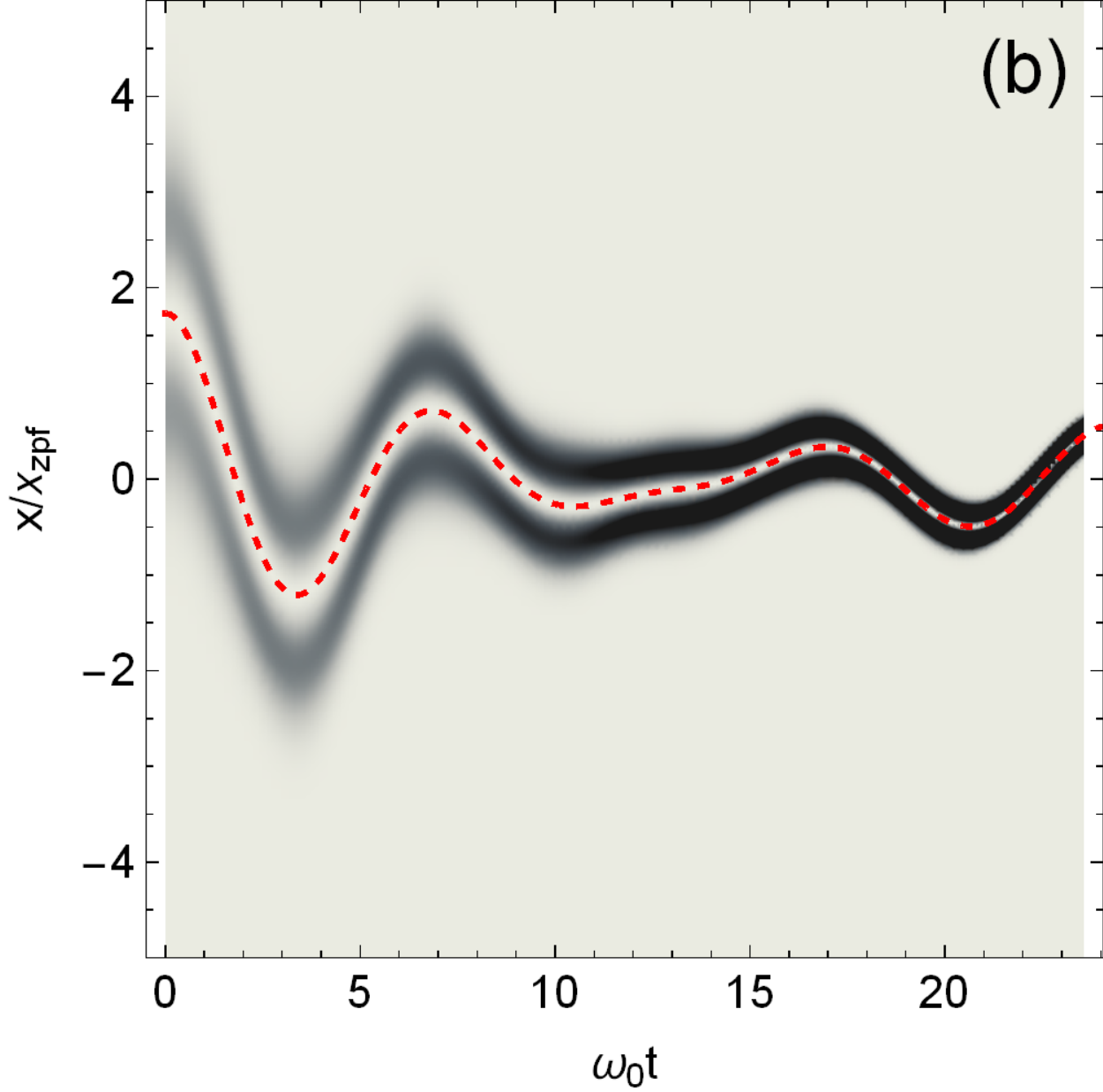}%
\includegraphics[width=0.33 \textwidth]{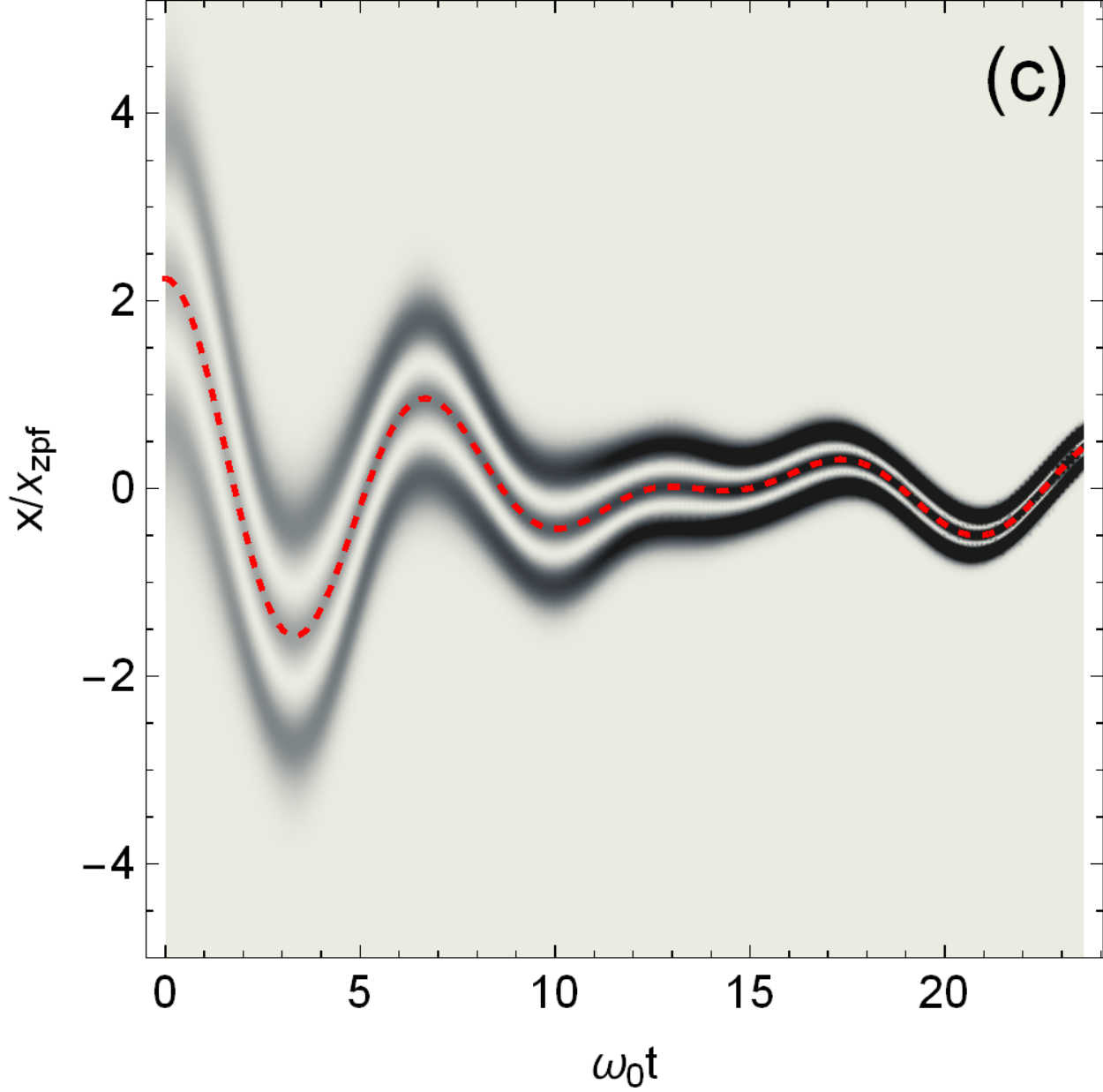}%
\end{center}
\caption{(Upper frames) The dynamical behaviors of the parameters of $\theta %
_{+}, \protect\theta _{0}, \protect\theta _{-}$ and the fluctuations
of position $\Delta X_h$ and momentum
$\Delta P_h$ as well as the uncertainty of
$\Delta X_h\cdot\Delta P_h$ for $\Omega_0=1$. (Lower frames) The evolutions
of the wave packets starting from the initial states of (a) $|0\rangle$, (b) $|1\rangle$
and (c) $|2\rangle$. The other parameters are the same as that in Fig.\protect\ref{figure6}.}
\label{figure9}
\end{figure}

In this case, the dynamics of $\theta $s (left), the variances $\Delta X_h, \Delta P_h$ and
the position-momentum uncertainty $\Delta X_h \Delta P_h$ (right) of the wave packet starting
from the ground state $|0\rangle$ are shown in the upper frames of Fig.\ref{figure9}. The
lower frames of Fig.\ref{figure9}(a)-(c) demonstrate the probability distributions
of the wave packets starting from different number states. Fig.\ref{figure9} reveals that
a squeezing of fluctuation in position is in the expense of a dilation
in momentum and the wave function will gradually go to a localized displaced number state in
real space, and then finally it explodes in momentum space due to $\Delta P_h\rightarrow \infty$
($\Delta X_h\rightarrow 0$). Unfortunately, a further introduced parameter $\Omega_0$ still fails
to stabilize the longtime evolution of the wave packet in this case. We find that only $\Omega_0=1$ is a better
choice for the damping case because Eq.(\ref{harmonic}) becomes unstable if $\Omega_0$ deviates a
little from 1. During a quantum state evolution, the damping rate $\gamma$ plays an important role
in compressing the wave packet in real space through $\theta$s and the state evolution
is exactly the same as that shown in Fig.\ref{figure8} by the complex variables of $\theta$s.
The highly compressed wave packet after a long time evolution is due to an unboundedly increasing
of the effective mass of the oscillator ($e^{2\gamma t}$) and clearly is unreasonable, but for a control problem,
the solution for a short time interval is still physically acceptable.

In order to give more insight into the state evolution based on Eq.(\ref{ho}),
we consider a specific model of a parametric oscillator with time-dependent
mass and frequency as
\begin{equation}
\hat{H}\left( t\right) =\frac{1}{2m\left( t\right) }\hat{p}^{2}+\frac{1}{2}%
m\left( t\right) \omega ^{2}\left( t\right) \hat{x}^{2}.  \label{har}
\end{equation}%
If $m(0)=m_{0}$ and $\omega \left( 0\right) =\omega _{0}$,
the scaled Hamiltonian in the units of Eq.(\ref{zpf0}) is
\begin{equation*}
\mathcal{\hat{H}}(t)=\frac{1}{2M\left( t\right) }\hat{P}^{2}+\frac{1}{2}M\left(
t\right) \Omega_{0} ^{2}\left( t\right) \hat{X}^{2},
\end{equation*}%
where $M\left( t\right) =m\left( t\right) /m_{0}$, $\Omega_0 \left( t\right)
=\omega \left( t\right) /\omega _{0}$ and $\mathcal{\hat{H}}$ is in unit of $%
\hbar \omega _{0}$.
Above Hamiltonian has a $su(2)$ algebraic structure and we can use
Lie transformations of
\begin{equation}
\hat{U}_{2}(t)=e^{-i\frac{\theta _{+}}{2}\hat{X}^{2}}e^{-i\frac{\theta _{0}}{2}%
\left( \hat{X}\hat{P}+\hat{P}\hat{X}\right) }e^{-i\frac{\theta _{-}}{2}\hat{P%
}^{2}}
\label{U2t}
\end{equation}%
to reverse it into the initial Hamiltonian of $\hat{\mathcal{H}}(0)$ as
\begin{equation*}
\mathcal{\hat{H}}_{U}=\hat{\mathcal{H}}(0)=\frac{1}{2}%
\hat{P}^{2}+\frac{1}{2} \hat{X}^{2},
\end{equation*}%
where $M(0)=\Omega_0(0)=1$. Therefore, in this case, the transformation
parameters must follow ($K_1=K_2=1$)
\begin{eqnarray}
\dot{\theta}_{+} &=&\frac{1}{M(t)}\theta _{+}^{2}-e^{-2\theta
_{0}}+M\left( t\right) \Omega_{0} ^{2}\left( t\right) ,  \notag \\
\dot{\theta}_{-} &=&\frac{1}{M(t)}e^{-2\theta _{0}}- \theta
_{-}^{2}-1 ,  \label{har2} \\
\dot{\theta}_{0} &=&\theta _{-}-\frac{1}{M(t)}\theta _{+},  \notag
\end{eqnarray}
which, clearly, is a specific case of Eq.(\ref{ho}) for $\Omega_0=1$. Actually, the condition
of Eq.(\ref{har2}) having a steady state is only for $\Omega_{0}^2(t)=1$.
The classical equation of a parametric oscillator
determined by Eq.(\ref{fm}) in this case is
\begin{equation}
\ddot{X}_{c}+\frac{\dot{M}(t)}{M(t)}\dot{X}_{c}+\Omega_{0} ^{2}\left( t\right)
X_{c}=0.
\label{harno}
\end{equation}%
%--Figure---
\begin{figure}[htp]
\begin{center}
\includegraphics[width=0.25 \textwidth]{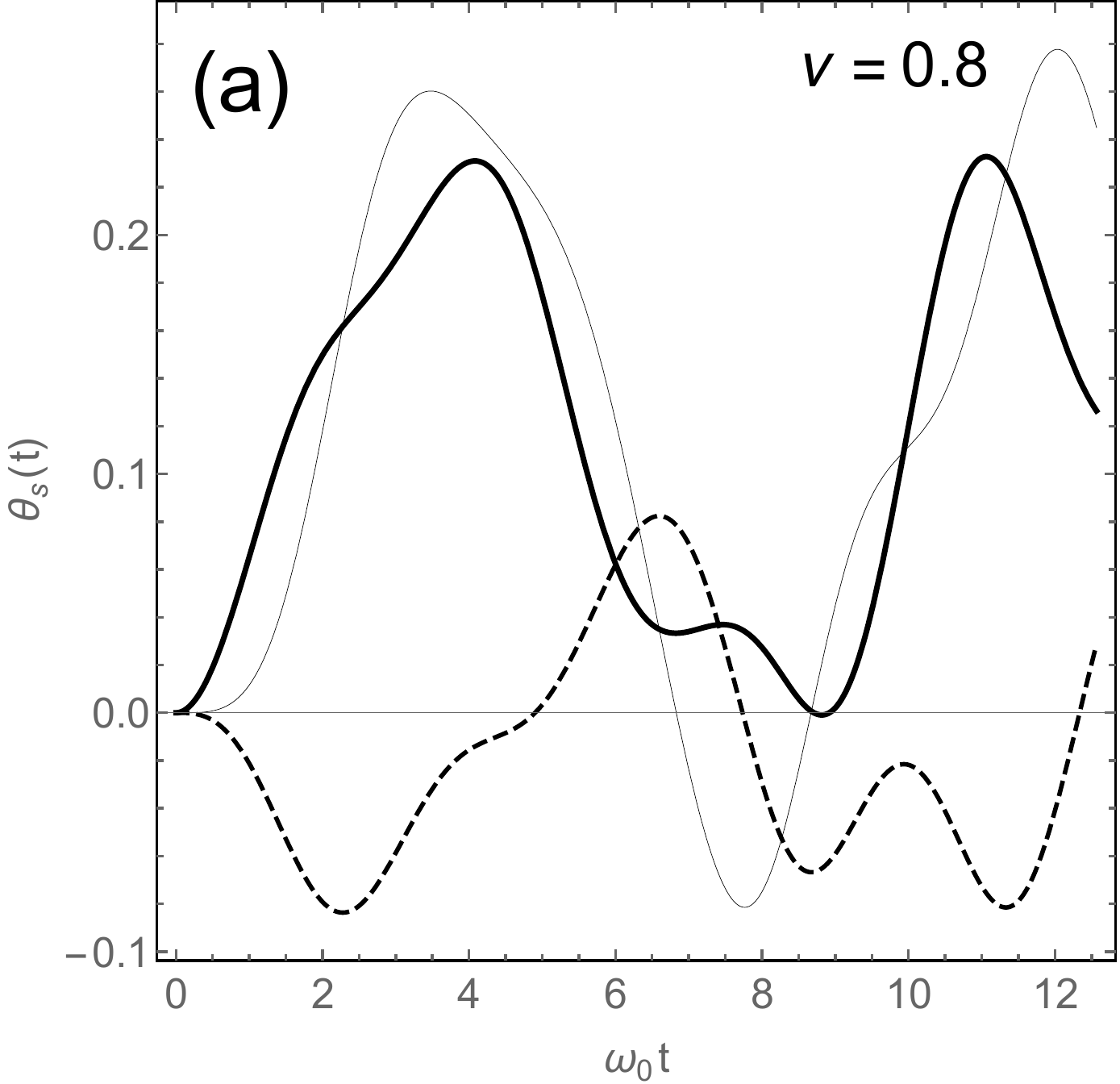}%
\includegraphics[width=0.25 \textwidth]{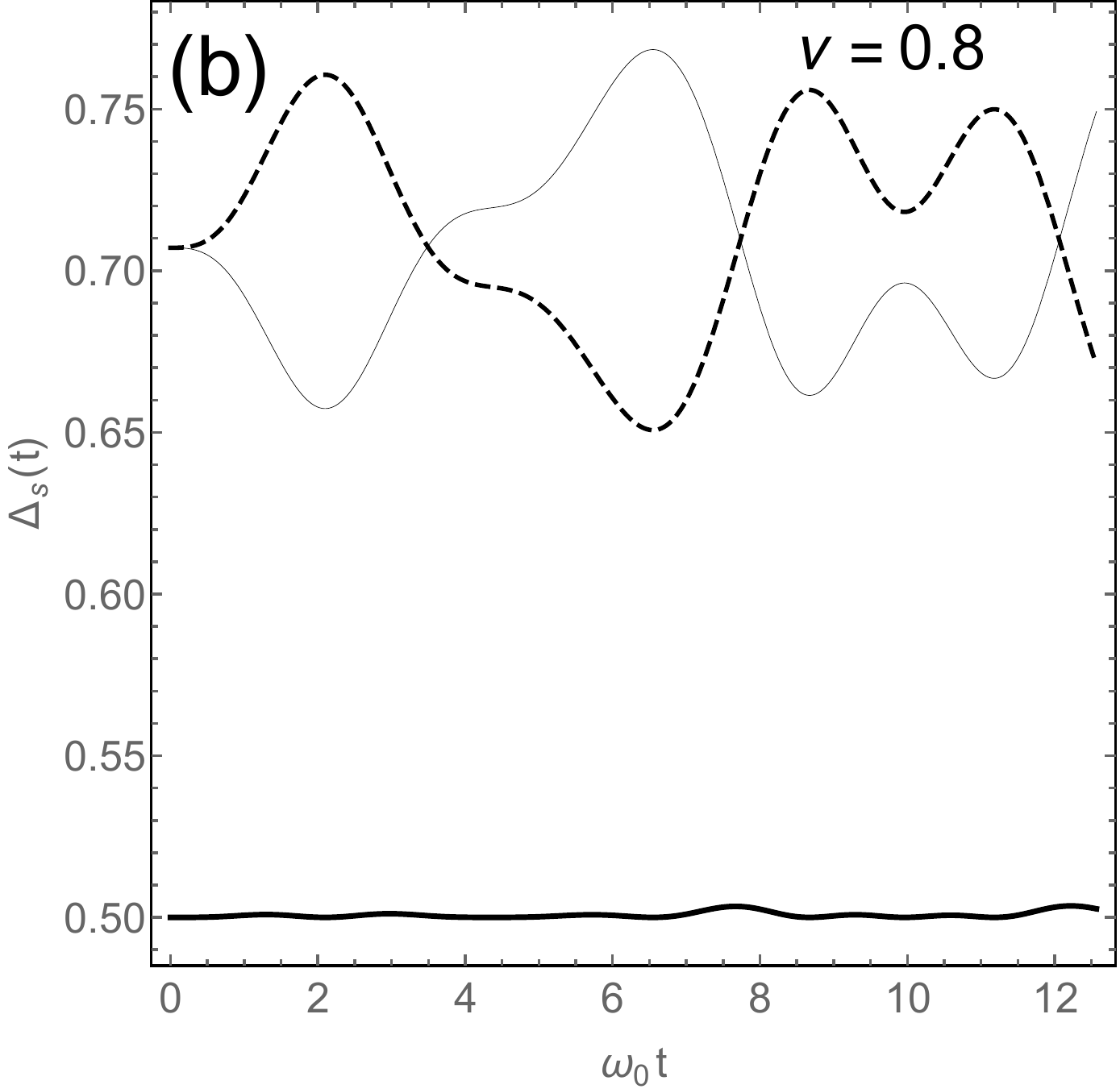}%
\includegraphics[width=0.25 \textwidth]{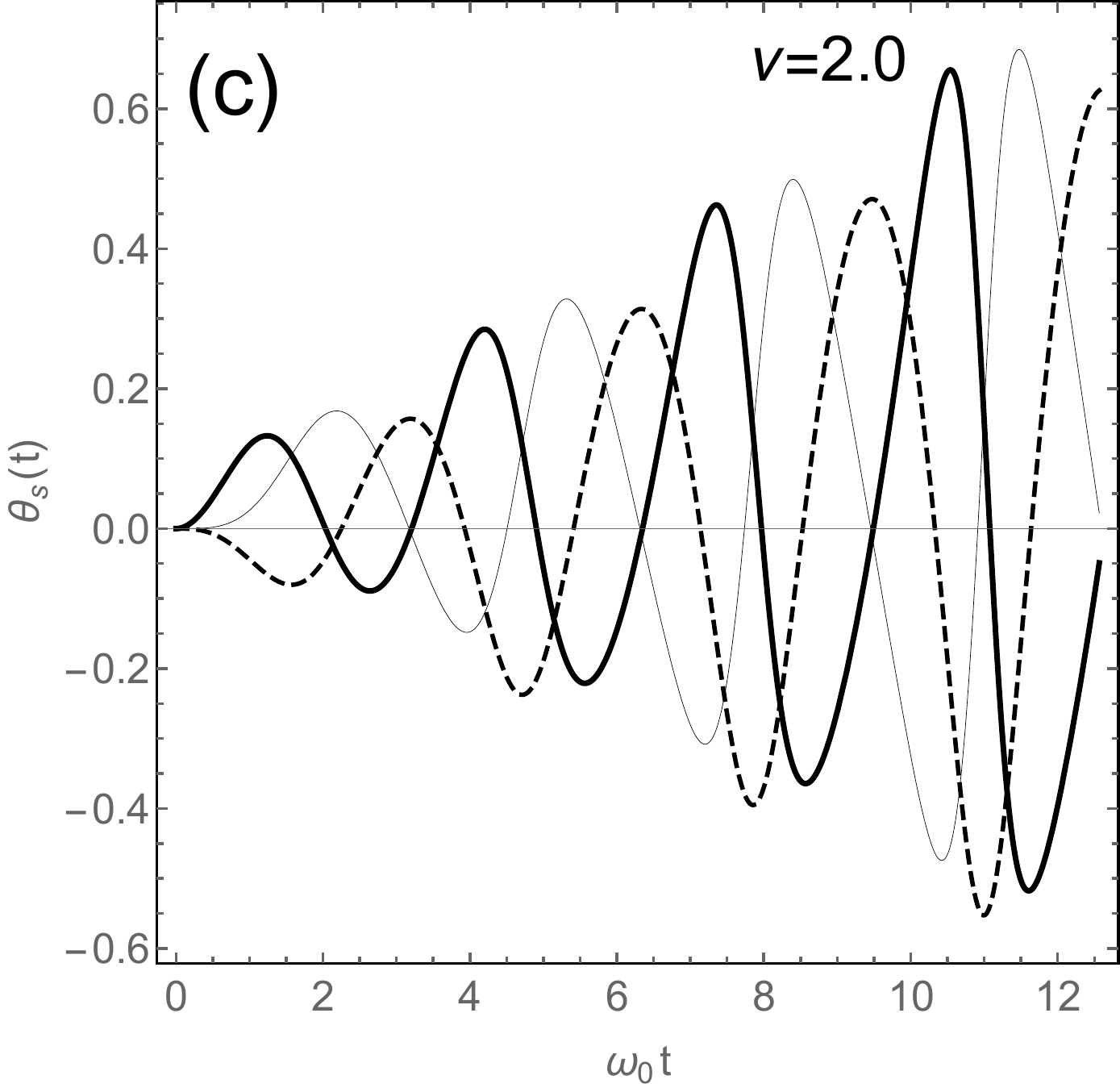}%
\includegraphics[width=0.25 \textwidth]{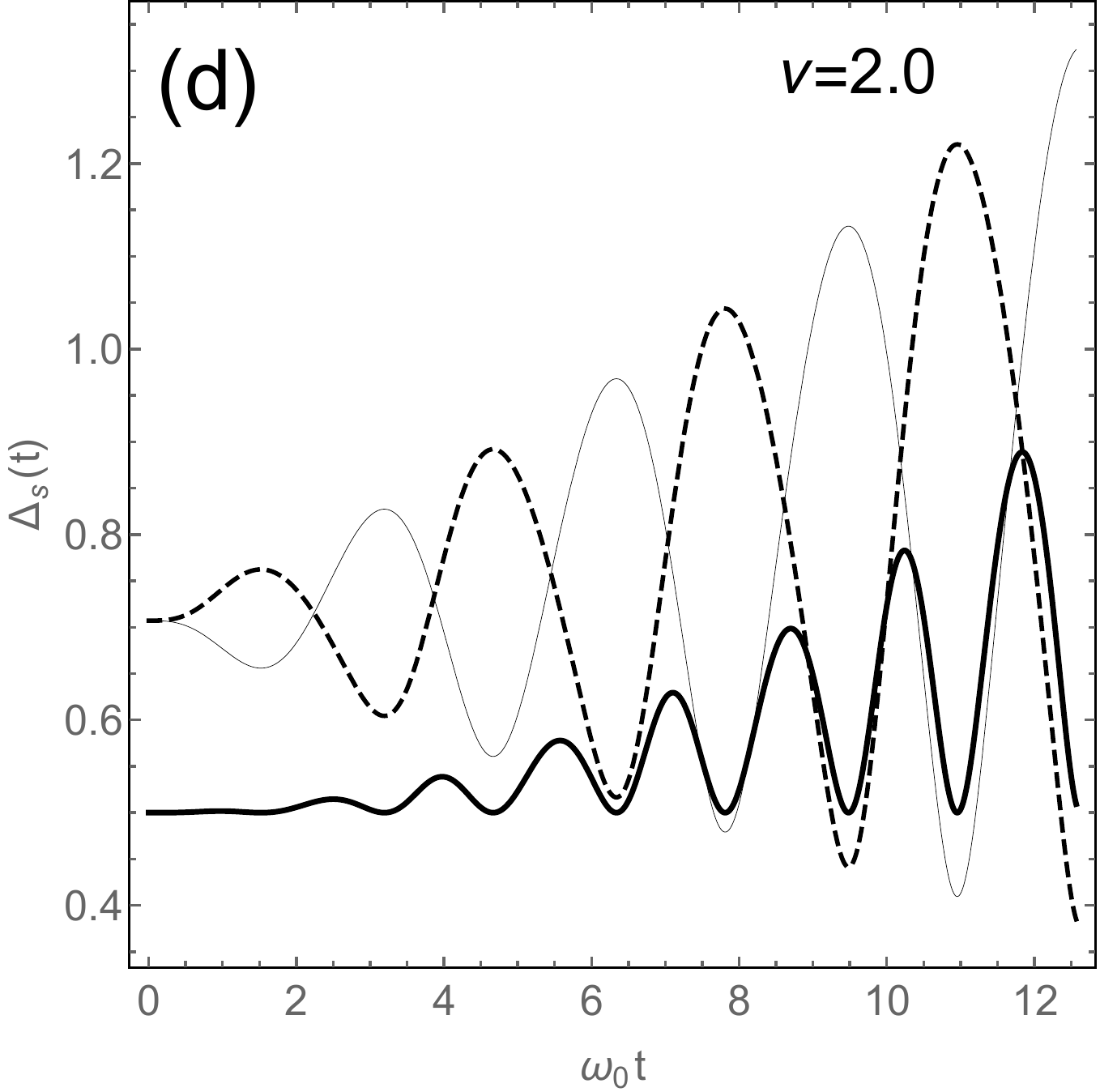}%
\end{center}
\caption{(a)(c)The dynamical behaviors of parameters $\theta %
_{+}$ (thick lines), $\theta _{0}$ (dashed lines), $\theta _{-}$
(thin lines) and (b)(d) the corresponding fluctuations
of position $\Delta X_h$ (thin lines), momentum $\Delta P_h$ (dashed lines) and the
uncertainty of $\Delta X_h\cdot\Delta P_h$ (thick lines) for the parametric harmonic
oscillator starting from ground state $|0\rangle$ under different modulating frequencies of $\nu$.
The modulating amplitude on the frequency is $\lambda=0.2$.}
\label{figure10}
\end{figure}

\begin{figure}[htp]
\begin{center}
\includegraphics[width=0.25 \textwidth]{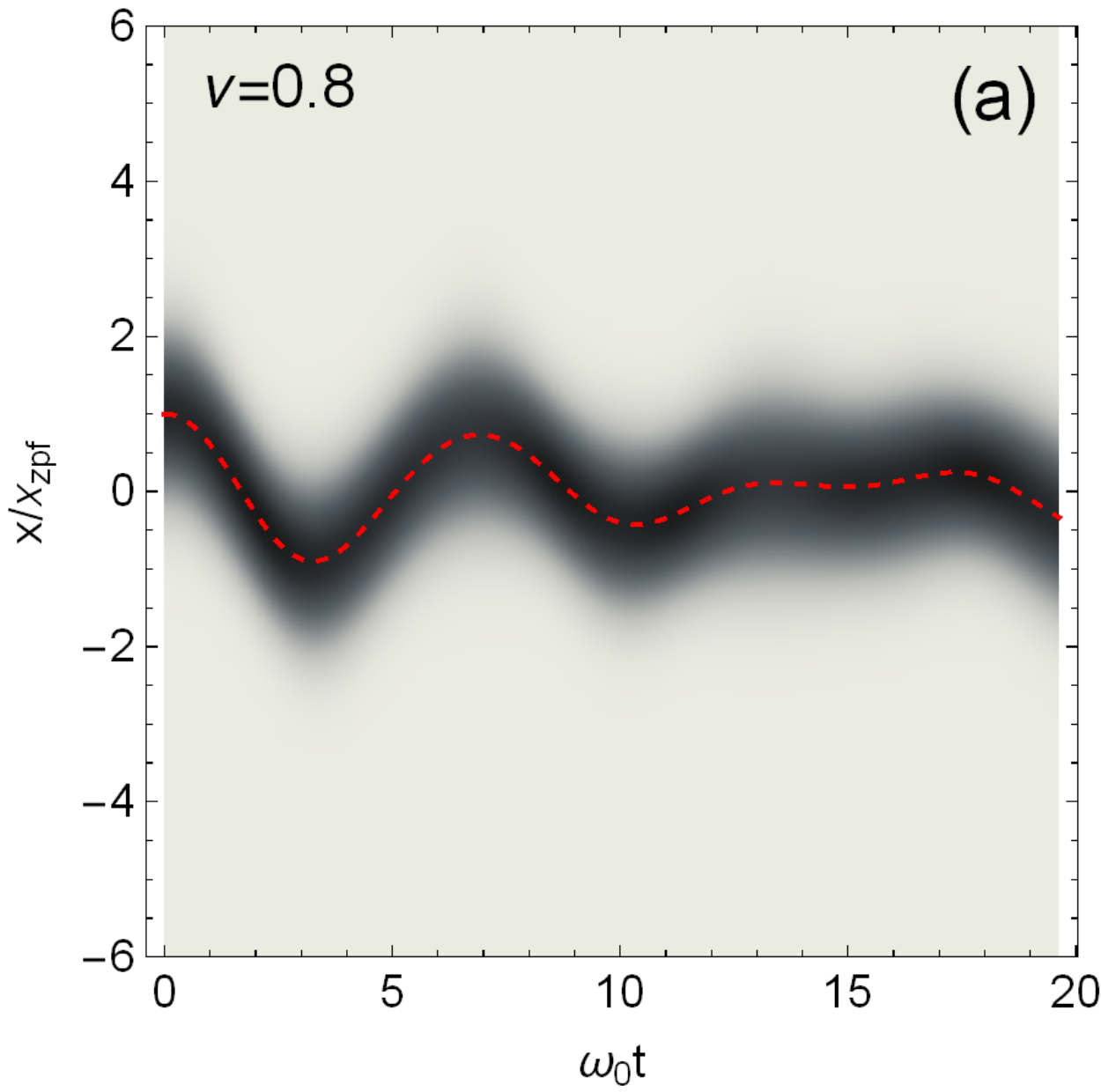}%
\includegraphics[width=0.25 \textwidth]{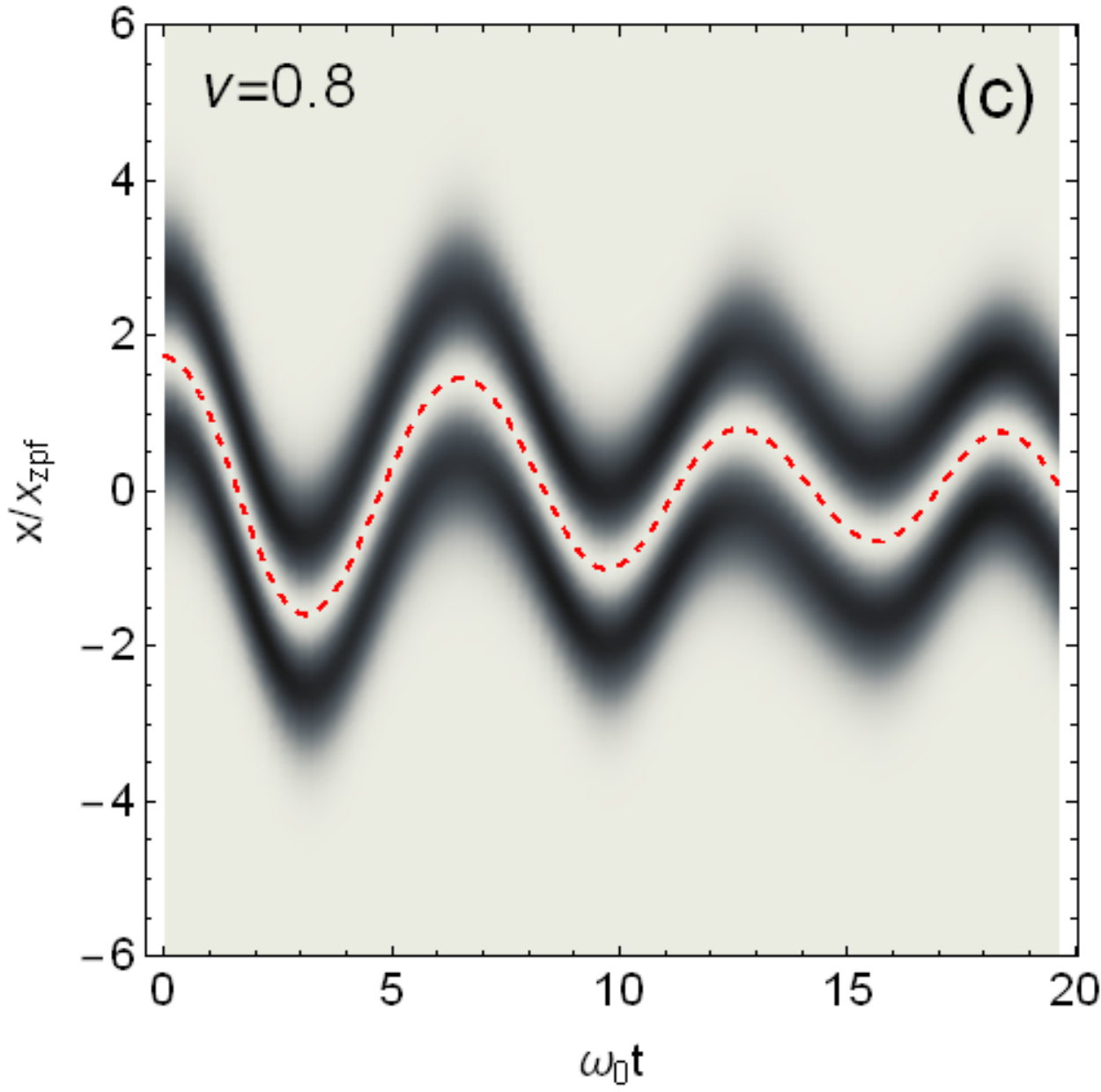}%
\includegraphics[width=0.25 \textwidth]{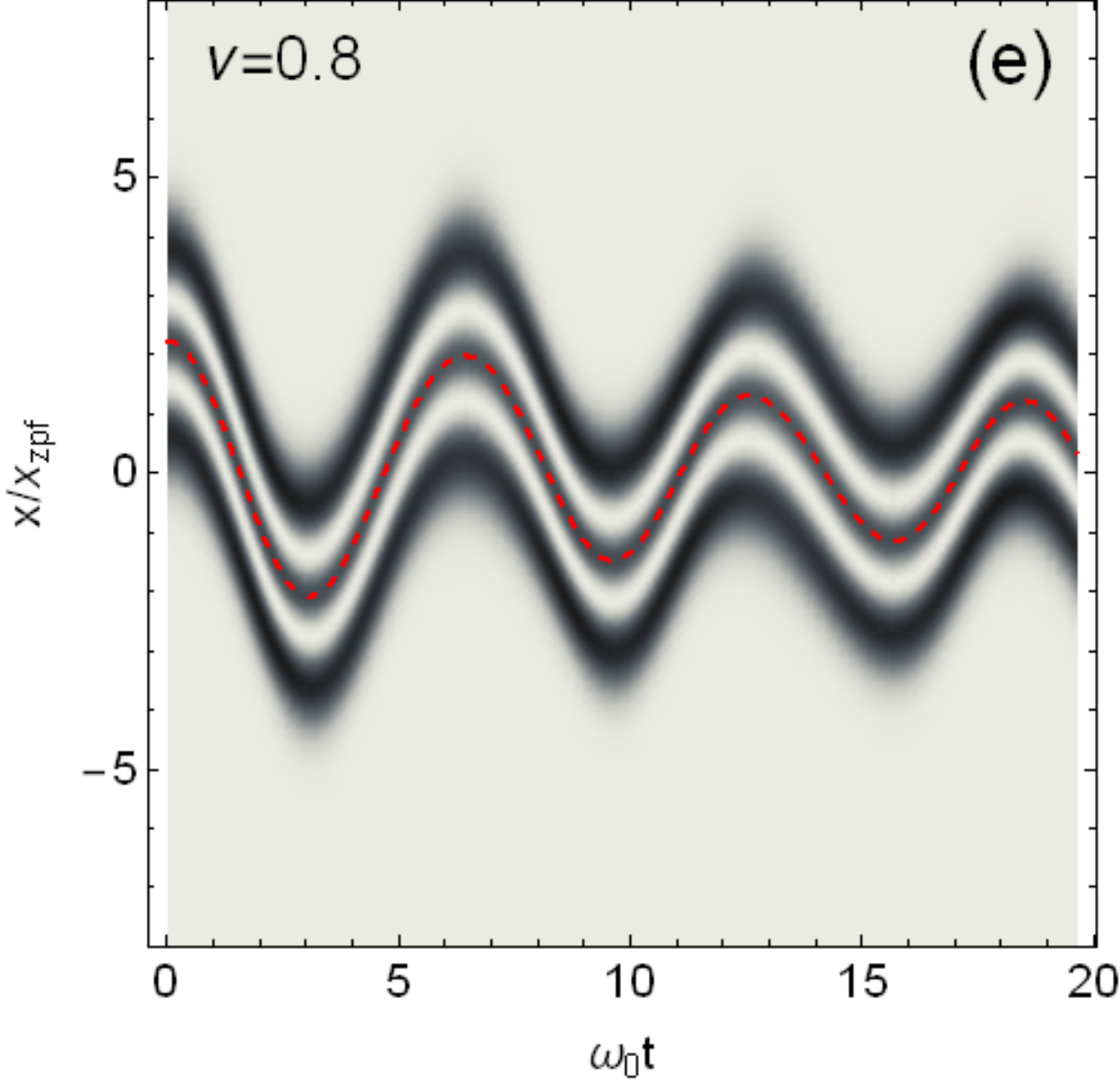}%
\includegraphics[width=0.25 \textwidth]{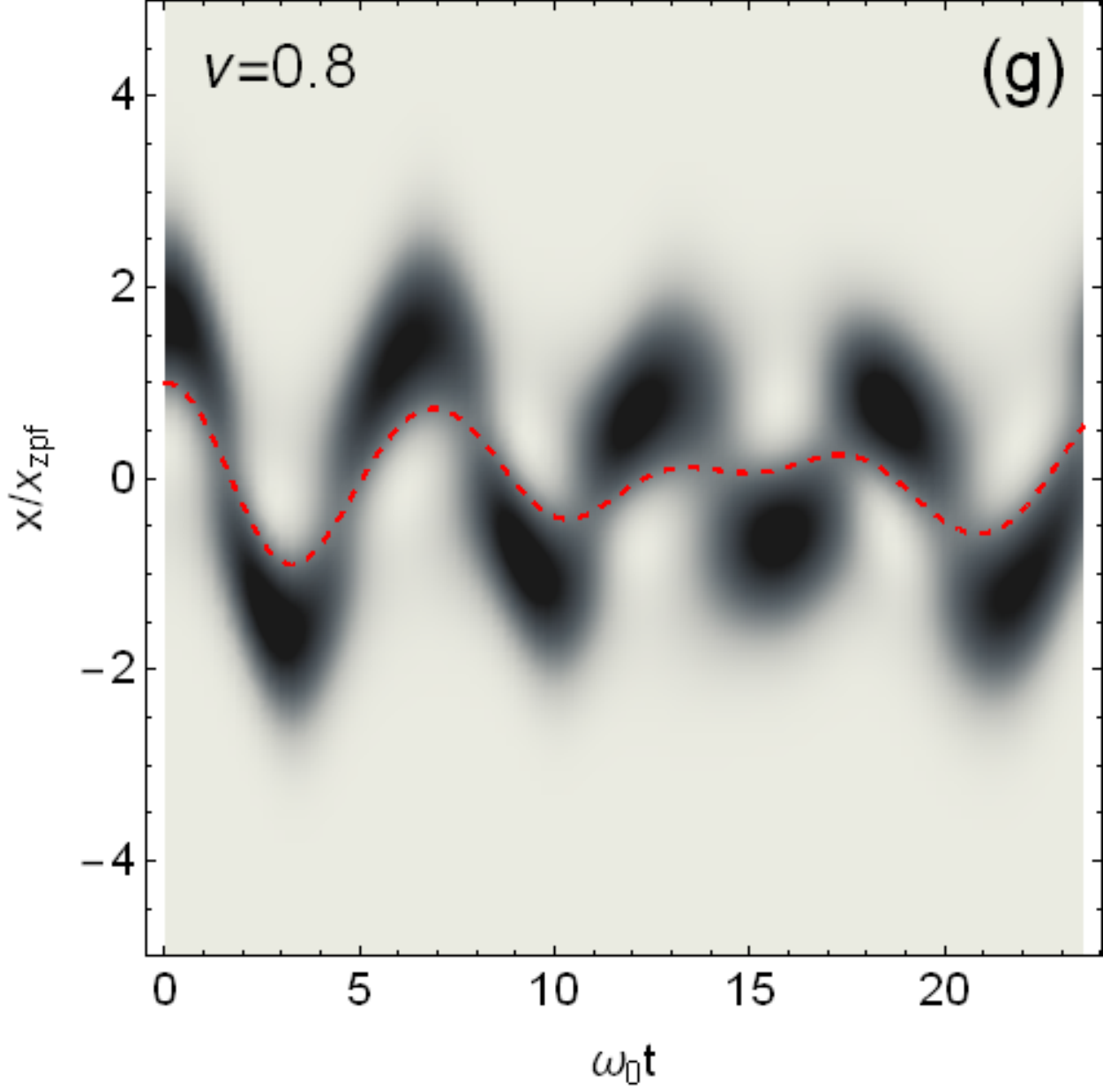}\\%
\includegraphics[width=0.25 \textwidth]{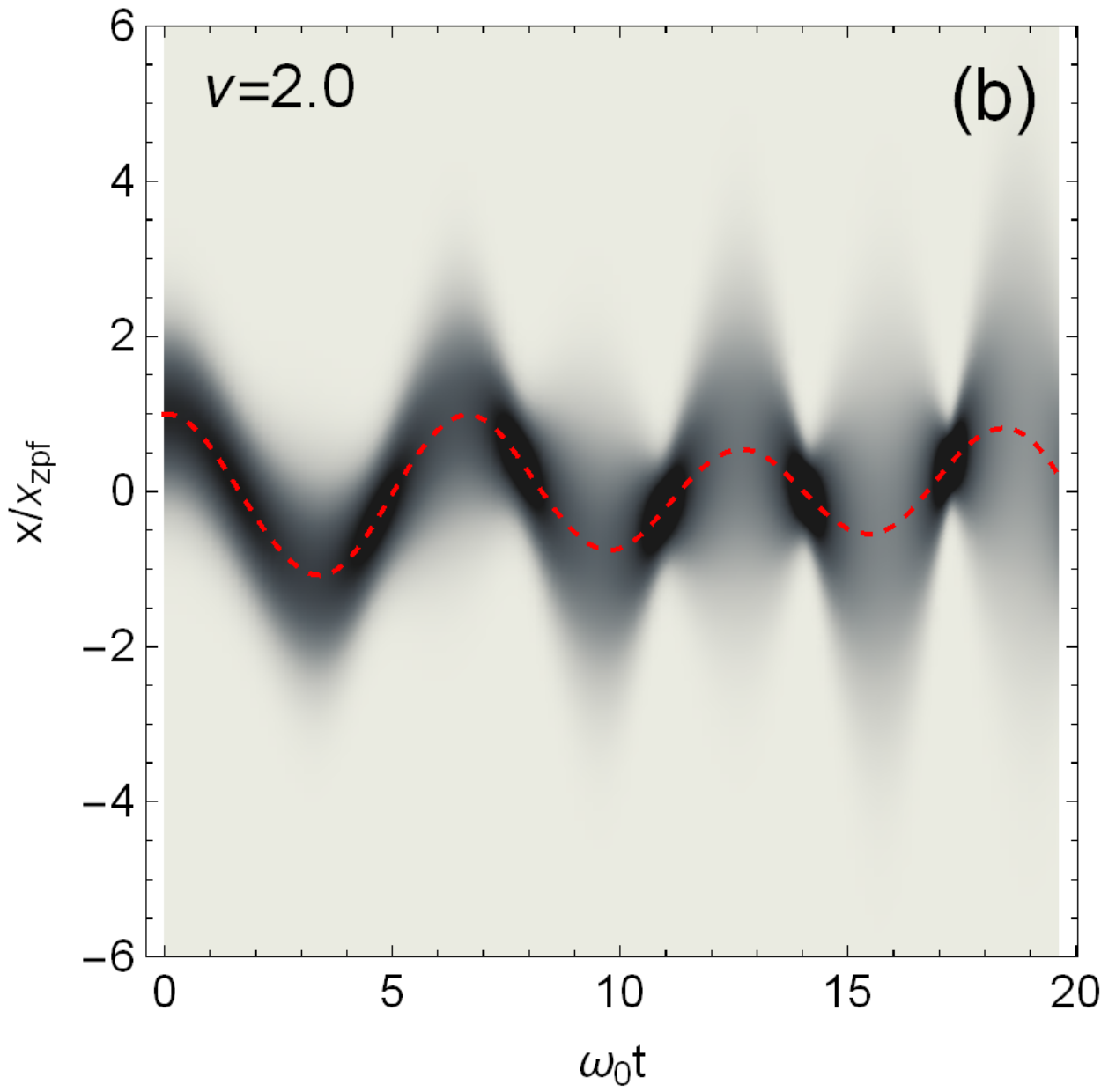}%
\includegraphics[width=0.25 \textwidth]{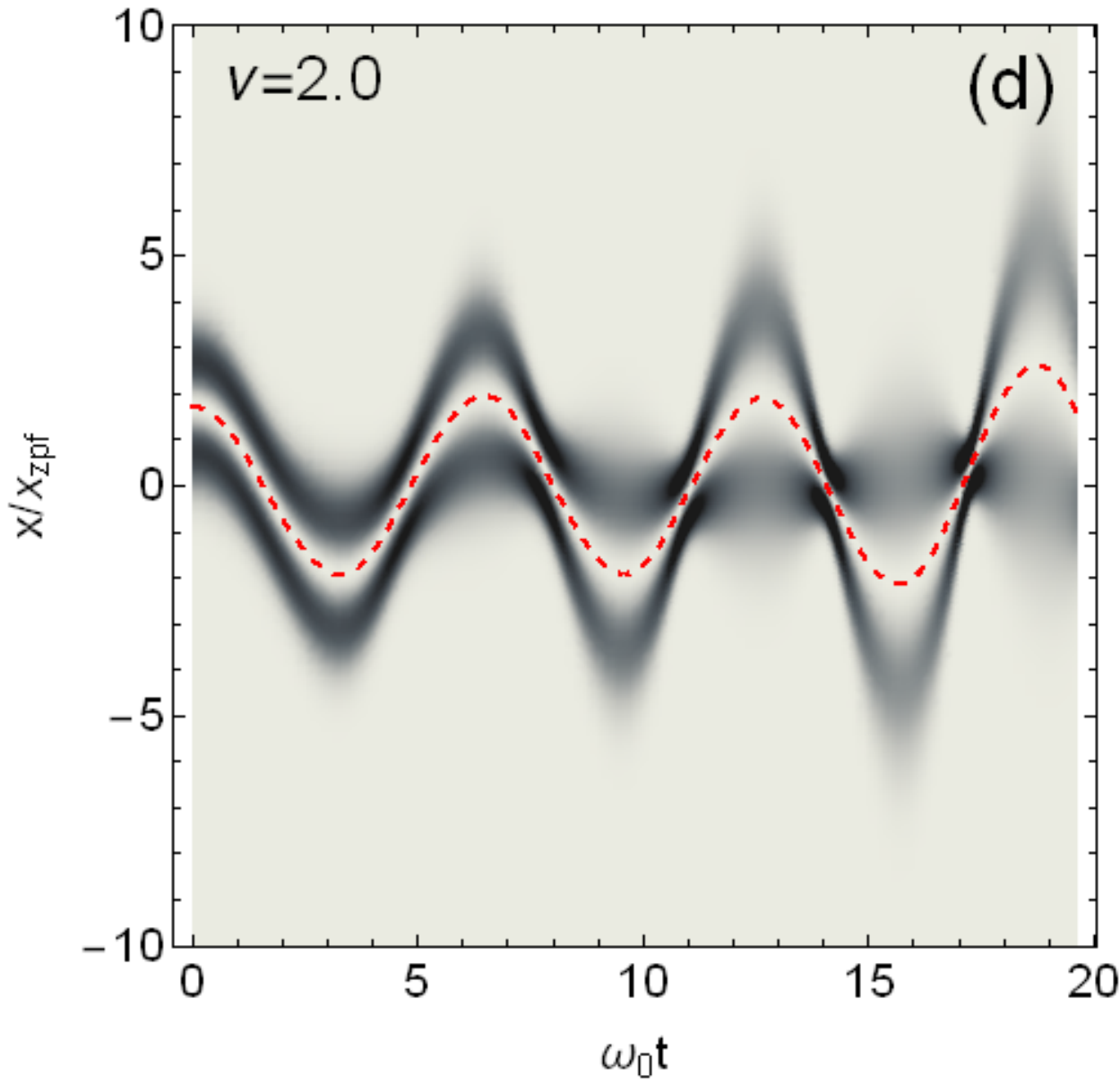}%
\includegraphics[width=0.25 \textwidth]{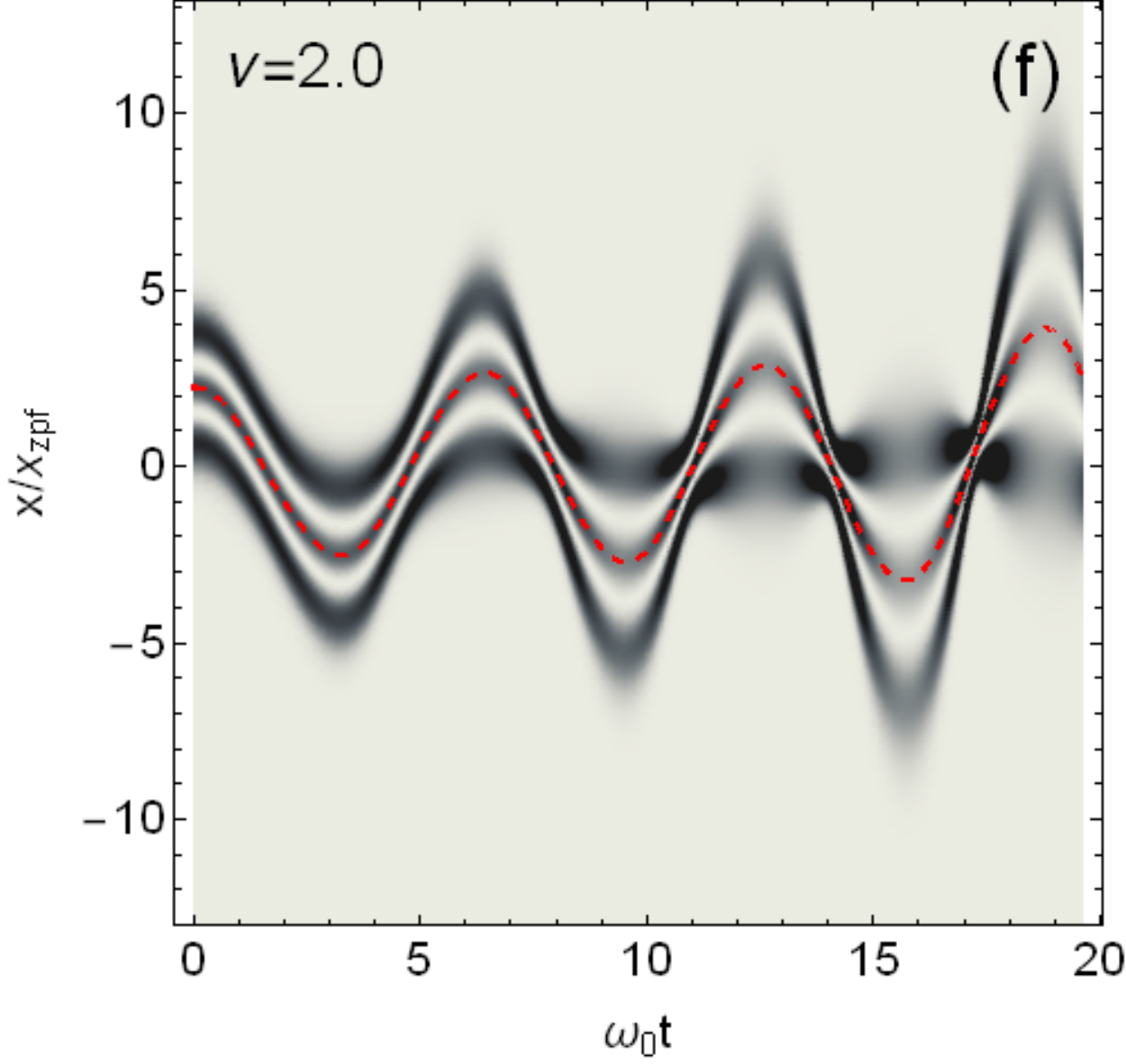}%
\includegraphics[width=0.25 \textwidth]{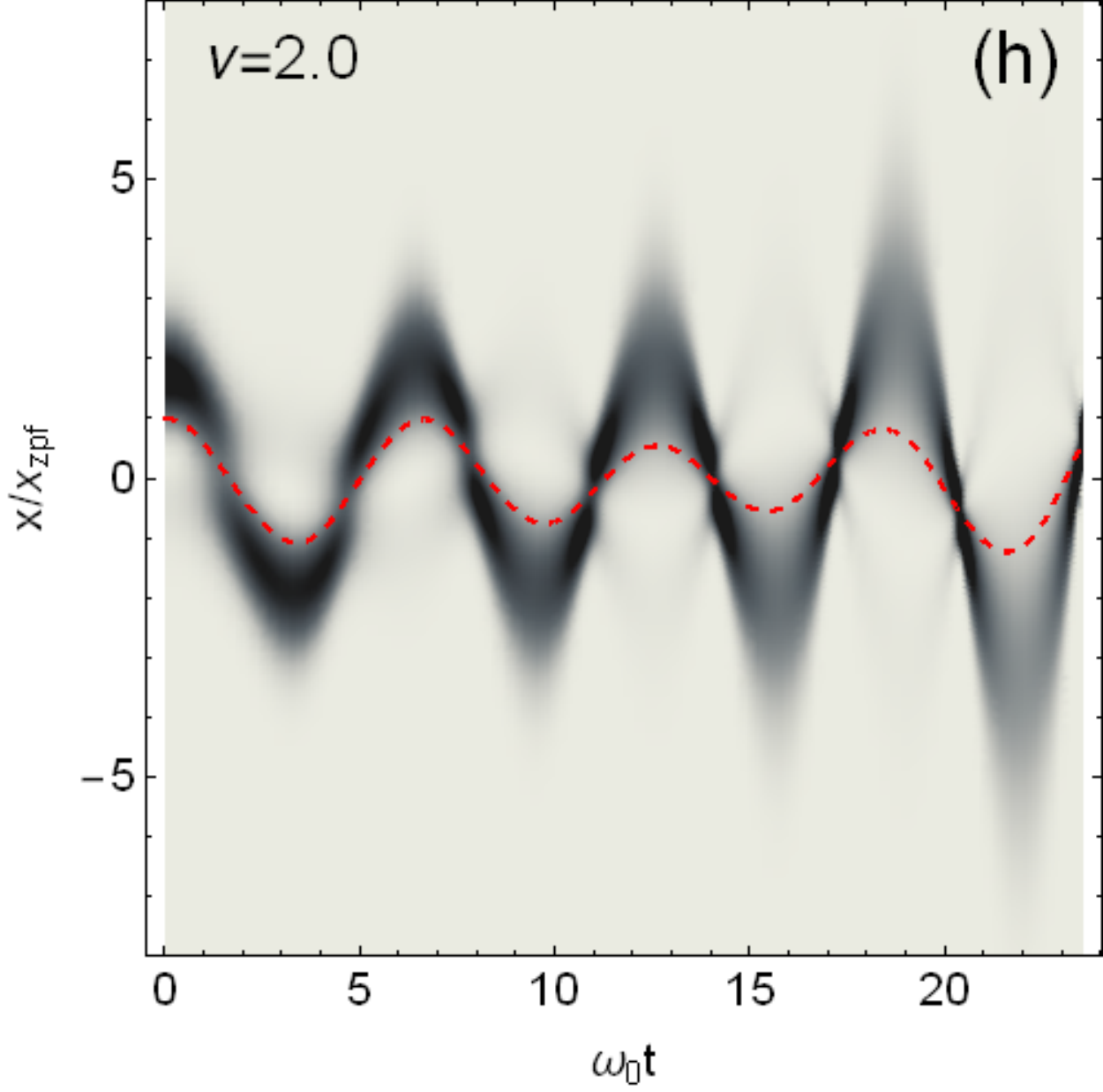}%
\end{center}
\caption{The evolution of wave packets of the driven parametric
oscillator starting from initial states of (a)(b)$|0\rangle$, (c)(d) $|1\rangle$,
(e)(f) $|2\rangle$ and (g)(h)$(|0\rangle+|1\rangle)/\sqrt{2}$.
The driving amplitude is $f_0=0.2$ and the driving frequency is $\Omega=0.8$. The
initial conditions of $X_c(0), P_c(0)$ for differen number states $|n\rangle$ are
the same as that in Fig.\ref{figure6} and the initial conditions for the superposition
state of $(|0\rangle+|1\rangle)/\sqrt{2}$ are $X_c(0)=1, P_c(0)=0$.}
\label{figure11}
\end{figure}

Certainly, as the original Hamiltonian (\ref{har}) only has an algebraic structure of $su(2)$,
the Lie transformation $\hat{U}_1$ of $h(4)$ is not necessarily needed
(no driving case) and the classical dynamical equation of Eq.(\ref{harno}) should not be included.
However, the model can be easily
generalized to solve a driven parametric oscillator with a classical dynamic equation as Eq.(\ref{dy})
\begin{equation}
\ddot{X}_{c}+\frac{\dot{M}(t)}{M(t)}\dot{X}_{c}+\Omega_0 ^{2}\left( t\right)
X_{c}=\frac{f_{0}}{M(t)}\cos \left( \Omega t+\phi \right) ,
\label{hard}%
\end{equation}%
which will lead to a wave function of%
\begin{equation*}
\Psi \left( X,t\right) =\hat{U}_{1}\left( t\right)\hat{U}_{2}\left( t\right) \psi_0 \left( X,t\right)
\end{equation*}%
with $\psi_0 (X,t)$ being defined by Eq.(\ref{psi0}).
Now we consider a specific control on the parametric functions of%
\begin{equation}
\label{mo}
M\left( t\right) =1,\quad \Omega ^{2}\left( t\right) =1+\lambda \sin \left(
\nu t\right),
\end{equation}%
to demonstrate the quantum evolution of a DDPO starting
from different number states. In Fig.\ref{figure10} we first show the temporal behaviors of
the transformation parameters of $\theta$s and the corresponding fluctuations
in position and momentum starting from the ground state $|0\rangle$.
We can see an alternative squeezing in position and momentum below $1/2$ for a
modulating frequency of $\nu=2\omega_0$ and the probability distributions of
the wave function display manifest distortions for the parametric resonant case
as shown in Fig.\ref{figure11}.
%--Figure---

For a harmonic oscillator with only a time-dependent mass of $M(t)$, we can simplify
the transformation of $\hat{U}_2$ in Eq.(\ref{U2t}) by setting $\theta_{-}=0$, and then
we get
\begin{eqnarray*}
\theta _{+}(t) =\frac{1}{2}\dot{M}(t), \quad
\theta _{0}(t) =-\frac{1}{2}\ln M(t),
\end{eqnarray*}%
and the mass should be controlled by an equation of%
\begin{equation*}
\ddot{M}-\frac{\dot{M}^{2}}{2M}+2M\left( 1-\Omega_{0} ^{2}\right) =0
\end{equation*}
in order to meet Eq.(\ref{har2}). Then the transformation in this case becomes%
\begin{equation*}
\hat{U}_{2}(t)=e^{-i\frac{1}{4}\dot{M}(t)\hat{X}^{2}}e^{i\frac{1}{4}\left[
\ln M(t)\right] \left( \hat{X}\hat{P}+\hat{P}\hat{X}\right) },
\end{equation*}%
and the solution without driving will be%
\begin{equation*}
\Psi \left( X,t\right) =M^{1/4}e^{-i\frac{1}{4}\dot{M}(t)X^{2}}\psi_0
\left( \sqrt{M(t)}X,t\right).
\end{equation*}
The wave function indicates a dilation of the position coordinate with an increase of the
mass of the parametric oscillator, which implying a localization of the wave packet as
shown in Fig.\ref{figure9}. Surely, another specific transformation of $\hat{U}_2$ to
control the evolution of the wave function can set $\theta_{+}=0$.

In order to produce more reliable wave functions for a long-time evolution, we can
combine Lie transformation with invariant operator to solve the problem based on Hamiltonian of Eq.(\ref{ho}).
If the controlled time-dependent system Eq.(\ref{Hs}) starts with a Hamiltonian of
$\hat{\mathcal{H}}(0)\equiv\hat{\mathcal{H}}_U$ as shown by Eq.(\ref{ini}), a general
invariant operator for system Eq.(\ref{Hddd}) can be constructed by an inverse transformation
of $\hat{U}(t)$ on $\hat{\mathcal{H}}_U$ by
\begin{equation*}
\hat{I}\left( t\right) =\hat{U}\hat{\mathcal{H}}_{U}\hat{U}%
^{-1}=\frac{1}{2}e^{2\theta _{0}}\left[ \left( \hat{P}+\beta \right) +\theta
_{+}\left( \hat{X}-\alpha \right) \right] ^{2}+\frac{\Omega_{0} ^{2}}{2}\left[
\left( e^{-\theta _{0}}-e^{\theta _{0}}\theta _{-}\theta _{+}\right) \left(
\hat{X}-\alpha \right) -e^{\theta _{0}}\theta _{-}\left( \hat{P}+\beta
\right) \right] ^{2}.
\end{equation*}
Clearly, if we set $\theta_{-}=0$ in Eq.(\ref{harmonic}), the invariant
operator reduces to%
\begin{equation}
\hat{I}\left( t\right) =\frac{1}{2}e^{2\theta _{0}}\left[ \left( \hat{P}%
+\beta \right) +\theta _{+}\left( \hat{X}-\alpha \right) \right] ^{2}+\frac{%
1}{2}\Omega_0 ^{2}e^{-2\theta _{0}}\left( \hat{X}-\alpha \right) ^{2},
\label{Ih}
\end{equation}%
which is equivalent to the invariant operator given by Ref.\cite{Choi,Choi2} when
we use a relation of $\theta _{0}=\ln \rho$. The dynamical invariant obtained here is also
equivalent to that derived from Eq.(\ref{It}), which we will investigate in detail
in a general case of $K$s by using Eq.(\ref{gtheta}) and Eq.(\ref{Kt}) simultaneously.

\subsubsection{The driven free particle in a well with moving boundary}

If the initial Hamiltonian of Eq.(\ref{Hs}), $\hat{\mathcal{H}}(0)$, describes a free particle
($a(0)=1, b(0)=1, c(0)=0$) or the system is initially prepared in a simple state of plane
waves,
then we can use the parameters $K_{1}=K(t)$ and $K_{2}=K_{3}=0$ to transform Eq.(\ref{Hs}) into
a free-particle like Hamiltonian of%
\begin{equation}
\hat{\mathcal{H}}_{U}=\frac{1}{2}K\left( t\right) \hat{P}^{2}.  \label{free}
\end{equation}%
Here the controlled parametric equations are
\begin{eqnarray}
\dot{\theta}_{+} &=&a\theta _{+}^{2}-2c\theta _{+}+b,  \notag \\
\dot{\theta}_{-} &=&ae^{-2\theta _{0}}-K(t),  \label{tf} \\
\dot{\theta}_{0} &=&c-a\theta _{+}.  \notag
\end{eqnarray}%
In this case, we should solve the eigenstate of $\hat{\mathcal{H}}_{U}$ by%
\begin{equation*}
i\frac{\partial }{\partial t}\varphi \left( X,t\right) =\frac{1}{2}K\left(
t\right) \hat{P}^{2}\varphi \left( X,t\right) ,
\end{equation*}%
and the separation variables of $\varphi(X,t)$
gives
\begin{equation*}
\varphi \left( X,t\right) =\frac{1}{\sqrt{2\pi \hbar }}e^{iK_0 X}e^{-iE_{0}\int_{0}^{t}K%
\left( \tau \right) d\tau },
\end{equation*}%
where $E_0$ is the separation constant (the kinetic energy) and
the scaled momentum $K_0$ is determined by the
eigenequation of the momentum operator
\begin{equation}
\hat{P}f_{K_{0}}\left( X\right) =K_{0}f_{K_0}\left( X\right) ,\qquad
f_{K_{0}}\left( X\right) =\frac{1}{\sqrt{2\pi \hbar }}e^{iK_{0}X},
\label{pwave}
\end{equation}%
where $K_0=\sqrt{2E_0}$.
Therefore, if the parameters $\theta$s can be properly controlled by Eq.(\ref{tf}),
the final wave function for DDPO in this case will be got by
the transformations of $\hat{U}_1$ and $\hat{U}_2$ as%
\begin{equation*}
\Psi \left( X,t\right) =\hat{U}_{1}\left( t\right) \hat{U}_{2}\left(
t\right) \varphi \left( X,t\right).
\end{equation*}

By using the system parameters of Eq.(\ref{Hd}), the parameters $\theta_+$
and $\theta_0$ have the solutions of
\begin{eqnarray}
\theta _{+}\left( t\right)  =-e^{2\gamma t}%
\dot{\rho}, \quad
\theta _{0}\left( t\right)  =\ln \rho
\label{solf}
\end{eqnarray}%
with an auxiliary equation of
\begin{equation}
\ddot{\rho}+2\gamma \dot{\rho}+\rho =0.
\label{f2}
\end{equation}%
Eq.(\ref{solf}) indicates that $\theta _{-}$ is decoupled
from $\theta _{+}$ and $\theta _{0}$ in Eq.(\ref{tf}), then we can simply set%
\begin{equation*}
K(t)=e^{-2\gamma t}e^{-2\theta _{0}}=\frac{e^{-2\gamma t}}{\rho ^{2}}
\end{equation*}%
to keep $\theta _{-}$ as a constant. Therefore the wave function in real space
will be%
\begin{equation}
\Psi \left( X,t\right) =\frac{1}{\sqrt{2\pi\hbar\rho }}e^{-is\left( t\right)
}e^{-iE_{0}\int_{0}^{t}\frac{e^{-2\gamma \tau }}{\rho ^{2}}d\tau
}e^{ie^{2\gamma t}\dot{X}_{c}\left( X-X_{c}\right) }e^{ie^{2\gamma t}\dot{%
\rho}\left( X-X_{c}\right) ^{2}/2}e^{-i\frac{\sqrt{2E_{0}}}{\rho }\left(
X-X_{c}\right) },
\label{plane}
\end{equation}%
where we have set $\theta _{-}=0$ for simplicity and $X_c$ satisfies the
classical dynamical equation of Eq.(\ref{dy}). As Eq.(\ref{f2}) is the same as
Eq.(\ref{dh}), the wave function of Eq.(\ref{plane}) is not optimal
because it has a coefficient of $1/\sqrt{\rho}$ which becomes divergent when $\rho\rightarrow 0$.
Although the wave function Eq.(\ref{plane}) is a particular solution of Eq.(\ref{Hs})
based on the plane wave basis, it can show the dynamic details of different phase
components during a state evolution starting from a plane wave function.

\begin{figure}[t]
\begin{center}
\includegraphics[width=0.25 \textwidth]{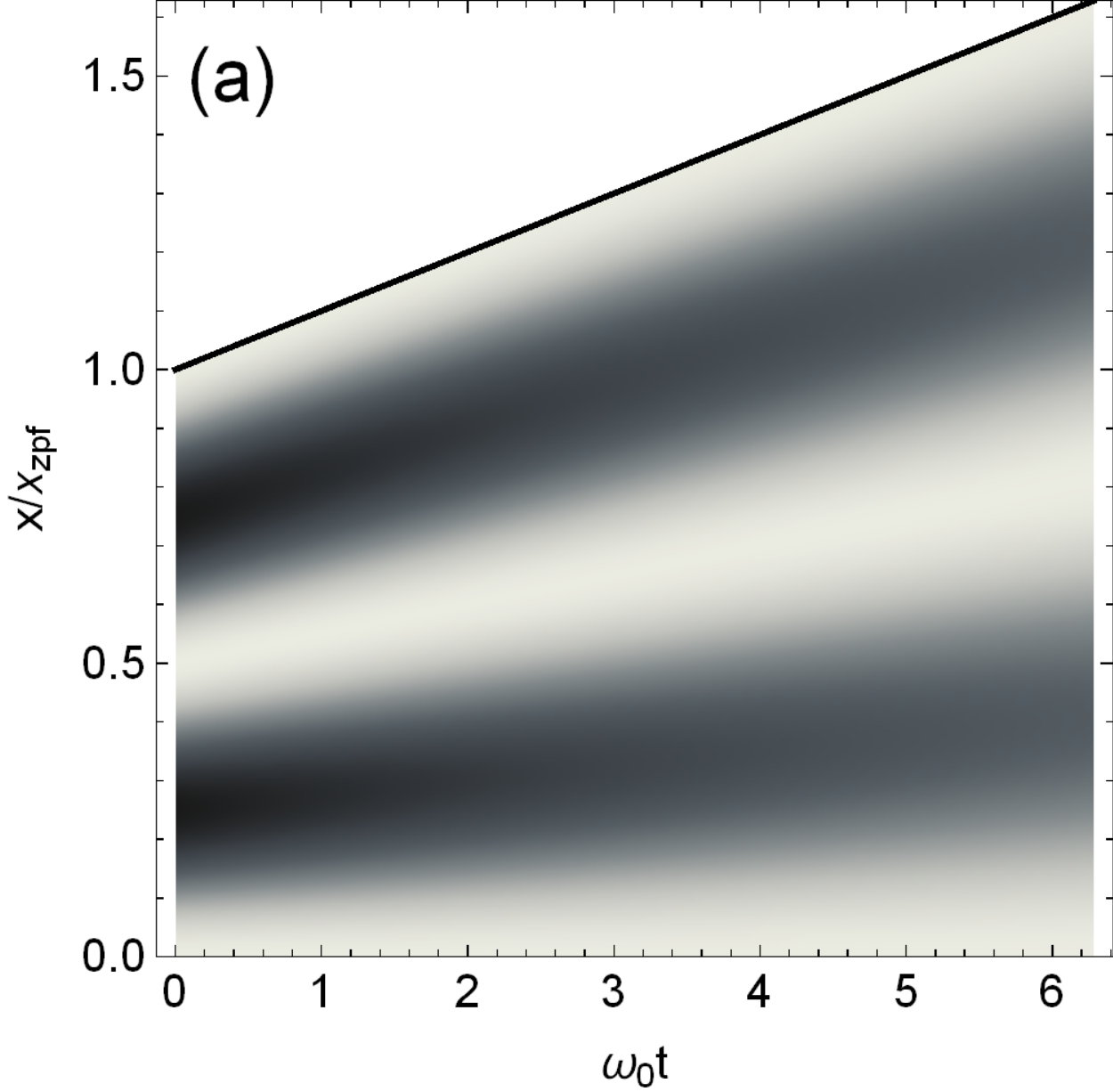}%
\includegraphics[width=0.25 \textwidth]{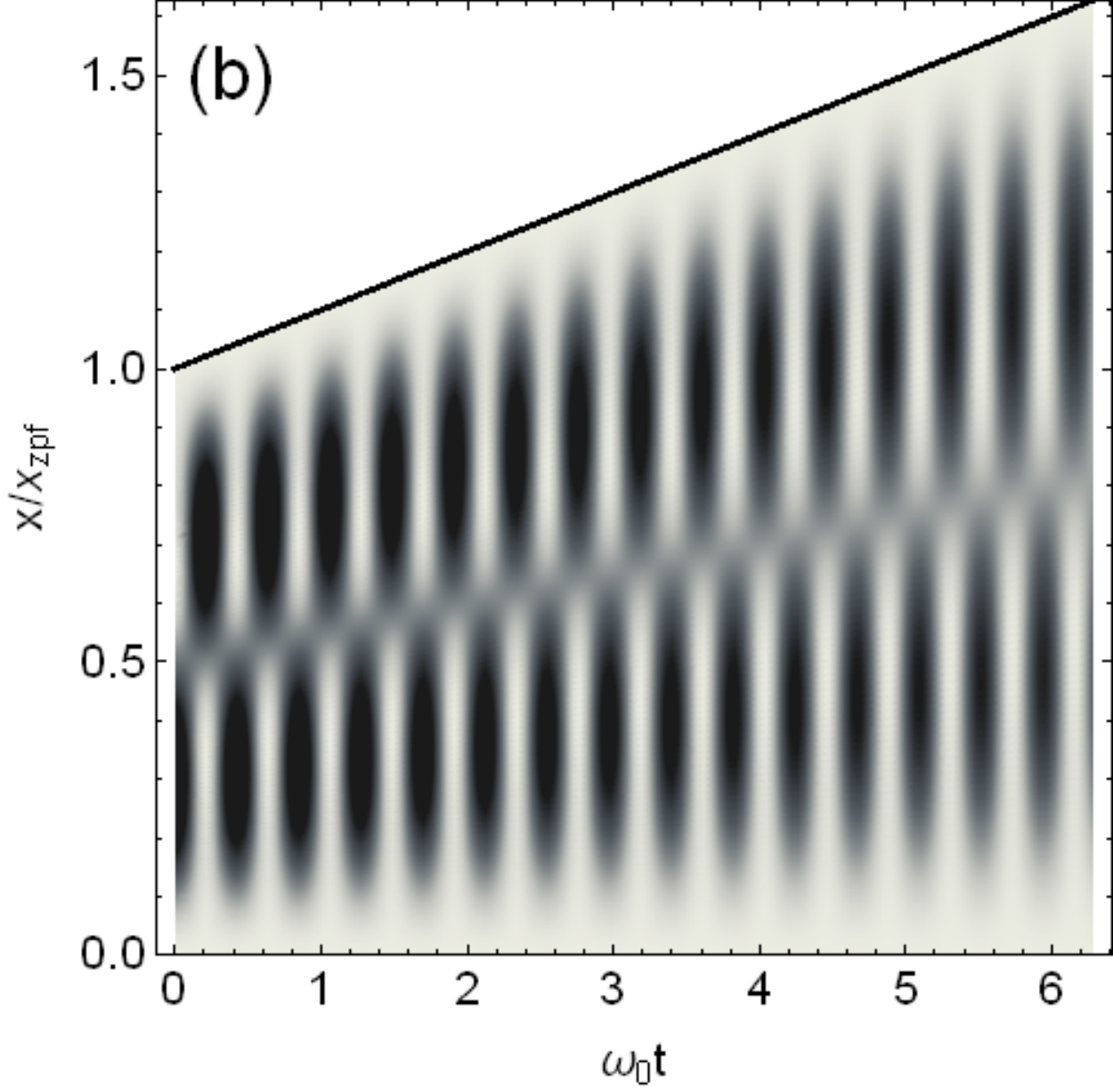}%
\includegraphics[width=0.25 \textwidth]{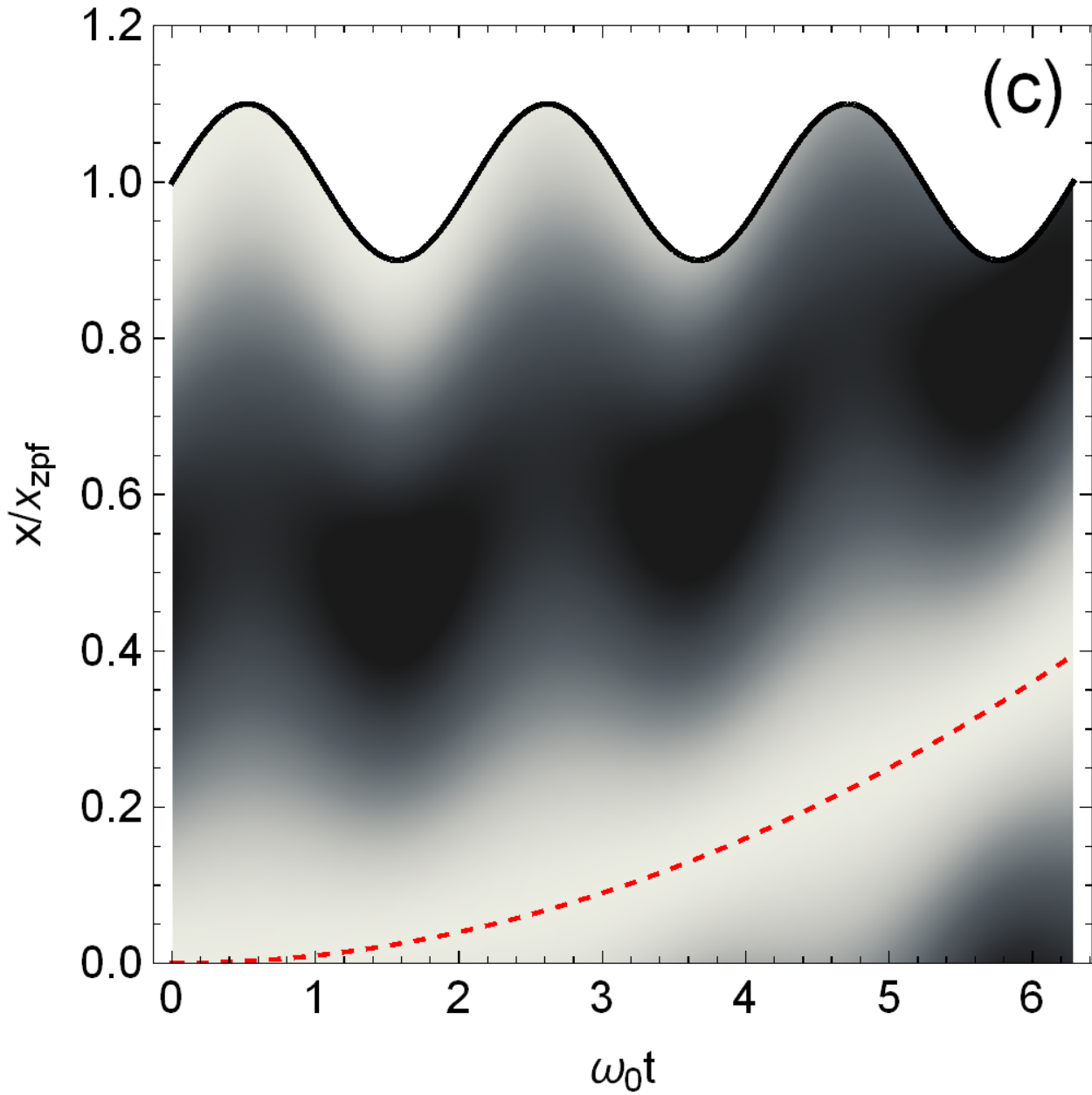}%
\includegraphics[width=0.25 \textwidth]{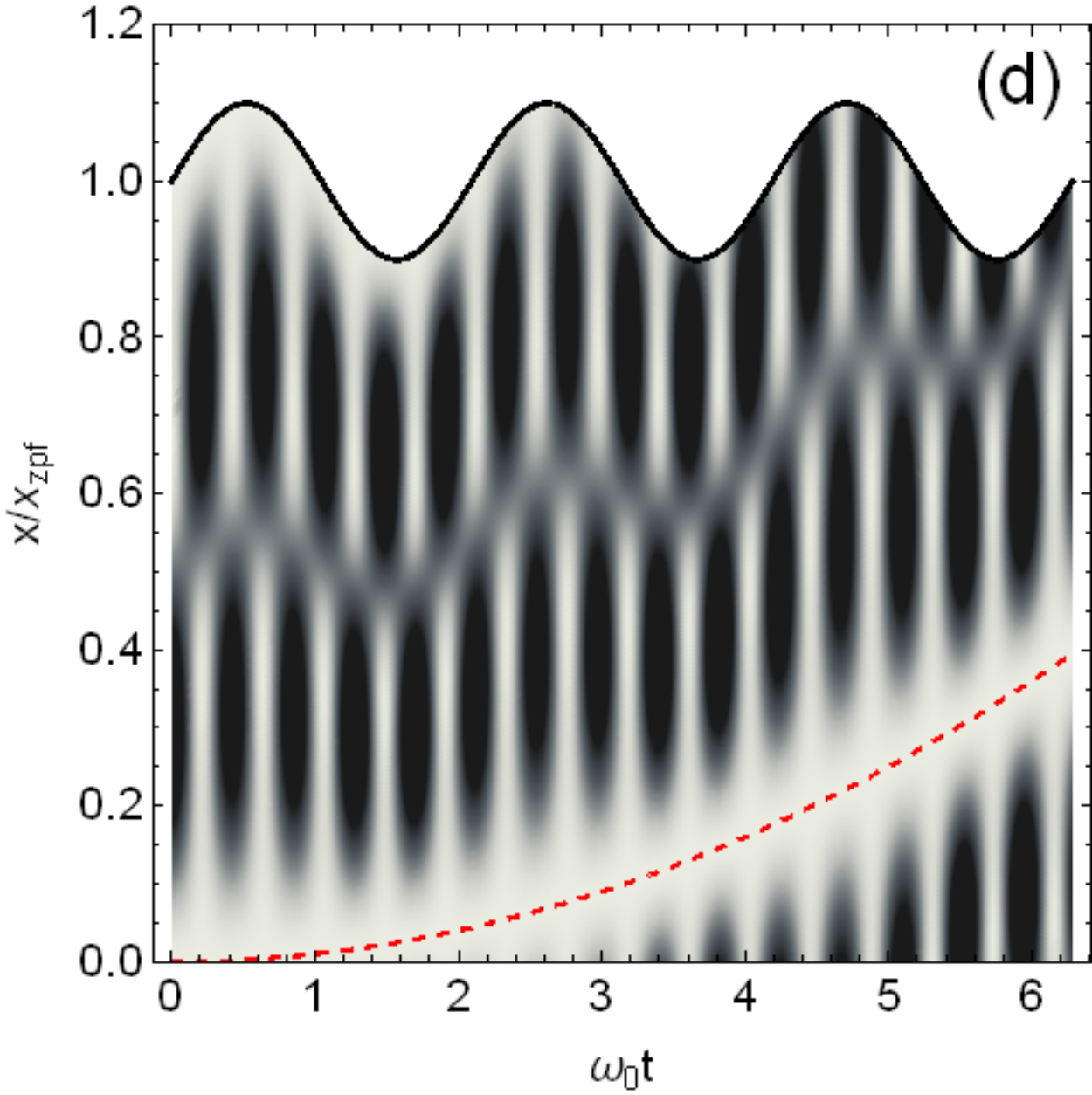}%
\end{center}
\caption{The evolution of the probability densities of a free particle in an
infinite well with expanding boundaries starting from (a) the ground state
$\varphi_1$ and (b) the superposition state of $(\varphi_1+\varphi_2)/\sqrt{2}$
with the parameters $L_0=1, v=0.1$. The corresponding cases for
the driven free particle in the infinite well with oscillating boundary are shown
by (c) and (d), respectively. The parameters are $L_0=1, \lambda=0.1, \nu=3$ and $f_0=0.02$.}
\label{figure12}
\end{figure}

The Lie transformation in the present case can provide a simple way to solve the problem
of a free particle in an infinite square well with a moving boundary if we use the
parameters $K$s to construct the invariant Hamiltonian for the free particle. The quantum
well with a moving boundary \cite{Lejarreta} is a popular model to simulate the quantum
piston in the quantum control problems \cite{Chen}. The position of the well boundary can be
described by a time-dependent function of $L(t)$ and a time-dependent scale transformation
of $x\rightarrow x/L(t)$ can convert the problem into a normal time-independent infinite
square well of unit width. Clearly, the Lie transformation corresponding to the scale
transformation in real space is
\begin{equation*}
\hat{U}\left( t\right) =e^{-i\ln L\left( t\right) \left( \hat{X}\hat{P}%
\right) },
\end{equation*}%
which gives $\hat{U}f(X)=f(X/L)$ for an arbitrary function of $f(x)$. Therefore, we can
choose a Lie transformation of%
\begin{equation}
\hat{U}_{2}=e^{-i\theta _{+}\hat{X}^{2}/2}e^{-i\theta _{0}\left( \hat{X}\hat{%
P}+\hat{P}\hat{X}\right) /2}
\label{uu2}
\end{equation}%
to solve this problem with $\theta _{0}=\ln L(t)$ and $\theta _{-}=0$.
After a FUT of $\hat{U}_2$, the problem becomes a free particle in an infinite square
well of unit width described by an initial Hamiltonian of
\begin{equation}
\hat{\mathcal{H}}_{U}\equiv\hat{\mathcal{H}}\left( 0\right) =\frac{1}{2}\hat{P}^{2},
\label{free2}
\end{equation}
which is a specific case of Eq.(\ref{ho}) for $\Omega_0=0$, and its
eigenstates are well known
\begin{equation*}
\varphi_n \left( X,t\right) =\sqrt{2}\sin \left( n\pi X\right)e^{-i(\pi^2 n^2 t/2)}.
\end{equation*}%
Since the wave function of Eq.(\ref{plane}) is bad at $\rho=0$ under the condition
of $K_3=1$ and $\theta_{-}=0$, we can use the invariant parameters $K$s instead to solve this
problem. By using Lie transformation of $\hat{U}_{2}$, we can construct an invariant
Hamiltonian by an inverse Lie transformation of $\hat{U}_2$ on Eq.(\ref{free2}) as
\begin{equation}
\hat{\mathcal{H}}\left( t\right) =\hat{U}_2\hat{\mathcal{H}}\left( 0\right)
\hat{U}_{2}^{-1}=\frac{1}{2}\left( e^{\theta _{0}}\hat{P}+\theta _{+}e^{\theta _{0}}%
\hat{X}\right) ^{2},
\label{If}
\end{equation}
where we have set $\alpha=0, \beta=0$ for the no-driving case. The condition of
Hamiltonian Eq.(\ref{If}) to be an invariant for the Hamiltonian Eq.(\ref{free2})
can give the transformation parameter of $\theta_{+}$ by
\begin{equation*}
\theta _{+}=-\frac{\dot{L}}{L}, \quad \ddot{L}=0.
\end{equation*}%
Alternatively, $\theta_{+}$ can also be determined by using the invariant operator
derived from Eq.(\ref{It}) for the free-particle Hamiltonian of Eq.(\ref{free2}) as
\begin{equation*}
\hat{I}_{1}(t)=\frac{1}{2}\left( \rho \hat{P}-\frac{\dot{\rho}}{a}\hat{X}\right)
^{2},
\end{equation*}
where $a=1, b=0, c=0$ and $\Omega_0 =0$ in this case. Compared the above invariant operator
with Eq.(\ref{If}), we have the same result as
\begin{equation*}
\rho=L(t), \quad \theta _{+}=-e^{-\theta _{0}}\frac{\dot{\rho}}{a}=-\frac{\dot{L}}{L}.
\end{equation*}%
Therefore, the final wave function will be%
\begin{equation*}
\Psi (X,t)=\hat{U}_{2}\left( t\right) \varphi _{n}\left( X,t\right) =e^{i%
\frac{\dot{L}}{2L}X^{2}}\sqrt{\frac{2}{L\left( t\right) }}\sin \left[
\frac{n\pi X}{L\left( t\right) }\right] e^{-in^{2}\pi ^{2}t/2},
\end{equation*}
where the width of the well should have a constant expanding speed with a constraint of
$\ddot{L}=0$. Fig.\ref{figure12}(a) and (b) display the probability distributions
of a free particle in an infinite well for a linear expansion of the well width, that
is $L(t)=L_0+v t$, where $v$ is the expanding speed.
%--Figure---
The time evolution of the wave functions starting from differen initial
states shown in Fig.\ref{figure12}(a)(b) reveals that the free particle
in an infinite well with a constant expanding width only exhibits adiabatic
dynamics no matter how fast the expanding speed is. This result is due to the
fact that the invariant Hamiltonian of Eq.(\ref{If}) for a free particle in an
infinite well maintains a constant ``energy" during the well expanding.
The restrict condition of $\ddot{L}=0$ on the above solution excludes an acceleration
of boundary but it can be easily generalized to this case with this method.
For a driven free particle in an infinite well with varying boundary $L(t)$ beyond
$\ddot{L}=0$, we can construct a general invariant Hamiltonian from Eq.(\ref{ho})
with $\Omega_0\neq0$ as
\begin{equation}
\hat{I}(t)=\hat{U}\hat{\mathcal{H}}_{U} \hat{U}^{-1}=\frac{1}{%
2}\left[ e^{\theta _{0}}\left( \hat{P}+\beta \right) +\theta _{+}e^{\theta
_{0}}\left( \hat{X}-\alpha \right) \right] ^{2}+\frac{1}{2}\Omega_{0}
^{2}e^{-2\theta _{0}}\left( \hat{X}-\alpha \right) ^{2},
\label{Ih2}
\end{equation}
where the parameters $\alpha$ and $\beta$ are for the classical dynamics induced
by the driving force.
In this case the transformation parameters can be determined by
\begin{eqnarray*}
\theta _{+} =-\frac{\dot{L}}{L}, \quad
\Omega_0  =\pm\sqrt{\frac{\ddot{L}}{L}}\frac{L^{2}}{\sqrt{1-L^{4}}},
\end{eqnarray*}%
under the conditions of $\theta _{0}=\ln L$ and $\theta _{-}=0$. As $\Omega_0
\neq 0$ in this case, a further harmonic potential is presented here to
validate the invariant operator of Eq.(\ref{Ih2}) with the driving force. Therefore the quadratic
Hamiltonian used to solve the problem of a driven particle in an infinite square well with an
accelerated expansion is%
\begin{equation*}
\mathcal{\hat{H}}(0)=\frac{1}{2}\left( \hat{P}-P_{c}\right) ^{2}+\frac{1}{2}%
\frac{\ddot{L}}{L}\frac{L^{4}}{1-L^{4}}\left( \hat{X}-X_{c}\right) ^{2},
\end{equation*}%
where the classical position of the driven particle follows the Newton equation of
$\ddot{X}_{c}=f\left( t\right) $ with $f(t)$ being the driving force.
We find that the auxiliary harmonic potential is undefined for $L=1$ due to
the initial scaled width of the well being $L(0)=1$, and it is
repulsive for $L<1$ but attractive for $L>1$
if $\ddot{L}>0$.
Therefore, the wave function starting from the initial state of
$\varphi_n(X,0)=\sqrt{2} \sin(n\pi X)$ can be determined by the Lie
transformation as%
\begin{equation*}
\Psi \left( X,t\right) =e^{i\int_{0}^{t}X_{c}\left( t^{\prime }\right)
f\left( t^{\prime }\right) dt^{\prime }}e^{-in^{2}\pi ^{2}t/2}e^{i\frac{\dot{%
L}}{2L}\left( X-X_{c}\right) ^{2}}\sqrt{\frac{2}{L\left( t\right) }}\sin %
\left[ \frac{n\pi }{L\left( t\right) }\left( X-X_{c}\right) \right] .
\end{equation*}
Fig.\ref{figure12}(c) and (d) demonstrate the evolutions of the probability distributions of a
free particle located in an infinite well with an oscillating boundary,
$L(t)=L_0+\lambda \sin(\nu t)$, and under a constant driving force of $f_0$.
We can see a clear probability modulation of the wave packet by the classical
motion of the particle (red dashed lines) and the varying boundary of the well
(thick black lines). In this case, the oscillating boundary can stimulate
quantum transitions between different quantum states which should enable emissions
or absorptions of the well. Under the frame work of Lie transformation method, the other types
of driving and width modulation for this problem can also be similarly solved.

\subsubsection{The position-momentum (posmom) model}

If we set $K_{1}(t)=K_{2}(t)=0$ and $K_{3}(t)=K\left( t\right) $, then we
reach a transformed Hamiltonian as%
\begin{equation}
\hat{\mathcal{H}}_{U}=\frac{1}{2}K\left( t\right) \left( \hat{X}\hat{P}+\hat{%
P}\hat{X}\right) ,  \label{XPh}
\end{equation}%
and the parametric equations leading to Eq.(\ref{XPh}) are
\begin{eqnarray}
\label{thexp}
\dot{\theta}_{+} &=&a\theta _{+}^{2}-2c\theta _{+}+b,  \notag \\
\dot{\theta}_{-} &=&ae^{-2\theta _{0}}+2K\left( t\right) \theta _{-}, \\
\dot{\theta}_{0} &=&c-a\theta _{+}-K\left( t\right). \notag
\end{eqnarray}
%--Figure---
\begin{figure}[t]
\begin{center}
\includegraphics[width=0.25 \textwidth]{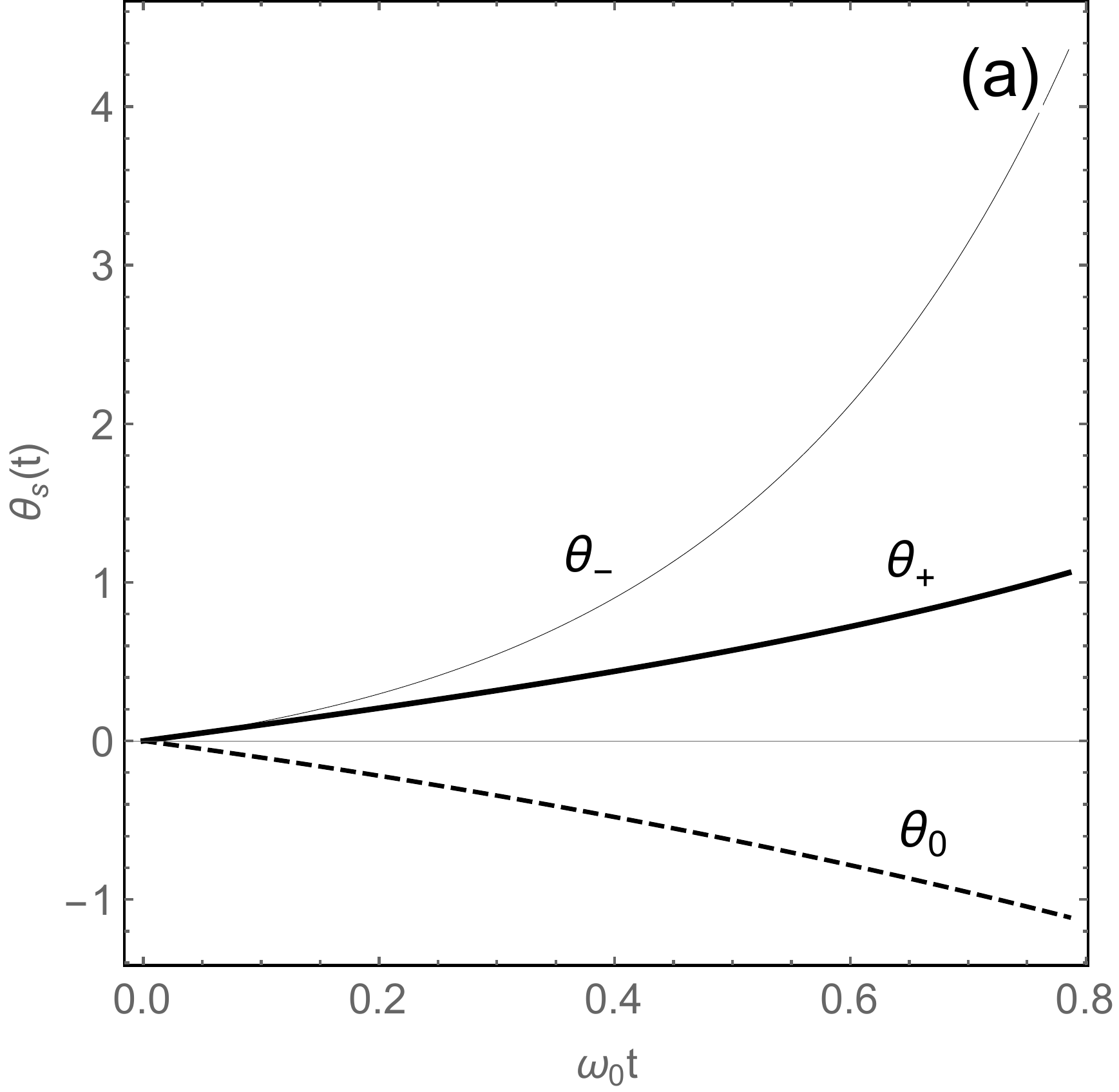}%
\includegraphics[width=0.25 \textwidth]{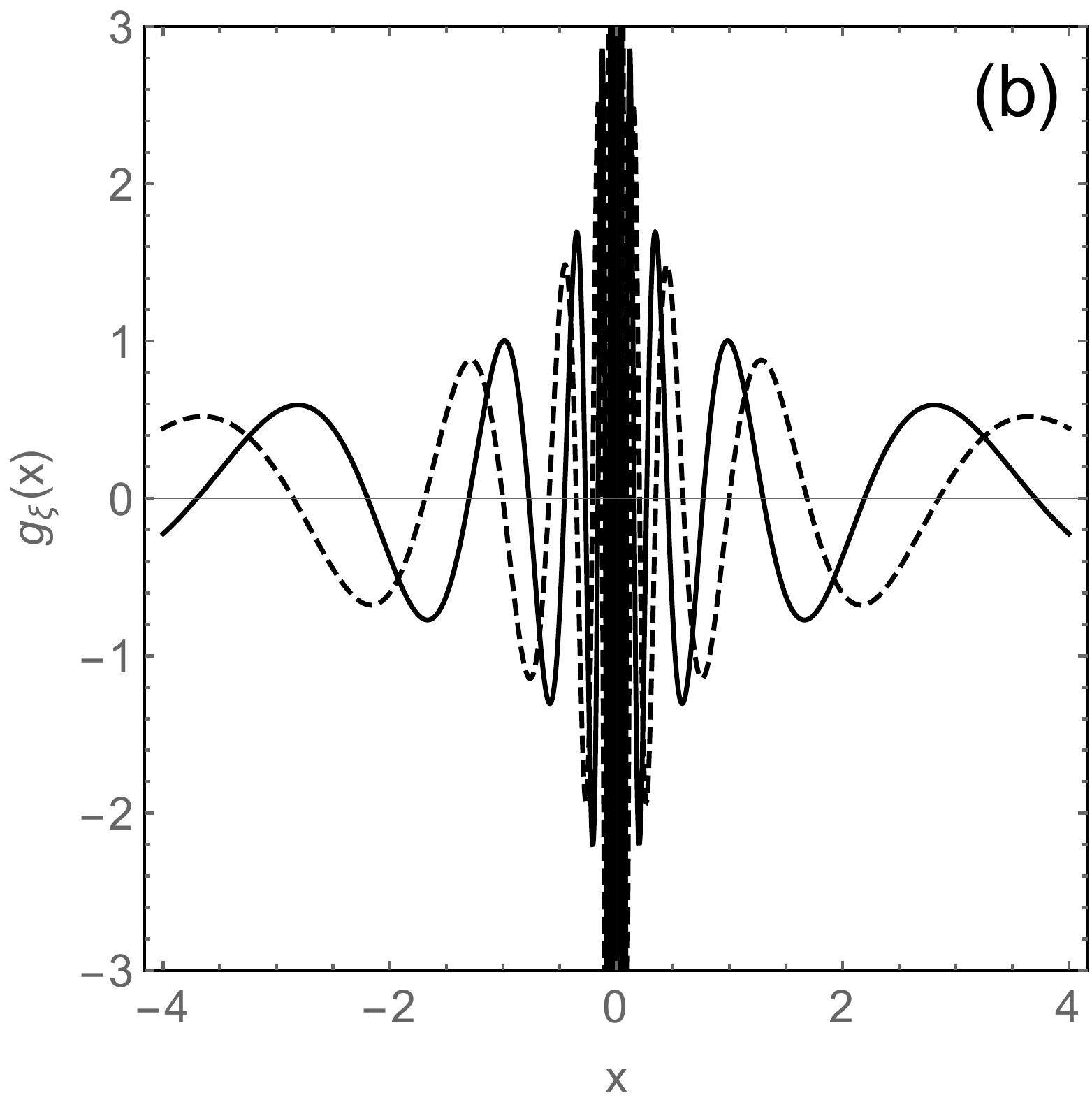}%
\includegraphics[width=0.25 \textwidth]{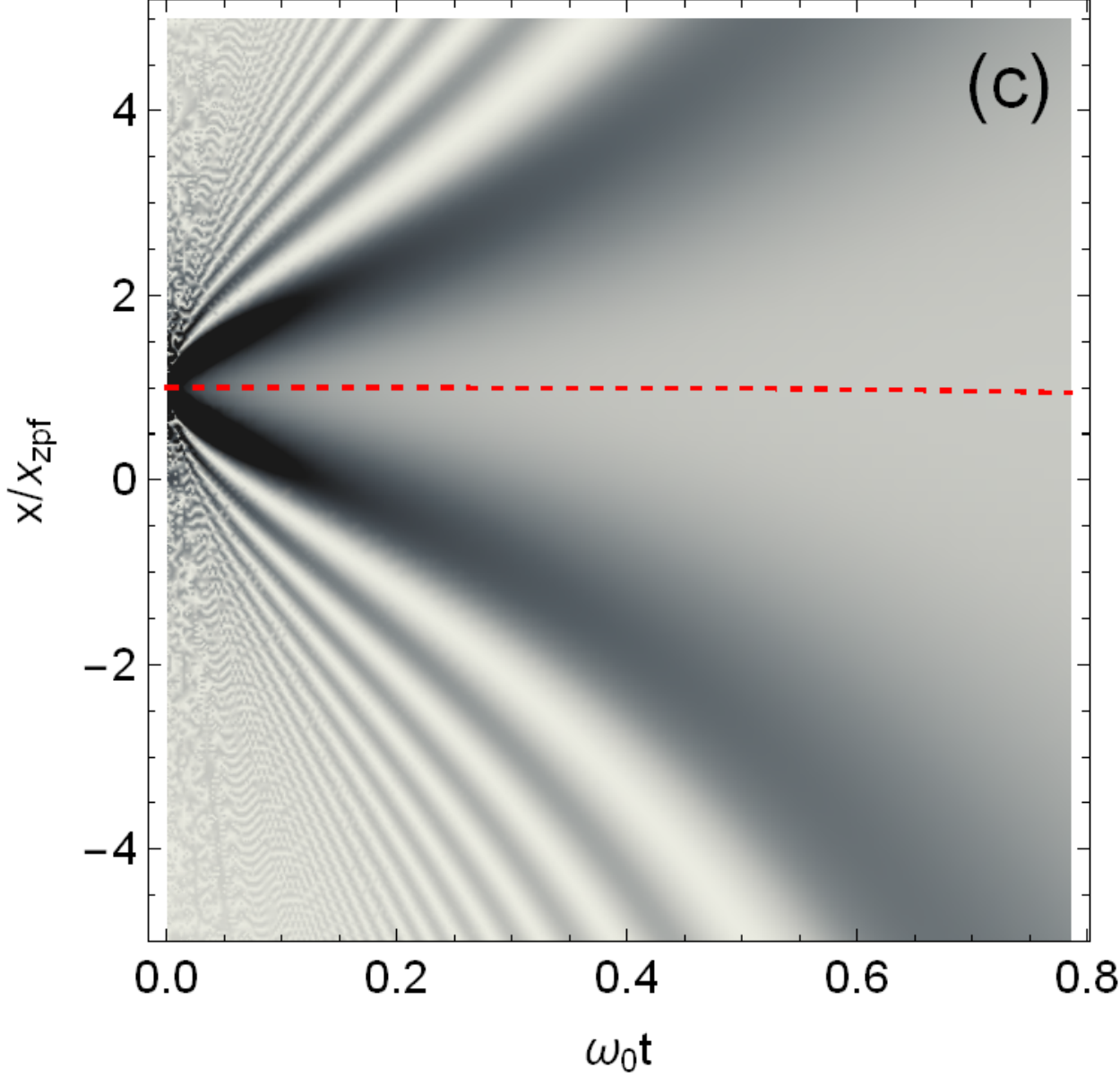}%
\includegraphics[width=0.25 \textwidth]{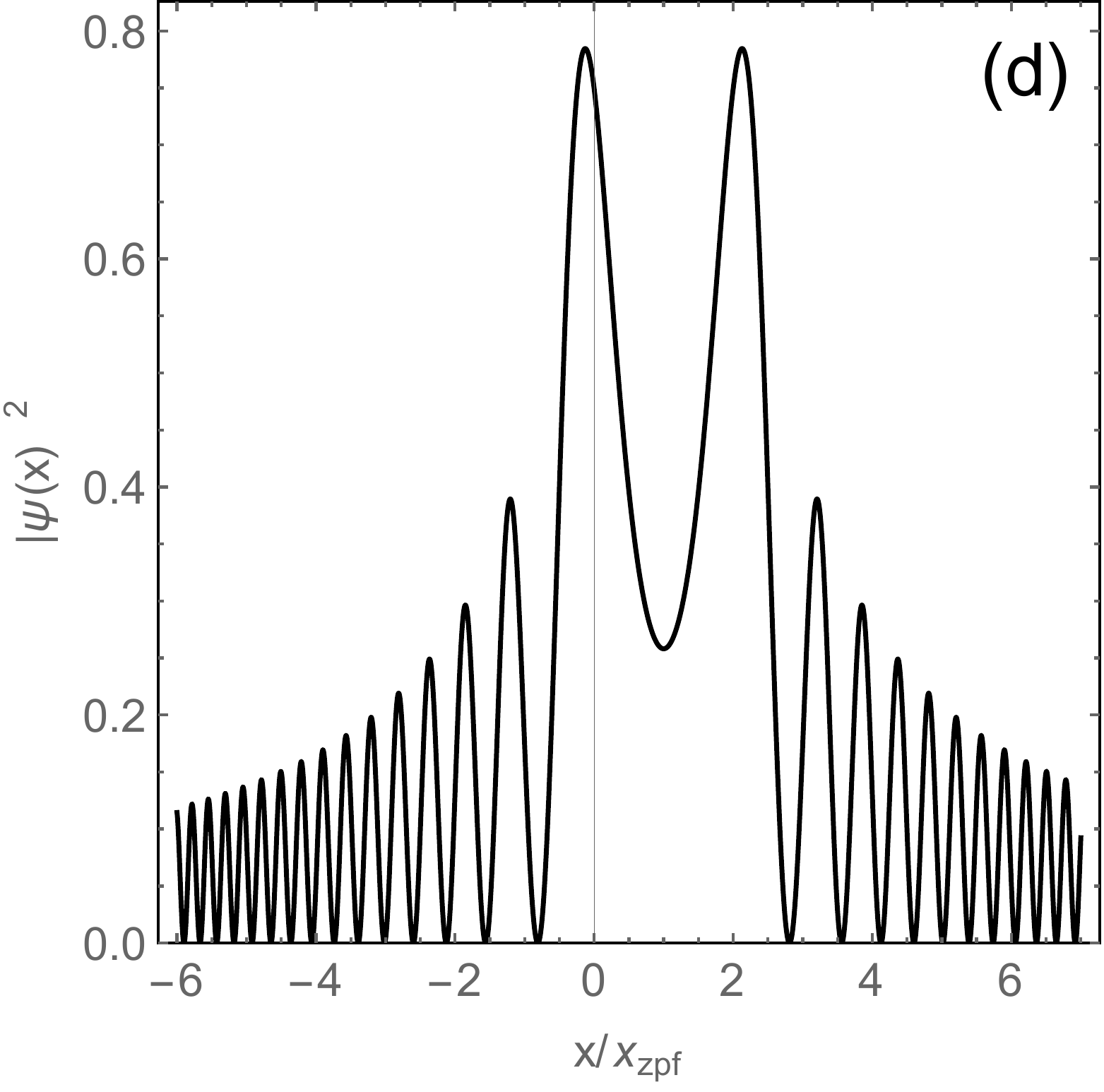}%
\end{center}
\caption{(a) The temporal evolution of parameters $\theta$s.
(b) The real part (solid line) and imaginary part (dashed line) of
the wave function of Eq.(\ref{gx}) for the posmom operator
with $\xi=6$. (c) The evolution of the probability density of the driven and damped
oscillator starting from eigenstate of posmom operator for $\xi=0.5$ (the red dashed line
is the classical orbit). (d) A distribution profile of the wave function at time $t=0.2$.
The other parameters are $X_0=1, \gamma=0.1, \Omega=2$ and $f_0=1$.}
\label{figure13}
\end{figure}

The parameters $\theta$s following Eq.(\ref{thexp}) can turn the original
Hamiltonian into a famous $\hat{X}\hat{P}$ Hamiltonian, which has been intensively
studied due to a close relation with the Riemann hypothesis \cite{Berry,Srednicki,Sierra2}.
Now we should solve the wave function of Hamiltonian Eq.(\ref{XPh}) by
\begin{equation*}
i\frac{\partial }{\partial t}\varphi \left( X,t\right) =\frac{1}{2}K\left(
t\right) \left( \hat{X}\hat{P}+\hat{P}\hat{X}\right) \varphi \left(
X,t\right).
\end{equation*}
By using variable separation method, the wave function will be
\begin{equation}
\varphi \left( X,t\right) =\frac{1}{\sqrt{2\pi \hbar }}e^{-i\xi\int_{0}^{t}K%
\left( \tau \right) d\tau }\frac{1}{\sqrt{\frac{X}{X_{0}}}}e^{i\xi \ln
\left( \frac{X}{X_{0}}\right) },  \label{xp}
\end{equation}%
where $X_0=X(0)$ and the solution is defined on
$X/X_0>0$. Surely, a solution for $X/X_0<0$ can also be constructed \cite{Srednicki,Bernard}.
The parameter $\xi$ is the eigenvalue determined
by the eigenequation of posmom operator \cite{Bernard}
\begin{equation*}
\frac{1}{2}( \hat{X}\hat{P}+\hat{P}\hat{X}) g(X)
=\xi g\left( X\right).
\end{equation*}
The eigenfunction of $g_{\xi}(X)$ is \cite{Bernard}
\begin{equation}
g_{\xi }\left( X\right) =\frac{1}{\sqrt{2\pi \hbar }}\frac{e^{i\xi \ln
\left\vert X\right\vert }}{\sqrt{\left\vert X\right\vert }}, \label{gx}
\end{equation}
which also forms a complete continuous basis to expand any initial state.
If we consider a simple case for $X>0, X_0>0$, the wave function will be
\begin{equation*}
\Psi \left( X,t\right) =A\sqrt{X_{0}}e^{-i\xi \ln X_{0}}e^{-i\xi
\int_{0}^{t}K\left( \tau \right) d\tau }e^{-is}e^{-i\alpha \hat{P}%
}e^{-i\beta X}e^{-i\theta _{+}X^{2}/2}e^{-i\theta _{0}\left(
\hat{X}\hat{P}+\hat{P}\hat{X}\right) /2}e^{-i\theta _{-}\hat{P}^{2}/2}\frac{1%
}{\sqrt{X}}e^{i\xi \ln X}.
\end{equation*}%
Fig.\ref{figure13} demonstrates the solutions of a DDPO
with $a(t)=e^{-2\gamma t}$, $b(t)=e^{2\gamma t}, c(t)=0$ and $K=1$
starting from the eigenstate of Eq.(\ref{gx}). The driving force $f_c(t)$ is defined as
that in Eq.(\ref{dy}).
Surely, we can further simplify the transformation by setting $\theta _{-}=0$
and $\theta _{+}, \theta_0$ can also be solved by the differential equation of
$\dot{\theta}_{+}=a\theta _{+}^{2}-2c\theta _{+}+b$.
The solution in this case is omitted here.

\subsection{A general solution solved by invariant parameter Ks}

Generally, we can safely solve DDPO of Hamiltonian Eq.(\ref{Hs}) by
using any invariant parameters of $K$s.
In this case, the evolution of a wave packet starting from a specific initial state
can be parametrically controlled by the dynamical equations of Eq.(\ref{h4}),
\begin{eqnarray}
\dot{\alpha} &=&c\alpha -a\beta +d, \notag \\
\dot{\beta} &=&b\alpha -c\beta +e, \label{cdy}\\
\dot{s} &=&\frac{1}{2}b\alpha ^{2}-\frac{1}{2}a\beta ^{2}+e\alpha +f,\notag
\end{eqnarray}%
and by Eq.(\ref{gtheta}) combined with Eq.(\ref{Kt}) or Eq.(\ref{Ks}) as
\begin{eqnarray}
\dot{\theta}_{+} &=&a\theta _{+}^{2}-2c\theta _{+}-K_{2}e^{-2\theta _{0}}+b,
\notag \\
\dot{\theta}_{-} &=&ae^{-2\theta _{0}}-K_{2}\theta _{-}^{2}+2K_{3}\theta
_{-}-K_{1},\notag \\ %
\dot{\theta}_{0} &=&K_{2}\theta _{-}-a\theta _{+}+c-K_{3}, \label{gdy} \\
\dot{K}_{1} &=&2cK_{1}-2aK_{3}, \notag \\
\dot{K}_{2} &=&2bK_{3}-2cK_{2}, \notag \\
\dot{K}_{3} &=&bK_{1}-aK_{2}. \notag
\end{eqnarray}%
%--Figure---
\begin{figure}[b]
\begin{center}
\includegraphics[width=0.25 \textwidth]{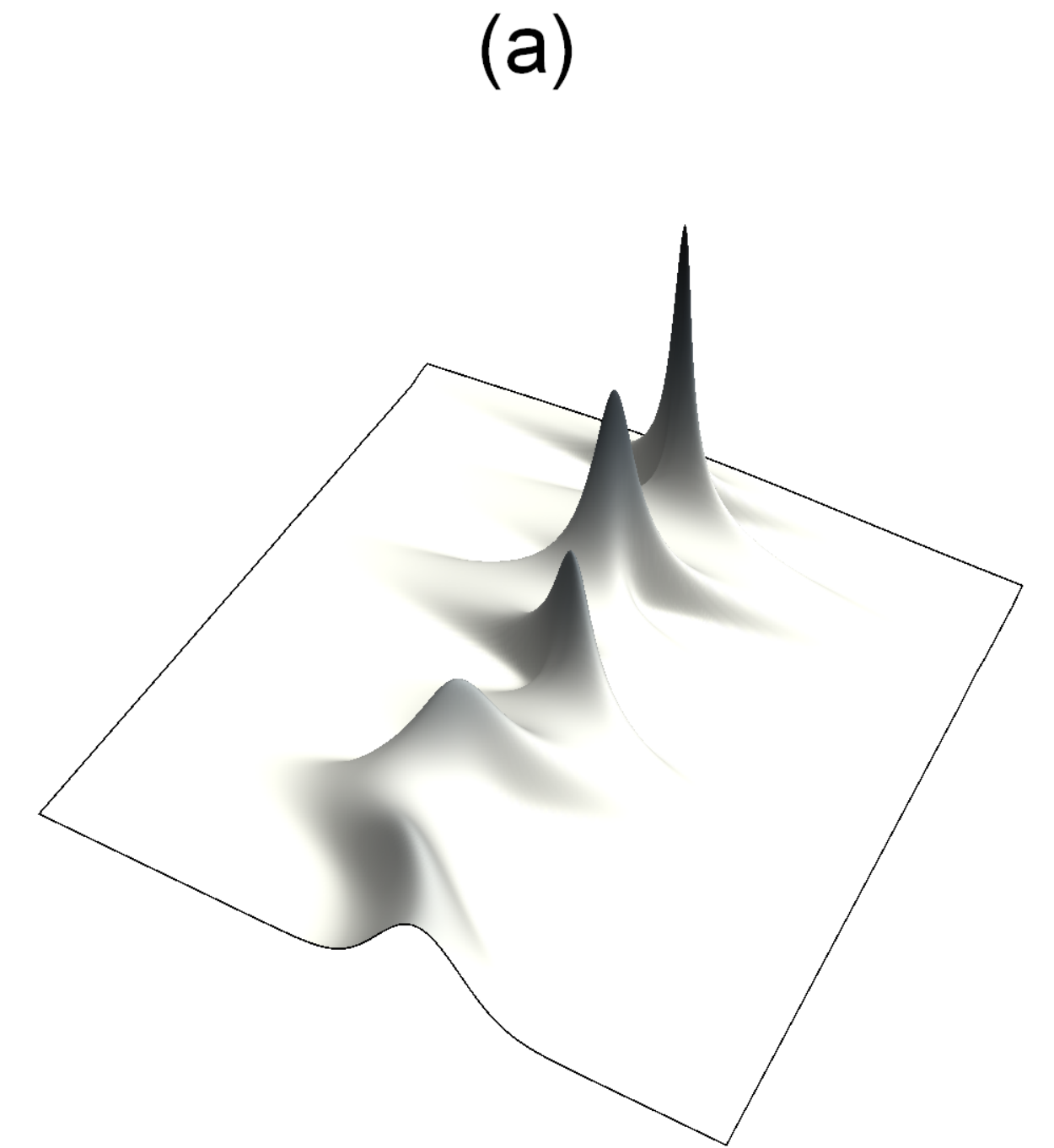}%
\includegraphics[width=0.25 \textwidth]{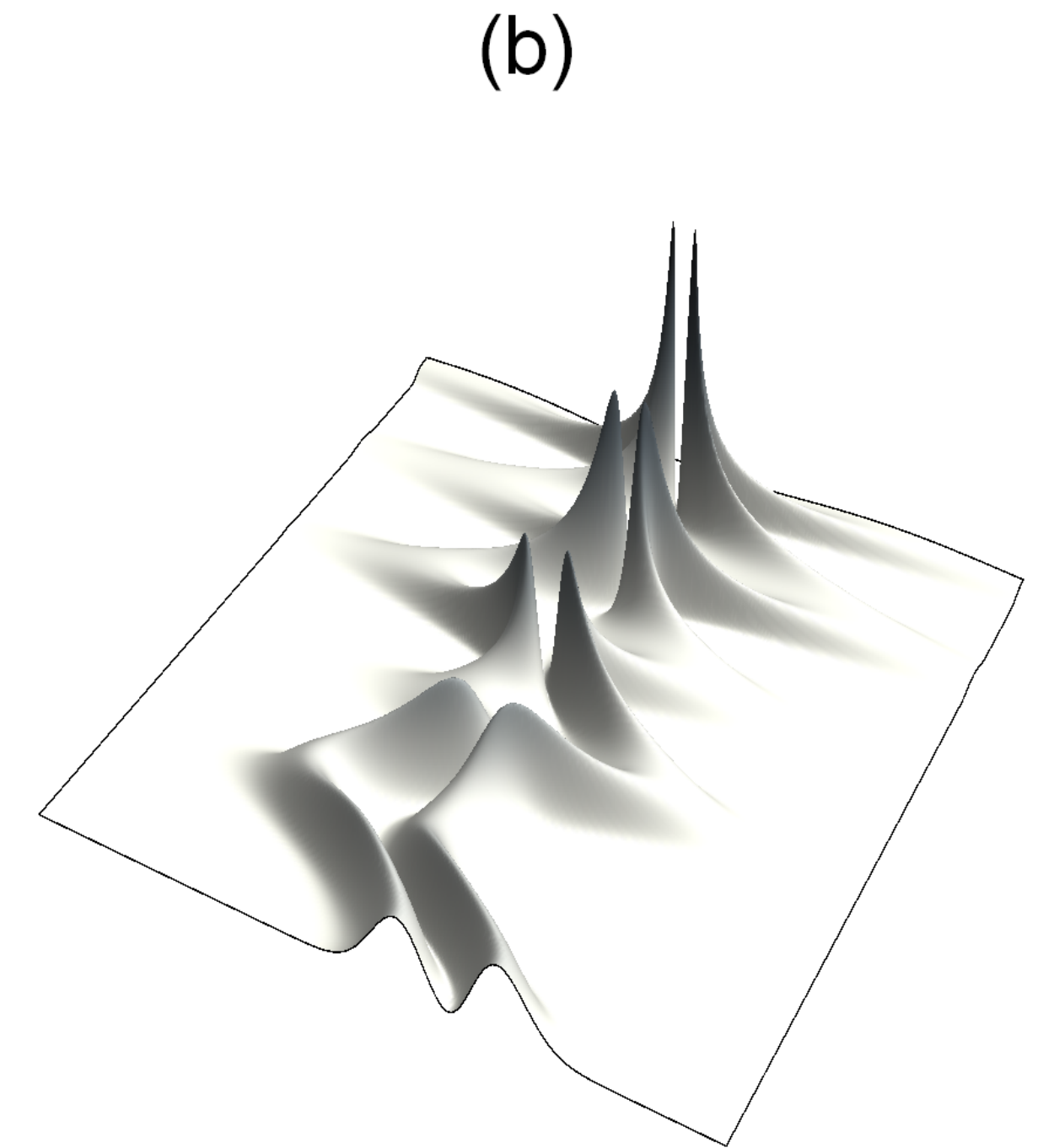}%
\includegraphics[width=0.25 \textwidth]{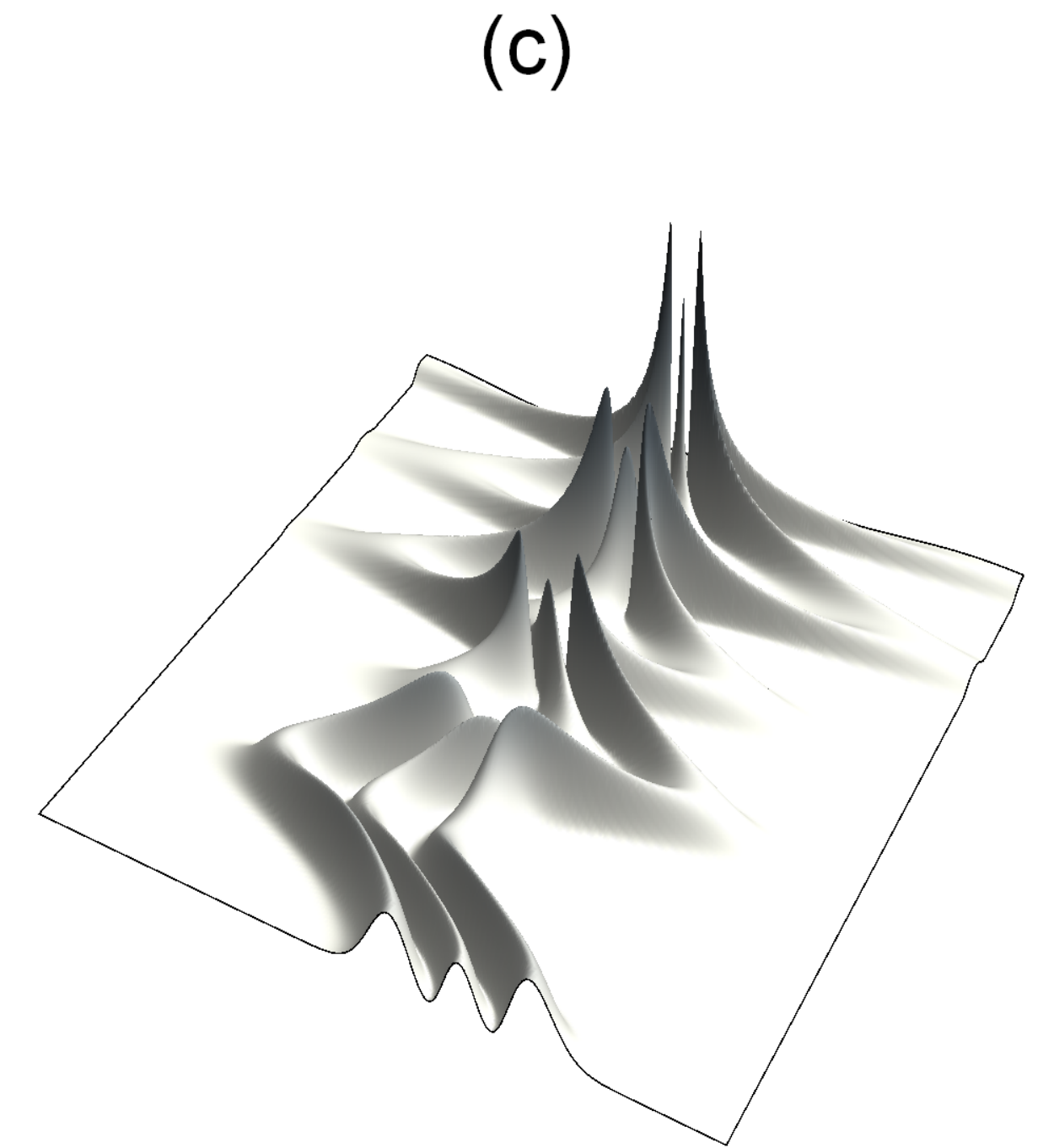}%
\includegraphics[width=0.25 \textwidth]{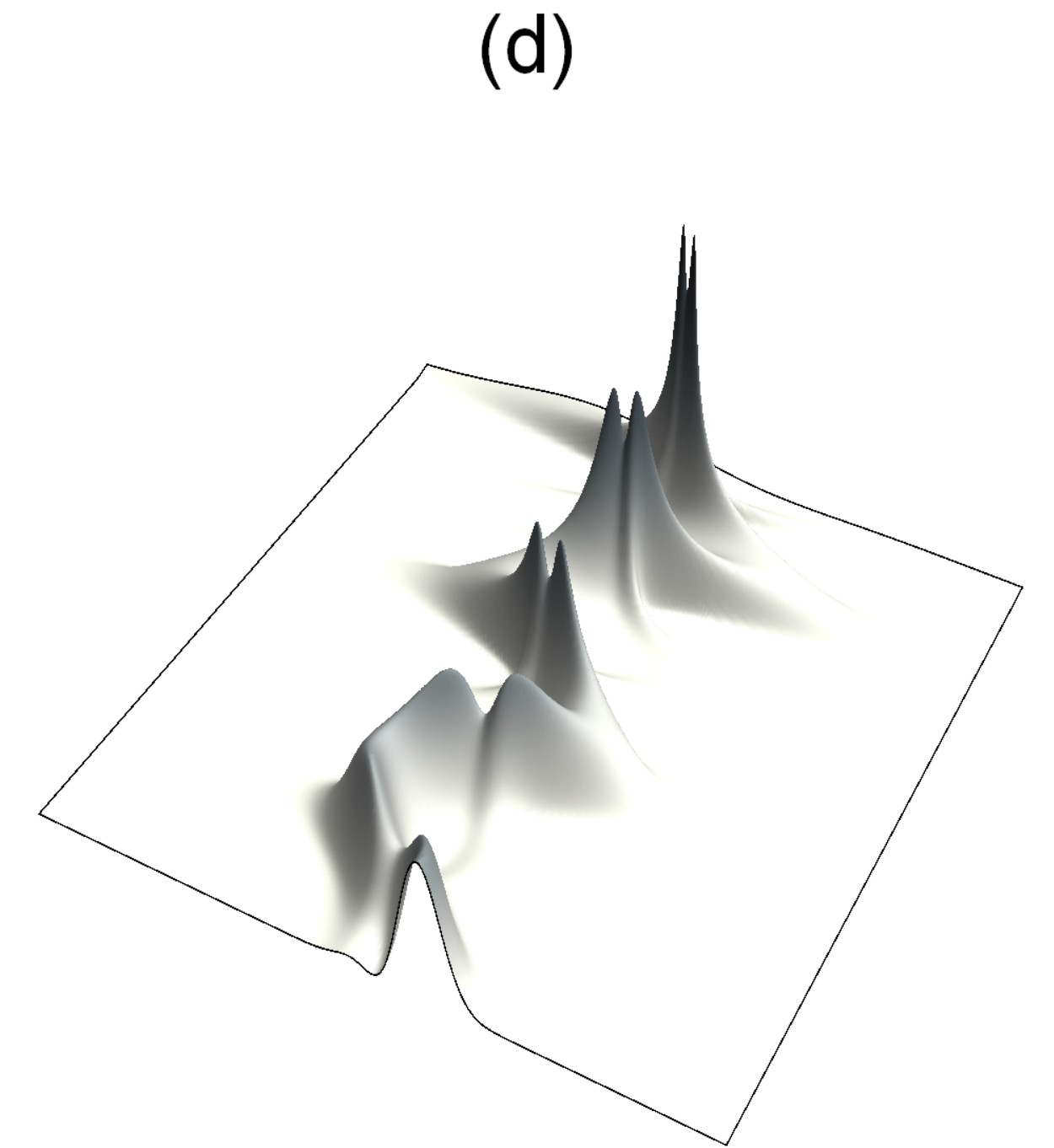}\\%
\includegraphics[width=0.25 \textwidth]{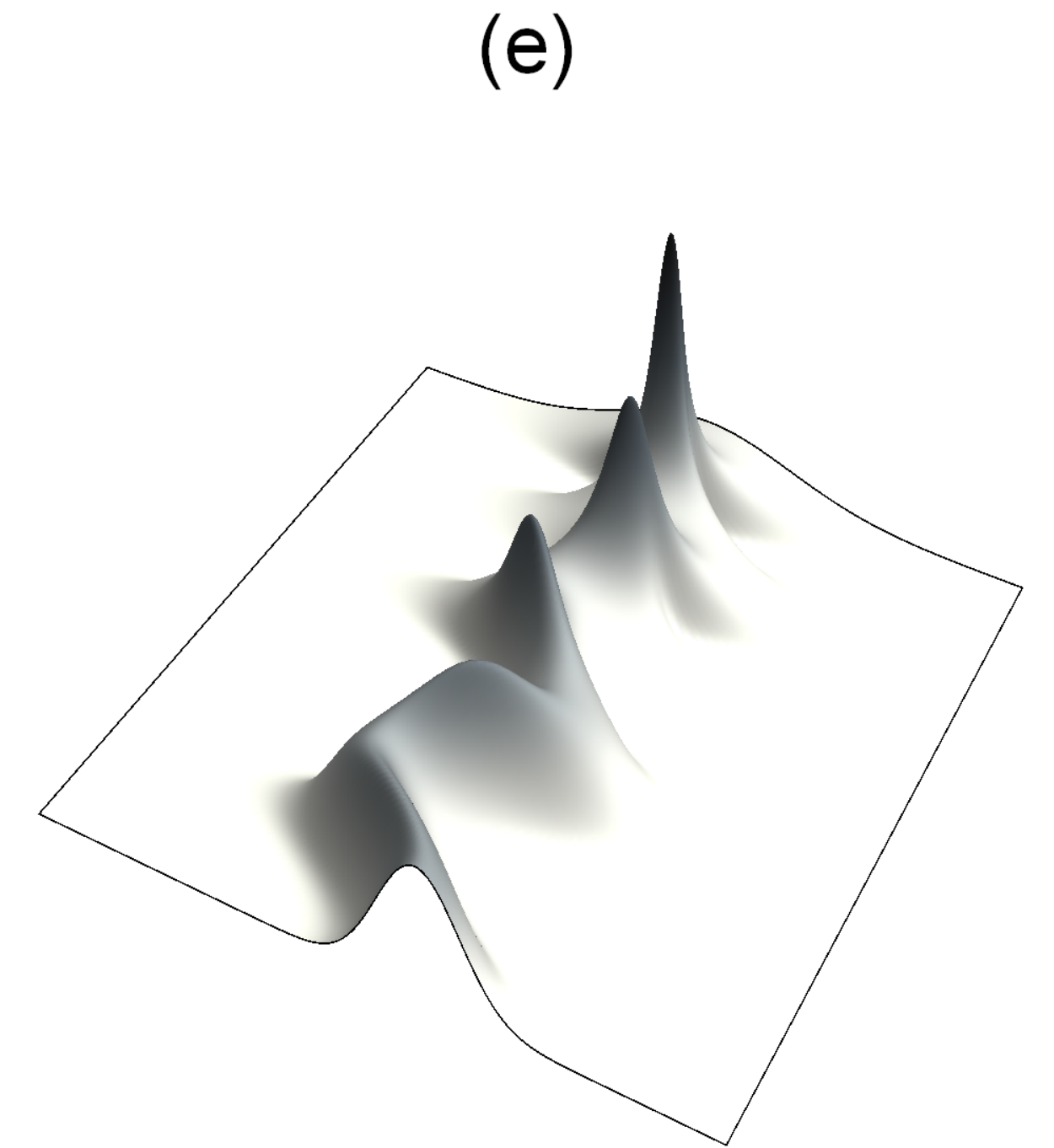}%
\includegraphics[width=0.25 \textwidth]{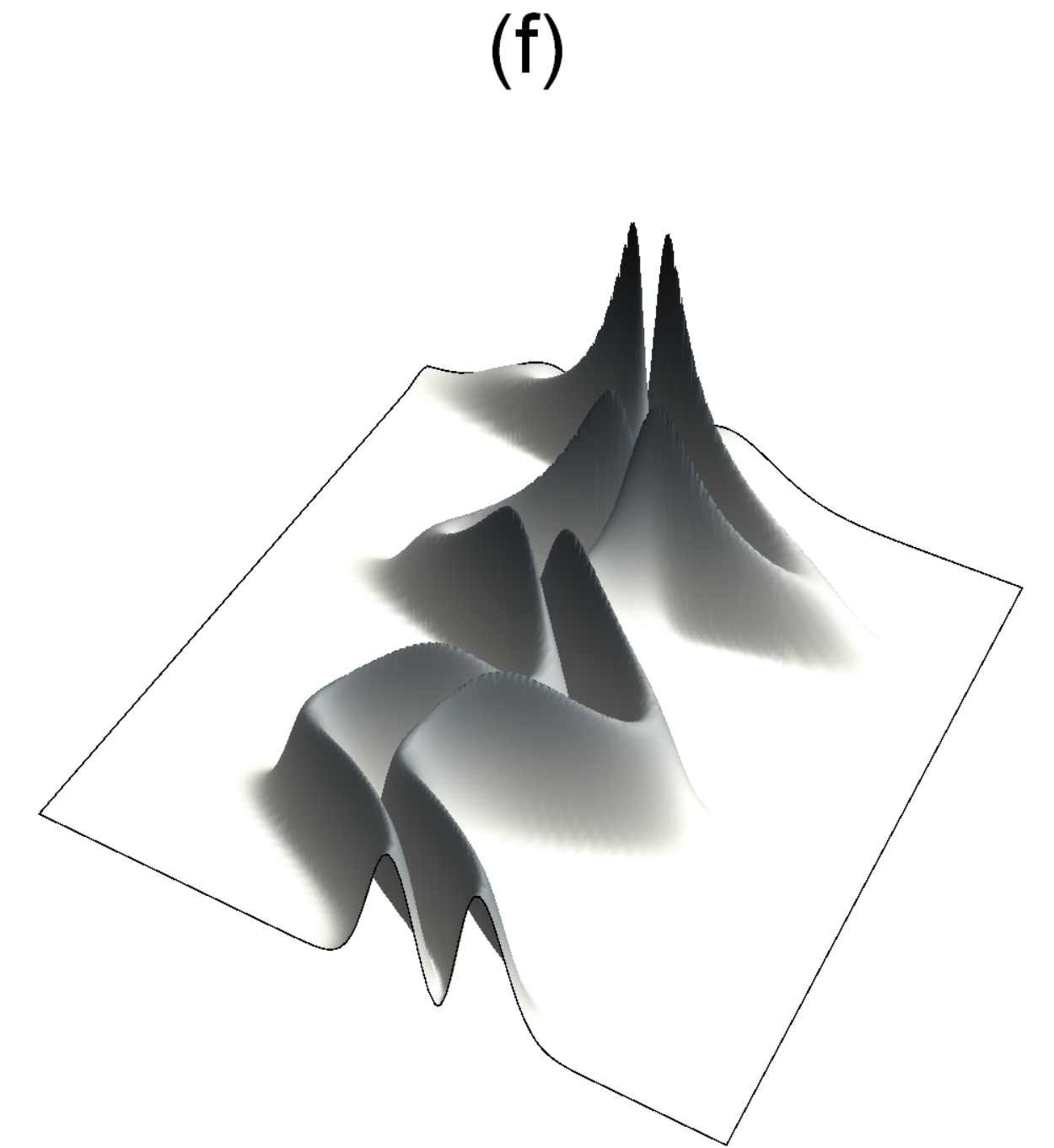}%
\includegraphics[width=0.25 \textwidth]{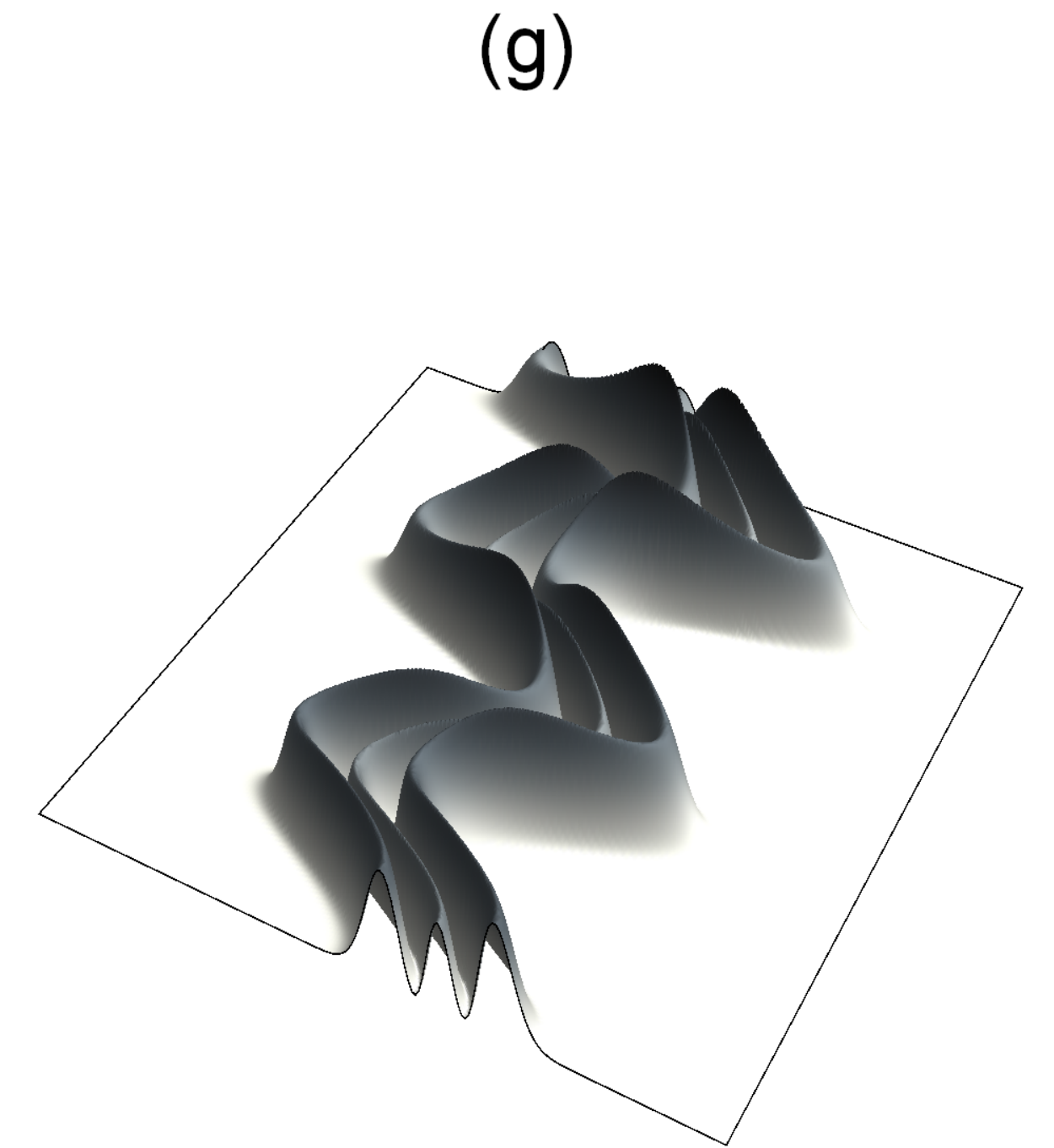}%
\includegraphics[width=0.25 \textwidth]{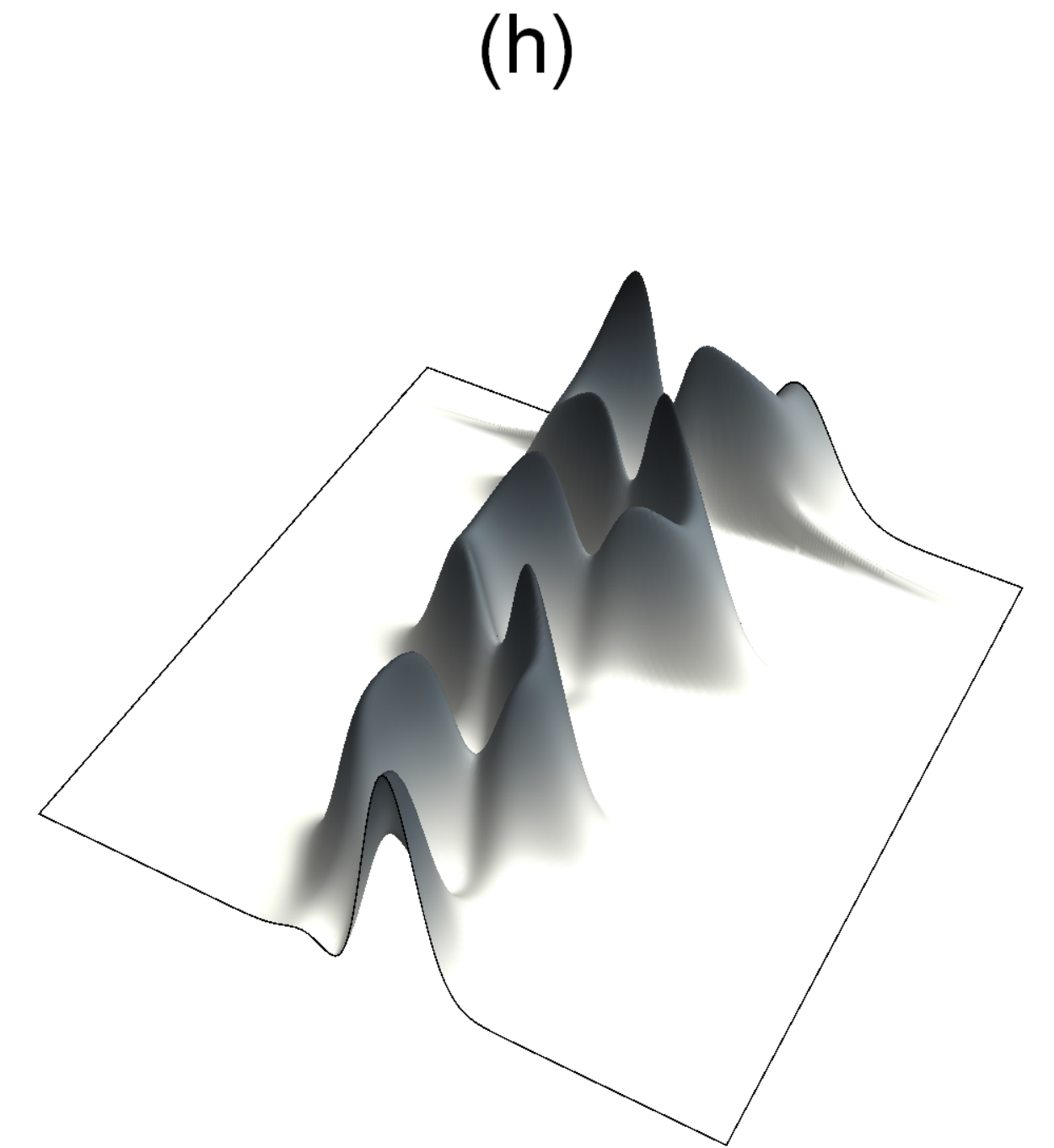}%
\end{center}
\caption{The evolutions of the wave packets of the driven parametric
oscillator starting from initial states of (a)(e)$|0\rangle$, (b)(f) $|1\rangle$,
(c)(g) $|2\rangle$ and (d)(h)$(|0\rangle+|1\rangle)/\sqrt{2}$.
The driving amplitude is $f_0=0.2$ and the driving frequency is $\Omega=0.8$. The
other parameters are the same as that in Fig.\ref{figure11}.}
\label{figure14}
\end{figure}
Therefore, the wave function determined by the above dynamic
parameters can be written as
\begin{equation*}
\Psi \left( X,t\right) =e^{-is\left( t\right) }e^{-i\alpha \left( t\right)
\hat{P}}e^{-i\beta \left( t\right) \hat{X}}e^{-i\theta _{+}\hat{X}%
^{2}/2}e^{-i\theta _{0}\left( \hat{X}\hat{P}+\hat{P}\hat{X}\right)
/2}e^{-i\theta _{-}\hat{P}^{2}/2}\psi_0 \left( X,t\right),
\end{equation*}
where the initial state $\psi_0(X,t)$ can be expanded by any basis of Eq.(\ref{ini}).
As discussed in previous sections, the classical dynamics of Eq.(\ref{cdy}) only
causes the overall shifts and Eq.(\ref{gdy}) will lead to the deformations of an initial wave packet
through the operations of translation, rotation and squeezing or dilation in the real space.

As an example, in Fig.\ref{figure14}, we demonstrate the time evolution of the wave packets for a DDPO
of Eq.(\ref{Hs}) with $a(t)=e^{-2\gamma t}$, $b(t)=e^{2\gamma t}$, $c(t)=0$ and $e(t)=-e^{2\gamma t}f_c(t)$
in the upper frames of Fig.\ref{figure14}(a)-(d). This example is still based on the model of Eq.(\ref{Hddd})
and the shortcoming of this model is that the damping will induce an increasing effective mass
which will bring nonphysical wave functions. The evolutions of the wave function for a DDPO
with varying mass of $a(t)=1/M(t)$, the time-dependent frequency of $b(t)=M(t)\Omega_{0} ^{2}(t)$,
$c(t)=0$ and $e(t)=f_c(t)$ are shown in the lower frames of Fig.\ref{figure14}(e)-(h). This model
adopts the same controlling parametric functions of $M(t)$ and $\Omega_{0}(t)$ as that defined in Eq.(\ref{mo}).
The time evolutions of the wave packets starting from different number states shown in Fig.\ref{figure14}
clearly exhibit the geometric influences of the parameters on the wave packets through the operations of
translation, rotation and squeezing or dilation. The classical resonance of the driving force for a
damped parametric oscillator with $f_c=f_0\cos(\Omega t+\phi)$ ($\Omega\sim 1$) is only governed by
Eq.(\ref{cdy}) (which decouples from Eq.(\ref{gdy})) and it doesn't directly induce quantum transitions
between different quantum states as normally expected. Since Eq.(\ref{cdy}) and Eq.(\ref{gdy})
are two systems of nonlinear dynamical equations, their solutions will be very sensitive to the initial
states in a chaotic parametric region. Therefore, the affiliated parametric equations of Eq.(\ref{cdy})
and Eq.(\ref{gdy}) can be used to define the quantum chaos for a quantum system when its parametric
equations are chaotic. Generally, the initial forms of the parameters $\alpha, \beta, s$, $\theta$s and $K$s can be randomly
chosen, but for a practical control problem, they often depend on
the initial parameters of the Hamiltonian, i.e., $a(0), b(0), c(0), d(0),
e(0)$ and $f(0)$, or they can be properly designed through the initial states and the target states
in a controlled system.

\section{Conclusions}

As the time-dependent system has attracted considerable interest due to its potential
applications in many fields for the quantum control problems \cite{Sierra},
we intensively investigate the model of DDPO described by a general time-dependent
quadratic Hamiltonian in a unified algebraic framework of Lie transformation. In order
to facilitate a comparison with the quantum dynamics, we first give a brief review on
the classical behaviors of DDHO to lay down a basic picture for the classical dynamics.
Then a time-dependent Lie transformation, Floquet U-transformation (FUT), is
introduced to provide a simple but consistent approach to solve the time-dependent
quadratic Hamiltonian system beyond the time-dependent perturbation theory and the
quantum adiabatic theory.

Based on a closed Lie algebra of the Hamiltonian, our study shows that
the state evolution of a time-dependent system of Eq.(\ref{H0}) can be
mapped into a 6-dimensional parametric space of $h(4)\bigoplus su(2)$ and
naturally be decomposed into a classical part and a quantum part
governed by their independent parametric equations of Eq.(\ref{h4}) and
Eq.(\ref{dyth}), respectively. This method reveals a close connection between
the classical and quantum dynamics of a wave-packet in a driven and damped system.
We demonstrate that the transformation parameters of Lie operators can selectively
control the evolution of a wave function in a separated parametric spaces,
$h(4)$ and $su(2)$, through the dynamical operations of translation, rotation and
squeezing on the wave packet. The Lie transformation method can easily discriminate
the classical resonance from a quantum one to reveal that the classical
resonance is governed by the dynamical equations of Eq.(\ref{h4}) or
Eq.(\ref{fm}) which only introduces an overall translation of the wave packet, while
the quantum resonance induces quantum transitions between different internal
states which dramatically modify the shape of the wave packet.

Although the Lie transformation method can exactly solve the general time-dependent
quadratic Hamiltonian, it can not always give an optimal solution to a dissipative system.
For the sake of eliminating the pathological properties of the solutions
obtained by this method, we combine the time-dependent Lie transformation with the
Lewis-Riesenfeld invariant method by introducing new dynamic parameters of $K$s
and this technique enables us to freely transform the Hamiltonian belonging to certain algebra into another.
In order to convert the transformed Hamiltonian Eq.(\ref{Hf}) into an invariant of the
system, a dynamical equation of Eq.(\ref{Kt}) is derived to determine $K$s.
Anyway, as the invariant of the system is not unique, the parametric
functions of $K$s can be properly chosen for different physical problems,
or $K$s can be consistently constructed to design different parametric controls on a
quantum system to get specific target states. In order to illustrate the applications of this
approach, we proceed with some typical examples to select proper
combinations of $K$s in order to find optimal solutions to different systems.
Our study not only recovers the main solutions of these typical models, but
also presents new features on the state dynamics through constructing invariant
Hamiltonians (a reverse Lie transformation) for these systems.
Surely, this Lie transformation method takes advantage of studying the quantum control
problems beyond the adiabatic theory (such as in the non-adiabatic shortcut control \cite{Chen}) and
can be easily generalized to other time-dependent Hamiltonians
beyond a quadratic form if proper algebraic realizations of
the system can be found.

\begin{acknowledgements}
This work is supported by the National Science Foundation (Grant No.11447025)
and the Scientific Research Foundation for Returned Overseas Chinese Scholars,
State Education Ministry.
\end{acknowledgements}

\appendix
\section{The FUT transformations for $\hat{\mathcal{H}}_U$}
\label{appA}
The following similarity transformations are used to derive Eq.(\ref{U2}):%
\begin{eqnarray*}
&&e^{i\theta _{+}\hat{X}^{2}/2}\hat{P}e^{-i\theta _{+}\hat{X}^{2}/2}=\hat{P}%
-\theta _{+}\hat{X}, \\
&&e^{i\theta _{-}\hat{P}^{2}/2}\hat{X}e^{-i\theta _{-}\hat{P}^{2}/2} =\hat{X}%
+\theta _{-}\hat{P}, \\
&&e^{i\theta _{0}\left( \hat{X}\hat{P}+\hat{P}\hat{X}\right) /2}\hat{P}%
e^{-i\theta _{0}\left( \hat{X}\hat{P}+\hat{P}\hat{X}\right) /2} =\hat{P}%
e^{-\theta _{0}}, \\
&&e^{i\theta _{0}\left( \hat{X}\hat{P}+\hat{P}\hat{X}\right) /2}\hat{X}%
e^{-i\theta _{0}\left( \hat{X}\hat{P}+\hat{P}\hat{X}\right) /2} =\hat{X}%
e^{\theta _{0}}.
\end{eqnarray*}

In order to give a general derivation of Eq.(\ref{U2}), we calculate
the FUT transformation of
\begin{equation*}
\mathcal{\hat{H}}_{U}=\hat{U}^{-1}\left( t\right) \hat{\mathcal{H}}(t)\hat{U}%
\left( t\right) -i \hat{U}^{-1}\left( t\right) \frac{\partial \hat{U}%
\left( t\right) }{\partial t}
\end{equation*}
for $e^{-i\theta _{+}\hat{X}^{2}/2}$, $e^{-i\theta _{-}\hat{P}^{2}/2}$ and
$e^{-i\theta _{0}\left( \hat{X}\hat{P}+\hat{P}\hat{X}\right) /2}$,
respectively, on the scaled standard
quadratic Hamiltonian of%
\begin{equation}
\mathcal{\hat{H}}=\frac{1}{2}a\hat{P}^{2}+\frac{1}{2}b\hat{X}^{2}+\frac{1}{2}%
c\left( \hat{X}\hat{P}+\hat{P}\hat{X}\right) .  \label{Hi}
\end{equation}

(1) The first transformation is for
\begin{equation}
\hat{U}_+=e^{-i\theta _{+}\hat{X}^{2}/2},
\end{equation}%
and the FUT on Hamiltonian Eq.(\ref{Hi}) gives%
\begin{equation}
\mathcal{\hat{H}}_{U_+}=\frac{1}{2}a\hat{P}^{2}+\frac{1}{2}\left( a\theta
_{+}^{2}-\dot{\theta}_{+}-2c\theta _{+}+b\right) \hat{X}^{2}+\frac{1}{2}%
\left( c-a\theta _{+}\right) \left( \hat{X}\hat{P}+\hat{P}\hat{X}\right) .
\end{equation}

(2) The second is for%
\begin{equation}
\hat{U}_0=e^{-i\theta _{0}\left( \hat{X}\hat{P}+\hat{P}\hat{X}\right) /2},
\end{equation}%
and the FUT on Hamiltonian Eq.(\ref{Hi}) results in
\begin{equation}
\mathcal{\hat{H}}_{U_0}=\frac{1}{2}ae^{-2\theta _{0}}\hat{P}^{2}+\frac{1}{2}%
be^{2\theta _{0}}\hat{X}^{2}+\frac{1}{2}\left( c-\dot{\theta}_{0}\right)
\left( \hat{X}\hat{P}+\hat{P}\hat{X}\right) .
\end{equation}

(3) The last one is%
\begin{equation}
\hat{U}_{-}=e^{-i\theta _{-}\hat{P}^{2}/2},
\end{equation}%
and FUT on the Hamiltonian Eq.(\ref{Hi}) leads to
\begin{equation}
\mathcal{\hat{H}}_{U_-}=\frac{1}{2}\left( a+2c\theta _{-}+b\theta _{-}^{2}-%
\dot{\theta}_{-}\right) \hat{P}^{2}+\frac{1}{2}b\hat{X}^{2}+\frac{1}{2}%
\left( b\theta _{-}+c\right) \left( \hat{X}\hat{P}+\hat{P}\hat{X}\right) .
\end{equation}
Then the combination of the above transformations can give any results of successive
FUTs on Hamiltonian Eq.(\ref{Hi}). For the successive transformations of
\begin{equation}
\hat{U}=\hat{U}_{+}\hat{U}_0\hat{U}_{-}=e^{-i\theta _{+}\hat{X}^{2}/2}
e^{-i\theta _{0}\left( \hat{X}\hat{P}+\hat{P}\hat{X}\right) /2}
e^{-i\theta _{-}\hat{P}^{2}/2},
\end{equation}%
we can easily combine the above three FUTs to obtain Eq.(\ref{U2}).

\end{document}